\documentclass[%
 reprint,
 amsmath,amssymb,
 aps,superscriptaddress
]{revtex4-2}

\usepackage[printonlyused]{acronym}
\usepackage[hidelinks]{hyperref}
\usepackage{graphicx}
\usepackage{subcaption}
\usepackage{dcolumn}
\usepackage{makecell}
\usepackage{array}
\usepackage{rotating}
\usepackage{tabularx}
\usepackage{booktabs}
\usepackage{braket}
\usepackage{bm}
\usepackage{framed}
\usepackage{amsmath}
\hypersetup{breaklinks=true}
\usepackage[english]{babel}

\usepackage{microtype}
\usepackage[table]{xcolor}
\usepackage{csquotes}
\usepackage{orcidlink}
\usepackage{endnotes}

\begin{document}

\preprint{APS/123-QED}

\title{Visualizing Quantum States: A Pilot Study on Problem Solving in Quantum Information Science Education}

\author{Jonas Bley}
\affiliation{%
Department of Physics and Research Center OPTIMAS, RPTU Kaiserslautern-Landau\\
Kaiserslautern, 67663, Germany
}%

\author{Eva Rexigel}
\affiliation{%
Department of Physics and Research Center OPTIMAS, RPTU Kaiserslautern-Landau\\
Kaiserslautern, 67663, Germany
}%

\author{Alda Arias}
\affiliation{%
Department of Physics and Research Center OPTIMAS, RPTU Kaiserslautern-Landau\\
Kaiserslautern, 67663, Germany
}%

\author{Lars Krupp}
\affiliation{%
Department of Computer Science and Research Initiative QC-AI, RPTU Kaiserslautern-Landau\\
Kaiserslautern, 67663, Germany
}%
\affiliation{%
Embedded Intelligence, German Research Center for Artificial Intelligence\\
Kaiserslautern, 67663, Germany
}%

\author{Steffen Steinert}
\affiliation{%
Faculty of Physics, Chair of Physics Education, Ludwig-Maximilians-Universit\"at M\"unchen (LMU Munich)\\
Munich, 80539, Germany
}%
\affiliation{%
Department of Electrical and Computer Engineering, RPTU Kaiserslautern-Landau\\
Kaiserslautern, 67663, Germany
}%

\author{Nikolas Longen}
\affiliation{%
Department of Computer Science and Research Initiative QC-AI, RPTU Kaiserslautern-Landau\\
Kaiserslautern, 67663, Germany
}%

\author{Paul Lukowicz}
\affiliation{%
Department of Computer Science and Research Initiative QC-AI, RPTU Kaiserslautern-Landau\\
Kaiserslautern, 67663, Germany
}%
\affiliation{%
Embedded Intelligence, German Research Center for Artificial Intelligence\\
Kaiserslautern, 67663, Germany
}%

\author{Stefan K\"uchemann}
\affiliation{%
Faculty of Physics, Chair of Physics Education, Ludwig-Maximilians-Universit\"at M\"unchen (LMU Munich)\\
Munich, 80539, Germany
}%

\author{Jochen Kuhn}
\affiliation{%
Faculty of Physics, Chair of Physics Education, Ludwig-Maximilians-Universit\"at M\"unchen (LMU Munich)\\
Munich, 80539, Germany
}%

\author{Maximilian Kiefer-Emmanouilidis}
\affiliation{%
Department of Physics and Research Center OPTIMAS, RPTU Kaiserslautern-Landau\\
Kaiserslautern, 67663, Germany
}%
\affiliation{%
Department of Computer Science and Research Initiative QC-AI, RPTU Kaiserslautern-Landau\\
Kaiserslautern, 67663, Germany
}%
\affiliation{%
Embedded Intelligence, German Research Center for Artificial Intelligence\\
Kaiserslautern, 67663, Germany
}%

\author{Artur Widera}
\affiliation{%
Department of Physics and Research Center OPTIMAS, RPTU Kaiserslautern-Landau\\
Kaiserslautern, 67663, Germany
}%

\date{\today}

\begin{abstract}
In the rapidly evolving interdisciplinary field of quantum information science and technology, a major obstacle is the need to understand advanced mathematics to solve complex problems. Current findings in educational research suggest that incorporating visualizations into problem-solving settings can have beneficial effects on students' performance and cognitive load compared to relying solely on symbolic problem-solving content. Visualizations like the (dimensional) circle notation enable us to represent not only single-qubit but also more complex multi-qubit states, entanglement, and quantum algorithms. In this pilot study, we aim to take an initial step toward identifying the contexts in which students benefit from the presentation of visualizations of single- and multi-qubit systems in addition to mathematical formalism. 
For this purpose, we propose a set of test items and a comprehensive methodology to assess students' performance and cognitive load when solving problems. This is a pilot investigation with a large breadth of questions intended to generate hypotheses and guide larger-scale, more focused studies in the future. Specifically, we compare two approaches: using the mathematical-symbolic Dirac Notation alone and using it in combination with the (dimensional) circle notation. In surveys in one-, two- and three-qubit systems, we gather qualitative data from five, five and two think-aloud interviews, identifying problems that students encounter and their problem-solving strategies. In addition, we analyze quantitative data (performance and cognitive load) from 23, 27 and 17 participants in surveys on one-, two- and three-qubit systems recruited mainly from our quantum computing lectures. We find that most of the test items are appropriate for a heterogeneous target group, as they can differentiate between participants in terms of performance and time taken. In general, the A-B crossover structure of the study is suitable for investigating the benefits of visualization for problem solving with regard to length and feasibility. Future studies should, however, be narrower in scope, given the observable dependence on student characteristics and context, with particular interest lying in the further investigation of the Hadamard gate, the CNOT gate, and Entanglement in multi-qubit systems.
\end{abstract}

\maketitle

\section{Introduction}\label{sec1}

The interdisciplinary nature of the field of \ac{qist} necessitates concerted efforts to educate a workforce with diverse academic backgrounds~\cite{PhysRevPhysEducRes.18.010150}. This implies that the quantum education community needs to find ways to introduce the field of \ac{qist} to a broad audience. As a foundational concept, qubit systems are inherently interdisciplinary and span all of \ac{qist}. The \enquote{qubit-approach} can therefore be used to abstractly introduce a new, heterogeneous audience to the field. In consequence, the question of how best to make this approach as accessible as possible is raised. Fortunately, existing didactical theory provides a foundation from which to draw ideas. 

Visualization has been shown to be a useful tool in \ac{stem} education~\cite{gilbert_visualization_2005,Evagorou2015,doi:10.1080/09500693.2022.2140020,rau2017conditions,opfermann2017multiple,munfaridah2021use}. In \ac{qist} education in particular, visualization has been proposed to be useful for learning~\cite{10.1145/3545945.3569836,goorney2023framework}. The field of \ac{qist} is based on the utilization of systems of multiple qubits, but the effectiveness of presenting visualizations of qubit states in problem solving, in terms of performance~\cite{Schewior2024} and cognitive load~\cite{Sweller1998}, is currently unknown. Understanding the contexts and conditions under which learners benefit from such visualizations would enable us to provide \ac{qist} educators with empirically grounded recommendations for how to most effectively integrate visualizations into teaching. This is an important step towards solving the challenge of high heterogeneity of target groups in \ac{qist} education. It remains unclear whether, and under which conditions, learning and problem solving are supported by providing visualizations alongside the mathematical–symbolic representation (e.g., the \ac{dn}) in \ac{qist} education. One theoretical construct that may offer explanatory insight is that of representational competence~\cite{Daniel2018}. We identify three overarching questions in the context of visualizations in \ac{qist} education: In what contexts does incorporating visualizations support students' performance in task solving on quantum operations and entanglement? How does the inclusion of visualizations affect cognitive load? How are performance and cognitive load affected by representational competence?

In order to answer these questions, appropriate and validated test instruments on task solving in \ac{qist} are needed. However, validated test instruments in the context of \ac{qist} education, and in particular on quantum operations and entanglement, are scarce in single-qubit systems and do, to our knowledge, not exist for multi-qubit systems. As an essential first step in the investigation of these questions, we conducted a pilot study validating relevant test instruments to investigate the overarching questions. In this way, we initiate an investigation into whether visualizations provide a benefit to solve problems in \ac{qist} when provided in addition to the \ac{dn} and under what circumstances such benefits occur. The instruments, survey design, and preliminary results presented here lay the groundwork for future, more targeted studies that can build on this foundation. 

The paper is structured as follows. First, we introduce the research background in quantum education and the visualizations used in \ref{sec:research_background}. We follow with a description of the study design in Section \ref{sec:res_ques} and describe the creation of the test items in Section \ref{sec:design}. The preliminary results are summarized, and the test instruments are discussed with respect to their suitability to the target group and indications of effects worth exploring further in Section \ref{sec:discussion}. We conclude and provide an outlook for future research in Section \ref{sec:conclusion}.

\section*{List of acronyms}





\begin{acronym}[ANOVA]
\acro{cf}[{CF}]{Competence Framework}

\acro{qist}[QIST]{Quantum Information Science and Technology}

\acro{ecl}[ECL]{Extraneous Cognitive Load}

\acro{gcl}[GCL]{Germane Cognitive Load}

\acro{cn}[CN]{Circle Notation}

\acro{dn}[DN]{Dirac Notation}

\acro{dcn}[DCN]{Dimensional Circle Notation}

\acro{icl}[ICL]{Intrinsic Cognitive Load}

\acro{deft}[DeFT]{Design, Functions, Tasks}

\acro{mer}[MERs]{Multiple External Representations}

\acro{clt}[CLT]{Cognitive Load Theory}

\acro{stem}[STEM]{Science, Technology, Engineering \& Mathematics}
\end{acronym}

\section{Research background}\label{sec:research_background}

The research background of this study includes the functions that \ac{mer} serve (Section \ref{sec:functions}), an overview of educational research on representations in quantum education and visualizations of multi-qubit quantum states and entanglement. This encompasses the two representations investigated in the present study, \ac{cn} and \ac{dcn} (Section \ref{sec:quantum_vis}), as well as the associated measures, including cognitive load, multiple-choice test design, and time taken (Section \ref{sec:tasks}).

\subsection{Learning and Problem solving with visualizations in \ac{qist}}\label{sec:functions}

Based on educational research, it is well known that learning and problem solving in \ac{stem} can be supported by focusing not only on text-based and mathematical-symbolic representations (e.g., written text and formulas) but also on visualizations (e.g., pictures and diagrams)~\cite{Ainsworth2021, Hu2021}. From a theoretical perspective, the benefit of learning with text accompanied by visualization can be explained by more efficient information processing and the dual-channel structure of the working memory~\cite{Mayer_2021,Schnotz.2021}. When learning with \acl{dn} alone, the processing load of information is solely on the verbal channel of working memory. On the other hand, when learning with text and visualization, both the verbal and visual channels of working memory are activated and integrated in the construction of coherent mental schemata~\cite{Schnotz.2021,Mayer_2021}. This distributed information processing across two channels reduces the load on each single channel and, thus, the risk of cognitive overload.

Within her \ac{deft} framework, Ainsworth~\cite{ainsworth2006} has formulated three main functions of \ac{mer} to support learning. First, \ac{mer} can \textit{complement each other} by containing different information or supporting different processes. Second, they can \textit{constrain each other}, e.g., by familiarity of the learners with a subset of the \ac{mer} or inherent properties. Third, incorporating \ac{mer} can \textit{construct deeper understanding} as learners are confronted with the abstraction of the underlying knowledge structures, the extension of knowledge to an unknown representation or improving the understanding of the relations between different representations.

However, incorporating additional representations can also impose additional cognitive load on students, as they must understand not only how each representation conveys information but also how to translate between them. Therefore, the learning effectiveness of \ac{mer} does not only rely on the learning material but also on the characteristics of the learners~\cite{ainsworth2006,cooper2018}. To learn and solve problems with \ac{mer} efficiently, students must possess representational competence~\cite{fredlund2014unpacking,rau2017conditions,Daniel.2018b}. Rau divides representational competence with multiple visual representations into three domains~\cite{rau2017conditions}:

\begin{description}
    \item[conceptual competencies] \textit{visual understanding}, the ability of connecting the representation to concepts and understanding of the main features, and \textit{connectional understanding}, the ability to connect and compare the representation to other representations,
    \item[perceptual competencies] \textit{visual fluency}, the efficiency of connecting the representation to concepts and the ability to work effectively in the representation, and \textit{connectional fluency}, the efficiency and flexibility of connecting multiple representations to each other and switching between them, and
    \item[meta-representational skills] the ability to choose suitable representations based on task demands and context, and own ability level and personal goals.
\end{description}

\subsection{Representations in \ac{qist} education}\label{sec:quantum_vis}

Previous research has focused on learning quantum physics with representations such as the Dirac and vector/matrices notation of single-qubit states and operations, considering meta-representational competencies like assessing the suitability of these representations~\cite{PhysRevPhysEducRes.16.020112}. The authors found that, rather than focusing solely on students’ \textit{learning} difficulties, examining how representations are used to solve \textit{tasks} can yield valuable insights into more effective teaching approaches. Extending this line of research to include visualization techniques represents a natural next step.

Because of the field's complexity and counterintuitive nature, visualizations are widely used in teaching and learning quantum physics. Several have been investigated for their educational efficacy and associated difficulties. Common visualizations of qubit states, such as the unit circle, the Bloch sphere, coin analogies, and quantum circuits, are summarized in~\cite{10.1145/3481312.3481348}. Different visualizations of a beam splitter were examined in~\cite{vis_mach}, leading to an interactive visualization tool of the Mach-Zehnder interferometer~\cite{10.1119/1.4913786}, an optical implementation of a single-qubit system. Student difficulties with the Mach-Zehnder interferometer when learning with an interactive visualization were also analyzed in~\cite{Marshman_2016}. One of the most prominent visualizations of single-qubit systems is the Bloch sphere, which represents quantum operations as rotations and projections in three-dimensional space. Difficulties related to the Bloch sphere, the concept of a basis and the measurement process were analyzed in~\cite{Hu_2024}. Their tutorial improved student performance regarding measurement probabilities and outcomes, demonstrating the Bloch sphere's utility in identifying equivalent states (up to a global phase) and visualizing single-qubit unitary operations as rotations. In~\cite{Hennig_2024}, a new learning sequence using the Mach-Zehnder interferometer was proposed, combining polarization visualizations, an interactive GeoGebra animation, a graphical experiment setup, and abstract representations. Such studies highlight the diversity of representations even within a narrow context like the Mach–Zehnder interferometer. 

\paragraph{Representations of multi-qubit system}

\ac{qist} relies heavily on linear algebra in complex Hilbert spaces~\cite{geometry_Quantum_States,Watrous2018-ck}, where quantum states can be represented as vectors, and quantum operations correspond to unitary operations or projections. As systems expand to multiple qubits, these relationships become increasingly abstract and difficult to grasp. To address this challenge, various visualization techniques have been developed. Advanced visualizations include geometric depictions of quantum states, entanglement~\cite{Brody_2001,PhysRevA.64.062307,geometry_Quantum_States,sym13040581} and multi-qubit states~\cite{Makela_2010,PhysRevA.95.032308,PhysRevA.93.062320,PhysRevResearch.4.023120}. The diagrammatic ZX-calculus~\cite{Backens_2014,7174913,ng2017universal,Backens_2019} describes operations and algorithms in a ``LEGO-like'' manner~\cite{COECKE20221}, simplifying standard quantum circuit notation. Color code visualizations~\cite{PhysRevLett.97.180501, Kubica_2015,2017CoTPh68285A,Vasmer_2022} depict quantum error correction algorithms on hypercubes or hypercube-like lattices. The DROPS representation visualizes quantum states aand entanglement via generalized Wigner functions~\cite{PhysRevA.91.042122,Leiner_2020,KOCZOR20191} and has evolved into to the BEADS representation~\cite{huber2025beadscanonicalvisualizationquantum}, a more intuitive extension of the Bloch sphere to multi-qubit systems.

While these visualizations are basis-independent, our work focuses on the computational basis, which is commonly used for introductory purposes \cite{9781107002173}. Dynamic visualizations of quantum states and circuits, such as the “quantum puzzle game” based on state diagrams \cite{doi:10.1080/02635143.2021.1920905,felloni2009diagramsstatesquantuminformation,nita2021inclusivelearningquantumcomputing}, exemplify this trend. The \acf{cn} is a visual representation of the complex amplitudes of states in the computational basis using circles with gauges~\cite{johnston_harrigan_gimeno-segovia_2019}. While in \ac{cn}, amplitudes are aligned in a row as you would when writing states in the \acl{dn}, qubits can also be assigned axes in space~\cite{just_2021,Hellstern2024}. This dimensional approach has the possible benefit of making quantum operations more intuitive and can be combined with \ac{cn} (then called \acf{dcn}). It can be used to visualize the entanglement properties of qubit systems~\cite{bley2024visualizing}.

Currently, there is little empirical evidence on the effectiveness of different visualizations of multi-qubit systems for learning or problem solving. Because learning and problem solving share cognitive processes~\cite{Schewior2024}, differing only in the decision-making stage~\cite{LINDNER201791}, and in a similar line of thought to~\cite{PhysRevPhysEducRes.16.020112}, insights from problem-solving studies may improve instruction by identifying when and how to employ visualization techniques. Enhanced problem-solving performance could, in turn, lead to deeper conceptual understanding of underlying concepts like superposition and entanglement, which are known to show intrinsic susceptibility to misunderstandings~\cite{Marshman_2016,Brang2024}.

Referring to the \ac{deft} framework, using visualizations to support problem solving leverages their complementary function, adding visual access to qubit characteristics and gate operations. When students are presented with both visual and a symbolic representations, they gain multiple new routes to understanding \ac{qist} concepts. In addition to this, the use of visualization facilitates the understanding of the corresponding symbolic descriptions. When using both visual and symbolic representations, students can be supported in extending existing knowledge structures to mathematical descriptions. For example, in the context of entanglement, dimensional visualizations like \ac{dcn} provide new representations of the separability of qubit systems or even subsystems~\cite{bley2024visualizing} (see also Section \ref{sec:intro}). While \ac{cn} is merely a representation of complex numbers, \ac{dcn} adds an additional layer of information by spatially relating the basis states. When conducting a study comparing \ac{cn} and \ac{dcn}, one may be able to distinguish the benefits of both of these functions separately. If it is found out that benefits lie merely in the introduction of dimensionality, one could conclude putting complex numbers on the edges of a cube to be sufficient, instead of also using the circles.

However, visualizations also have limitations. In one-qubit systems, the Bloch sphere and the generalization to multi-qubit systems in form of the QuBeads representation~\cite{huber2024beadscanonicalvisualizationquantum} might make operations and entanglement more intuitive than \ac{cn} by relating them to rotations in space, and could therefore be used to complement mathematical notation more effectively. Secondly, these representations take up a lot of space and can display information only approximately, making mathematical symbolism and the circuit notation that can give an overview of a complete quantum algorithm the more concise choice to constrain interpretation. Especially in systems beyond three qubits, and especially \ac{dcn} takes up a large amount of space due to the additional axes that are presented and the need for spatially separating the cubes. \ac{cn} does not have this drawback to this extent, e.g., one can find eight-qubit systems being visualized on two thirds of a page in a book in~\cite{johnston_harrigan_gimeno-segovia_2019}. Furthermore, visualizations depict numeric relationships but cannot replace algebraic proofs, which remain essential for formal understanding. Thus, we advocate for their use in conjunction with, rather than as replacements for, symbolic representations, in order to fulfill the functions of the \ac{deft} framework.

\subsection{Relevant outcome measures and test construction}\label{sec:tasks}

\paragraph{Cognitive Load}

It is essential to consider the cognitive load imposed on learners when designing tasks and learning materials, as working memory capacity limits both learning and problem solving~\cite{Sweller1998}. According to \ac{clt}, cognitive load comprises three components: extraneous (\ac{ecl}), intrinsic (\ac{icl}), and germane (\ac{gcl}) load \cite{SWELLER1988257}. \ac{ecl} arises from presentation design; reducing it frees resources for goal-relevant processing. \ac{icl} reflects inherent task complexity, reducible only through simplification or scaffolding. \ac{gcl} refers to the cognitive effort invested in schema construction. Effective instructional design should therefore minimize \ac{ecl} and maximize \ac{gcl}.

When incorporating a visualization in addition to the \ac{dn}, learner characteristics can determine the imposed cognitive load and whether the visualization helps or not. For novices, visualizations may reduce \ac{icl} by providing concrete anchors, whereas for experts, redundant visuals may increase \ac{ecl}. This phenomenon is known as the \textit{expertise reversal effect} \cite{kalyuga_reversal}. Conversely, learners with strong representational understanding, e.g., of gate operations or separability in \ac{dcn}—may derive greater benefit. Such differences exemplify aptitude–treatment interactions \cite{Yeh2012}.

\paragraph{Multiple-choice test item design}

Multiple-choice items were chosen for this pilot study due to their practical advantages over open-ended questions~\cite{Downing2006}. They are efficient, easily scored, and suitable for analyzing response time without confounding factors such as writing speed. When validated, they offer reliable assessment while leveraging the testing effect, which enhances retention~\cite{Little2015}.

Designing effective multiple-choice items requires attention to distractors—plausible incorrect options that reflect common misconceptions~\cite{https://doi.org/10.1111/j.1745-3984.1989.tb00326.x,Little2015}. In this study, such misconceptions include confusion about qubit order, misinterpretation of measurement probability (as magnitude rather than squared magnitude)~\cite{styer1996common}, conflation of basis states with qubits~\cite{PhysRevPhysEducRes.20.020108}, and confusing separability with entanglement~\cite{Freericks_2023}. If distractors do not follow common misconceptions, they should share characteristics with the correct answer or the other distractors in order to be competitive~\cite{doi:10.3102/0034654317726529}. Distractors should be designed in a way that leaves no hint towards the correct answer without reading the question.

\paragraph{Time taken}

Response time provides additional insights into task difficulty. According to the distance–difficulty hypothesis~\cite{THISSEN1983179,ferrando2007item}, the logarithm of the response time is inversely proportional to the difference between a student’s ability and an item’s difficulty: students spend more time on tasks close to their ability level. For example, in~\cite{goldhammer2014time}, it was confirmed that in complex tasks, students who took longer also performed better. Similarly, it was found that children with higher levels of ability take longer to incorrectly solve the questions~\cite{jintelligence11040063}. When examining the effect of visualizations, longer response times for items with visualizations may indicate that the visualization engages additional problem-solving strategies and aligns task difficulty with the student’s ability.

When only considering the time taken for correctly solved questions, most of the results where the question was above the skill level of the participant are eliminated (except for those questions that were correctly solved by random chance). Therefore, analyzing the time taken exclusively for correctly solved questions focuses on students whose ability matches or exceeds the question’s difficulty. If a question becomes trivial for a student when presented with a visualization, they will take less time for correctly solved questions. Consequently, when requiring that both the question without and with visualization be solved correctly, the average time taken may be lower if the visualization indeed simplifies the question.

The time taken for task solving is a multidimensional construct. Clear results are expected to be very difficult to obtain, as there are many factors that play a role. Nevertheless, according to the distance-difficulty hypothesis~\cite{THISSEN1983179,ferrando2007item}, time could serve as an indicator of how well a task's difficulty aligns with student ability. For comparisons between questions to be meaningful, the questions should be of similar complexity. Furthermore, a detailed analysis of the relationship between question difficulty and the time taken for correct answers could offer additional insights into the benefits of providing visualization -- for instance, if a difficult question requires more time with visualization than without, it might indicate that the visualization supports the use of deeper or alternative problem-solving strategies. 

\section{Research Questions}\label{sec:res_ques}

Based on current educational research, we expect that the use of visualizations such as \acf{cn} and \acf{dcn} can leverage the known advantages of learning with \acf{mer} in \acf{qist} by constructing deeper understanding of entanglement and by complementing the mathematical-symbolic \ac{dn} by offering new strategies to predict the outcome of quantum gate operations. In this study, we take the first step in the design and validation of test items that can be used to verify this expectation by comparing the task-solving abilities on items presented in \ac{dn} with those in which \ac{dn} is accompanied by \ac{cn} or \ac{dcn}. Such tests should take into account the representational competence of the participants. This competence is divided into conceptual competencies related to visual and connectional understanding and the perceived cognitive load during tasks without and with visualization~\cite{rau2017conditions}. Here, we refer to representational understanding instead of visual understanding, including competence with complex numbers in mathematical-symbolic notation. It is not clear whether this is an adequate generalization, namely, whether Rau's construct of representational competence adapted to incorporating representations beyond those that are purely graphical still predicts performance increases. In this work, we focus on conceptual competencies rather than perceptual or meta-representational competencies for two reasons: we discard perceptual competencies for practical reasons, as it is unclear how to measure the fluency or efficiency described by Rau independently of the accuracy of solving the questions. Efficiency is conceived as performance over effort, and there are multiple possible ways to define both~\cite{HOFFMAN21012010}, while accuracy should also be part of this metric. As the representations used in this work do not differ in level of abstraction or in the information presented, we see meta-representational competencies as less important for solving the given problems. In the framework used, we remain with competencies involving connecting representations and understanding their functioning.

Another point that is unclear is whether this framework holds in task solving as well as in learning. However, there is a strong argument that task-solving and learning share many similarities, except for differences in the decision-making phase~\cite{LINDNER201791}. 

Because the supposed benefits, in addition to being likely learner-dependent, are also likely to be context-dependent, we explore the effects of presenting a visualization in different contexts, that is, measurements, various unitary operations, and entanglement \& separability. The aim of this pilot study is to create test items and a suitable study design to answer the overarching questions described in Section \ref{sec1} of the conditions for benefits of visualization for problem solving in \ac{qist}. To this end, we formulate the following research questions.

\begin{description}
    \item[\textup{\textbf{RQ1}}] Are the proposed test items suitable, in terms of difficulty, for a heterogeneous target group?

    \item[\textup{\textbf{RQ2}}] Which test items should be omitted or redesigned?

    \item[\textup{\textbf{RQ3}}] Is the study design suitable for answering the overarching questions?
\end{description}

In addition, we intend to identify indications of effects related to the overarching questions that warrant further exploration in future studies. 

\section{Methods}\label{sec:design}

In order to gain insights into the effects and relevant conditions of including visualizations like \ac{cn} and \ac{dcn} in students' problem solving, we decided to compare the performance and cognitive load of participants with and without visualization within participants (intrapersonally) and relate them to representational competence. As is typical for a pilot study, our aim is to learn about the suitability of the test items that are presented to the participants and the study design. All test items were reviewed by experts using the design principles described in Section \ref{sec:tasks}. They are provided in the supplementary material, with the correct answers highlighted. The order of the answers was randomly assigned to each participant. In Section \ref{sec:structure}, we define the overall structure of the pilot study. Some students volunteered for think-aloud interviews, the methodology of which we describe in Section \ref{sec:qual_part}. We then describe the group of participants in Section \ref{sec:sample}. In Section \ref{sec:intro}, we describe the introductory slides used. In Sections \ref{sec:rep_und} and \ref{sec:con_und}, we outline the design of the questions on representational and connectional understanding. In Section \ref{sec:quantum_ops}, we describe the test items of the main parts of the survey. We finish with the cognitive load test items in Section \ref{sec:cognitive_load}.

\subsection{Structure of the study}\label{sec:structure}

The main study consisted of three online multiple choice tests on one-, two-, and three-qubit systems, respectively. Each of the tests took about 45 minutes. 

In each survey, except for the first, the participants were randomly assigned to two groups, where one group was presented with \ac{cn} and the other group was presented with \ac{dcn} (in one-qubit systems, \ac{cn} and \ac{dcn} are identical).  Both \ac{cn} and \ac{dcn} were included in the study for the following reason: by comparing \ac{cn} and \ac{dcn} in a study such as this, one may be able to investigate the effects of visualizing complex numbers separately from those of introducing dimensionality.

\begin{figure*}[ht]
    \centering
    \includegraphics[width=\linewidth]{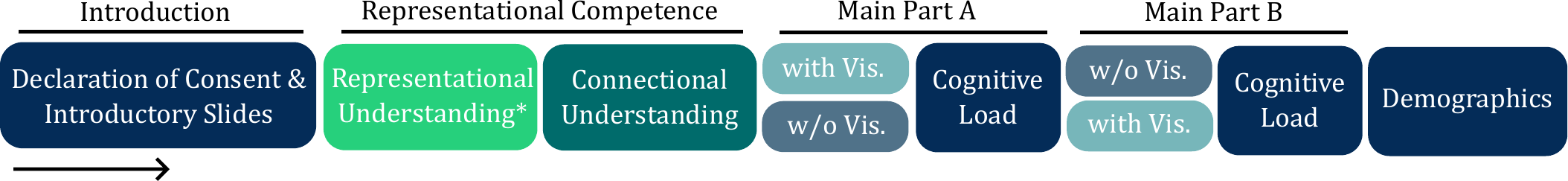}
\caption[Overall structure of the study.]{Overall structure of the study. In the two-, and three-qubit surveys, the participants were first grouped into \acl{cn} or \acl{dcn} randomly. During all surveys, the participant group(s) was/were split during the main parts of the studies into group vis-math or group math-vis, depending on whether they were presented with or without visualization first. The questions within one of the parts A or B only differed in whether they were with visualization or not. This results in two groups in the one-qubit survey and four groups in the two- and three-qubit surveys. $^*$Representational understanding was only tested in the one-qubit survey.}
    \label{fig:structure}
\end{figure*}

Each survey was separated into two parts, A and B, where questions using the same contexts (gate operations, measurements, entanglement, and separability) were asked in both parts, while all questions were identical within one part (with or without visualization). The participants were divided into two further groups, independent of the \ac{cn} and \ac{dcn} group: one group was presented with visualization in part A and without visualization in part B (group vis-math), while the conditions were reversed for the other group (group math-vis). The same questions were asked to all students in part A, but one group received them with visualization and one without; in part B, the roles were reversed. This resulted in two independent groups in the one-qubit survey and four independent groups in the two- and three-qubit surveys. After each part, we measured the perceived cognitive load. Before the surveys, participants viewed introductory slides explaining the respective visualization, operations, and entanglement concepts and answered questions assessing representational competence. At the end, we collected demographic information (age, level of education, and field of study). The overall structure of the surveys is shown in Figure \ref{fig:structure}.

\subsection{Qualitative part of the study and participants}\label{sec:qual_part}

To discover which problem-solving strategies students use and which problems they encounter with or without visualization, we conducted think-aloud interviews with students from our quantum computing lecture~\cite{WOLCOTT2021181}. This allowed us to identify reasons for differences in performance and cognitive load when visualization was provided. The performances of the interviewed students were included in the main survey results, but not their response times, as they required additional time to explain their thought processes.

Some students were recruited specifically for these think-aloud interviews. They were instructed to think about the question, select an answer, and then explain their reasoning. Participants received a compensation of 15€ for the one- and two-qubit surveys and 10€ for the three-qubit survey.

Five participants volunteered for think-aloud interviews in the one- and two-qubit surveys, two of whom also participated in the three-qubit survey. Of these participants, three were aged 22–25 and two were 26–30. Two identified as female and three as male. Three participants had an Information/Computer Science background and two came from Mathematics; one held a Master’s degree and four a Bachelor’s degree. All five had been introduced to both \ac{dn} and \ac{cn} before the survey. This group can be regarded as representative of the broader participant cohort in terms of academic background. The participants in the qualitative study were randomly assigned the conditions shown in Tab.~\ref{tab:interview_cond}.

\begin{table}[ht]
    \centering
        \caption{Assignment of conditions in the qualitative analysis to groups assigned questions with visualization first or second, and assigned to \acf{cn} or \acf{dcn} in the case of the two- and three-qubit surveys.}
    \begin{tabularx}{\linewidth}{@{}l|X|X|X@{}}
         \textbf{Student} & \textbf{One qubit} & \textbf{Two qubits} & \textbf{Three qubits}  \\ \hline
         \textbf{1} & vis. first & CN, vis. first & \mbox{-}  \\ \hline
         \textbf{2} & vis. first & CN, vis. first & DCN, no vis. first \\ \hline
         \textbf{3} & w/o vis. first & DCN, w/o vis. first & \mbox{-} \\ \hline
         \textbf{4} & w/o vis. first & DCN, vis. first & DCN, vis. first \\ \hline
         \textbf{5} & w/o vis. first & DCN, w/o vis. first & \mbox{-} \\
    \end{tabularx}
    \label{tab:interview_cond}
\end{table}

\subsection{Participants of the Quantitative part}\label{sec:sample}

Participants in the main study received 10€ compensation per test. Most participants were drawn from our quantum computing lecture. In the surveys on one-, two-, and three-qubit systems, there were 18, 22, and 15 participants, respectively (excluding the participants of the qualitative part). At the end of the first survey, demographic information was collected. Table \ref{tab:part_cond} shows the number of students assigned to the different conditions, including both general study participants and interview participants.

\begin{table}
\caption{Assignment numbers of students to the two or four conditions of the different surveys, to groups assigned questions without visualization first or second, and \acf{cn} or \acf{dcn} as visualization in the case of the two- and three-qubit surveys. Comparisons are made both between groups of the same column and within participants. In brackets are the numbers of the qualitative part of the survey. These are also included in the quantitative part.}
\begin{tabular}{@{}l|c|cc|cc@{}}
\toprule
\textbf{number of qubits} &  \textbf{one} & \multicolumn{2}{@{}c@{}}{\textbf{two}} & \multicolumn{2}{@{}c@{}}{\textbf{three}} \\
\cmidrule(r{0.75em}){1-1}\cmidrule(r{0.75em}){2-2}\cmidrule(lr{0.75em}){3-4}\cmidrule(lr{0.75em}){5-6}
\textbf{visualization} &  \textbf{CN} & \textbf{CN} & \textbf{DCN} & \textbf{CN} & \textbf{DCN} \\
\midrule
\textbf{w/o vis. first} & 12(3) & 6(1) & 8(1) & 5 & 3(1) \\
\textbf{vis. first} & 11(2) & 6(2) & 7(1) &  5 & 4(1) \\
\cmidrule(r{0.75em}){1-6}
\textbf{sum} & 23(5) & 12(3) & 15(2) & 10 & 7(2) \\
\bottomrule
\end{tabular}
\label{tab:part_cond}
\end{table}

Of the 18 participants in the first survey, eight reported that they had not been introduced to \ac{dn} before, and nine reported prior exposure to \ac{cn}. Among the participants in the one-, two-, and three-qubit studies, 1(1)(1) was aged 18–21, 12(11)(9) were 22–25, and 3(2)(3) were 26–30. Furthermore, 16(14)(13) were male and 1(1)(1) female. Regarding educational level, 4(3)(4) participants had completed high school, 9(7)(5) held a Bachelor’s degree, 3(3)(3) held a Master’s degree, and 1(1)(1) held a Ph.D. In terms of academic field, there was 1(1)(1) in Economics, 8(8)(7) in Information/Computer Science, 3(3)(2) in Computer Science and Engineering, 4(4)(4) in Mathematics, and 1(1)(1) in Physics. Of these 18 participants, 13 completed all three surveys, two completed only the first, two completed the first and second, and one completed the first and third. We do not have personal information for those who participated only in the second and/or third survey, but we assume that they were familiar with \ac{dn} and \ac{cn}, as they were also recruited from our quantum computing lecture. Most participants were new to quantum science, as it is not typically included in mathematics or computer science curricula, and the quantum computing lecture was introductory in nature. Except for the students in Physics and the Ph.D. holder, we assume that all participants had taken no more than one or two \ac{qist}-related courses. 

\subsection{Introductory slides}\label{sec:intro}

At the beginning of each survey, participants were presented with introductory slides that explained all key ideas necessary to solve the study questions, using both mathematical symbolism and the corresponding visualization, \ac{cn} or \ac{dcn}, depending on their assigned condition. Participants read through the slides at their own pace. The introductory slides covered the following topics:

\begin{itemize}
    \item Visualization of qubit states
    \item Action of single-qubit gates in single-qubit systems
    \item Action of gates in multi-qubit systems (only shown in the second and third surveys)
    \item Measurement probabilities and the resulting state after measurement
    \item Separability and Entanglement (only shown in the second and third survey)
\end{itemize}

All introductory slides of the one-, two-, and three-qubit survey are shown in the supplementary material. They were designed to introduce the visualization as a complement to the \ac{dn} and to demonstrate the strategies specific to each representation. The introductory slides are supposed to reduce the influence of previous knowledge on the test results and are designed to thoroughly explain ideas, making them accessible even to participants with only a basic mathematical background. This approach was chosen because the main purpose of the questions was to determine whether students benefit from visualization, with prior knowledge being uncontrolled for due to the lack of validated test instruments for this construct. In order to not hint some students towards important problem-solving strategies more than others, participants were not given the opportunity to ask questions for this or any part of the survey. The participants were instructed to read the introductory slides, answer the survey in good faith and to not use any additional tools. The participants took on average $55\pm33$, $127\pm68$ and $141\pm95$ seconds on slides of the one-, two- and three-qubit surveys, and at least 12, 49, and 56 seconds per slide. In general, the participants engaged with the introductory part of the study. The relevant concepts, aligned with the introductory slides -- with which, in principle, all survey questions could be answered correctly without prior quantum knowledge beyond basic mathematics -- are described in the following.

\paragraph{Visualization of qubit states}

\begin{figure*}[ht]
 \begin{minipage}{0.57\textwidth}\centering
    \includegraphics[width=\linewidth]{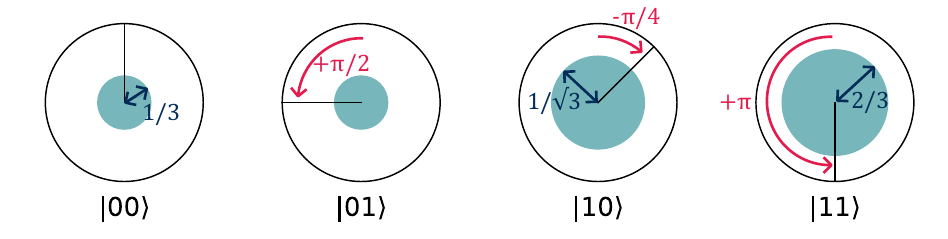}
\end{minipage}
\begin{minipage}{0.42\textwidth}\centering
    \includegraphics[width=\linewidth]{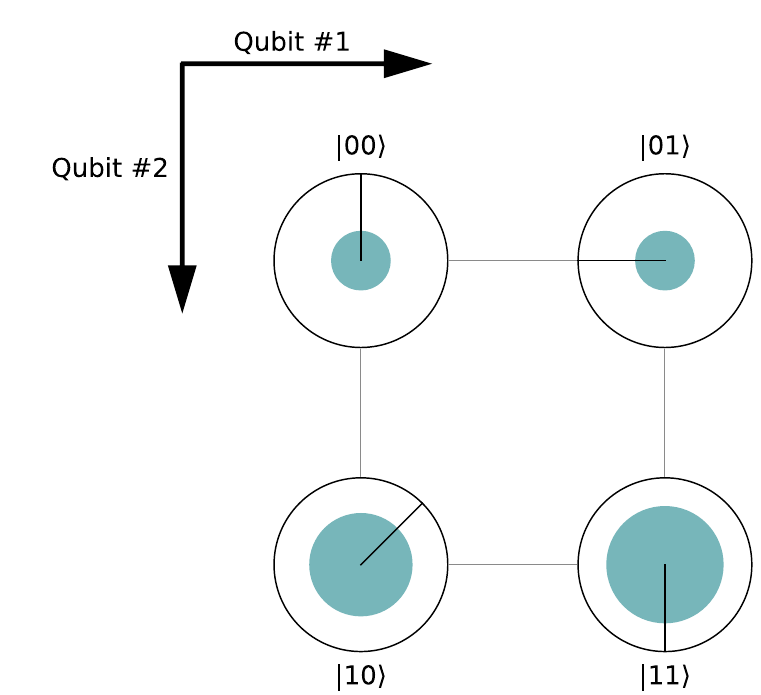}
\end{minipage}
\caption{The state $\ket{\psi}=\frac{1}{3}\ket{00}+\frac{1}{3}e^{i\pi/2}\ket{01}+\frac{1}{\sqrt{3}}e^{-i\pi/4}\ket{10}-\frac{2}{3}\ket{11}$ in \acl{cn} (left)~\cite{johnston_harrigan_gimeno-segovia_2019} and in \acl{dcn} (right). The qubits are ordered in from right to left, $\ket{\#2\#1}$.}
\label{fig:cn_dcn}
\end{figure*}

The \ac{cn} uses filled circles with pointers to visualize the amplitudes $c_i=r_i e^{i\varphi_i}$ of a quantum state $\ket{\psi}=\sum_i c_i\ket{i}$. We refer to qubit states using the computational basis, which means that $i$ is written in binary form, with the rightmost digit being the least significant and also the number corresponding to the first qubit. The magnitude $r_i$ is represented by the radius of the filling circle and the gauge at an angle $\varphi_i$, measured counterclockwise from the 12 o'clock axis. In dimensional notations, each qubit is assigned an axis in space~\cite{just_2021,bley2024visualizing}. Figure \ref{fig:cn_dcn} shows the same two-qubit state in \ac{cn} and in \ac{dcn}. Participants were shown a version of this image that corresponded to their assigned visualization type and the number of qubits in the respective survey.

\paragraph{Action of single-qubit states in single-qubit systems}

\begin{figure}[ht]
    \centering
    \includegraphics[width=\linewidth]{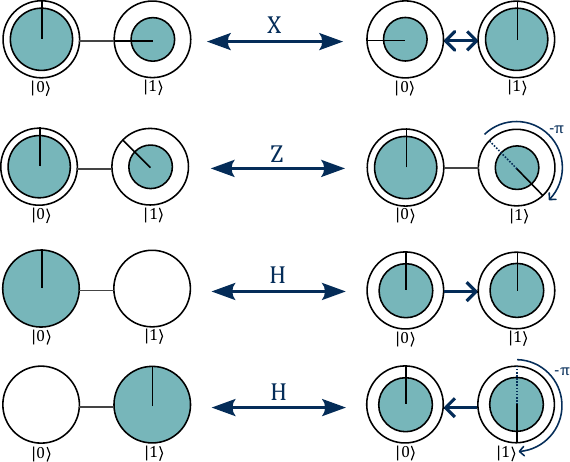}
    \caption{The X, Z, and H gate in a single-qubit system in \acl{cn}.}
    \label{fig:single_gates}
\end{figure}

The relevant gates to manipulate individual qubits shown in the study are the X, the Z, and the H gate. The X gate flips the qubit and swaps the two coefficients of the $\ket{0}$ and $\ket{1}$ states. The Z gate flips the relative phase by $\pi$, i.e., changes the sign in front of the coefficient of the $\ket{1}$-state. The H (Hadamard) gate creates equal superpositions from the $\ket{0}$ and $\ket{1}$ basis states but introduces a $\pi$ phase flip when acting on the $\ket{1}$ basis state. The X, Z, and H gate all act inversely in the same way. How these gates function is visualized in Figure \ref{fig:single_gates}.

\paragraph{Action of gates in multi-qubit systems}

\begin{figure*}[ht]
    \centering
    \includegraphics[width=\linewidth]{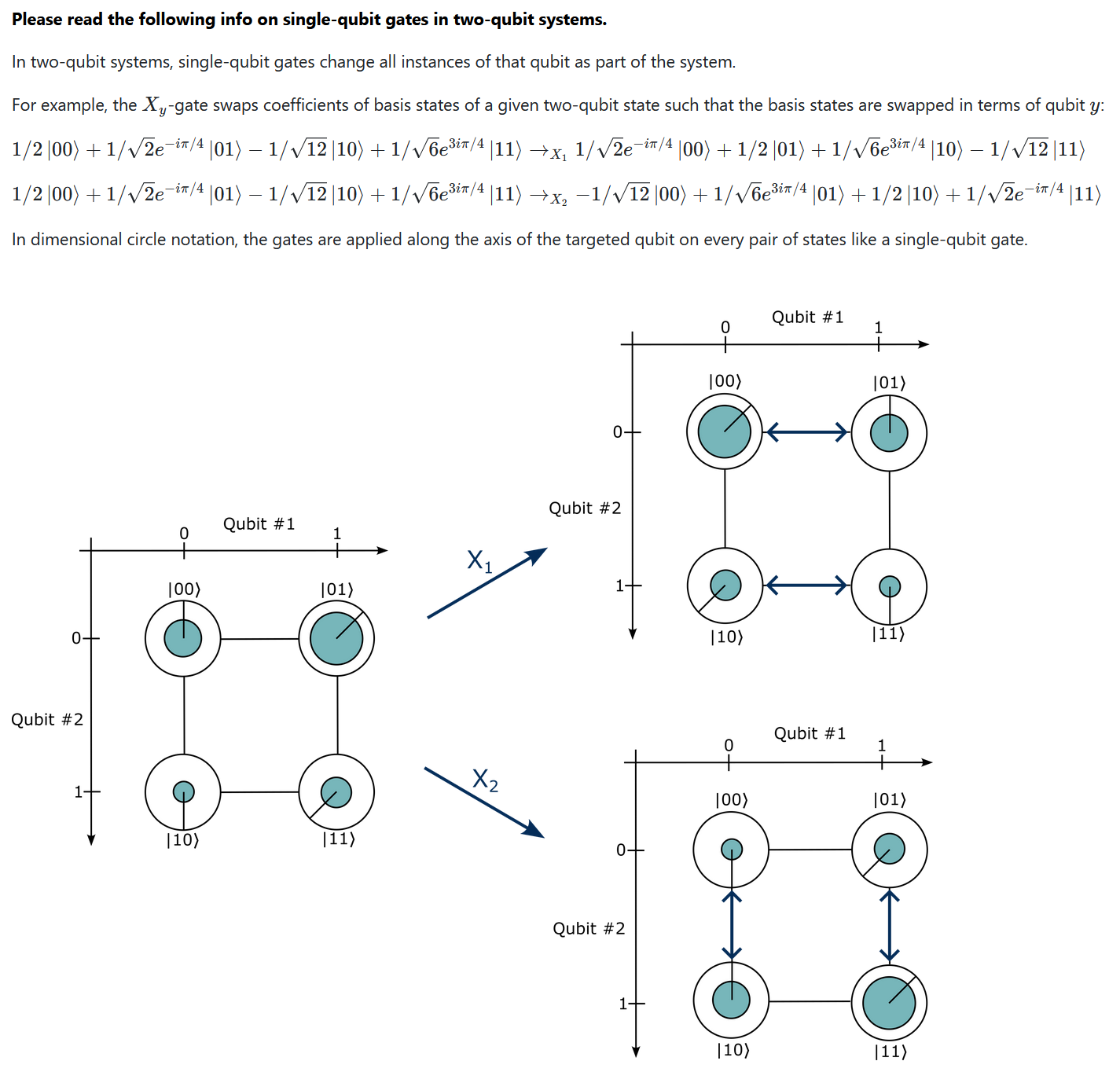}
    \caption{Introductory example slide for the action of single-qubit gates in two-qubit systems visualized in \acl{dcn}.}
    \label{fig:example_slide}
\end{figure*}

Single qubit gates act along the axis of one qubit as if they were operating on the individual pairs of basis states along that axis. The X$_y$ gate swaps coefficients of basis states of a given state such that the basis states are exchanged with respect to qubit $y$. An example slide used in the study, where this concept is visualized in \ac{dcn}, is shown in Figure \ref{fig:example_slide}.

\paragraph{Measurement probabilities and the resulting state after measurement}

\begin{figure*}[ht]
    \centering
    \includegraphics[width=0.75\linewidth]{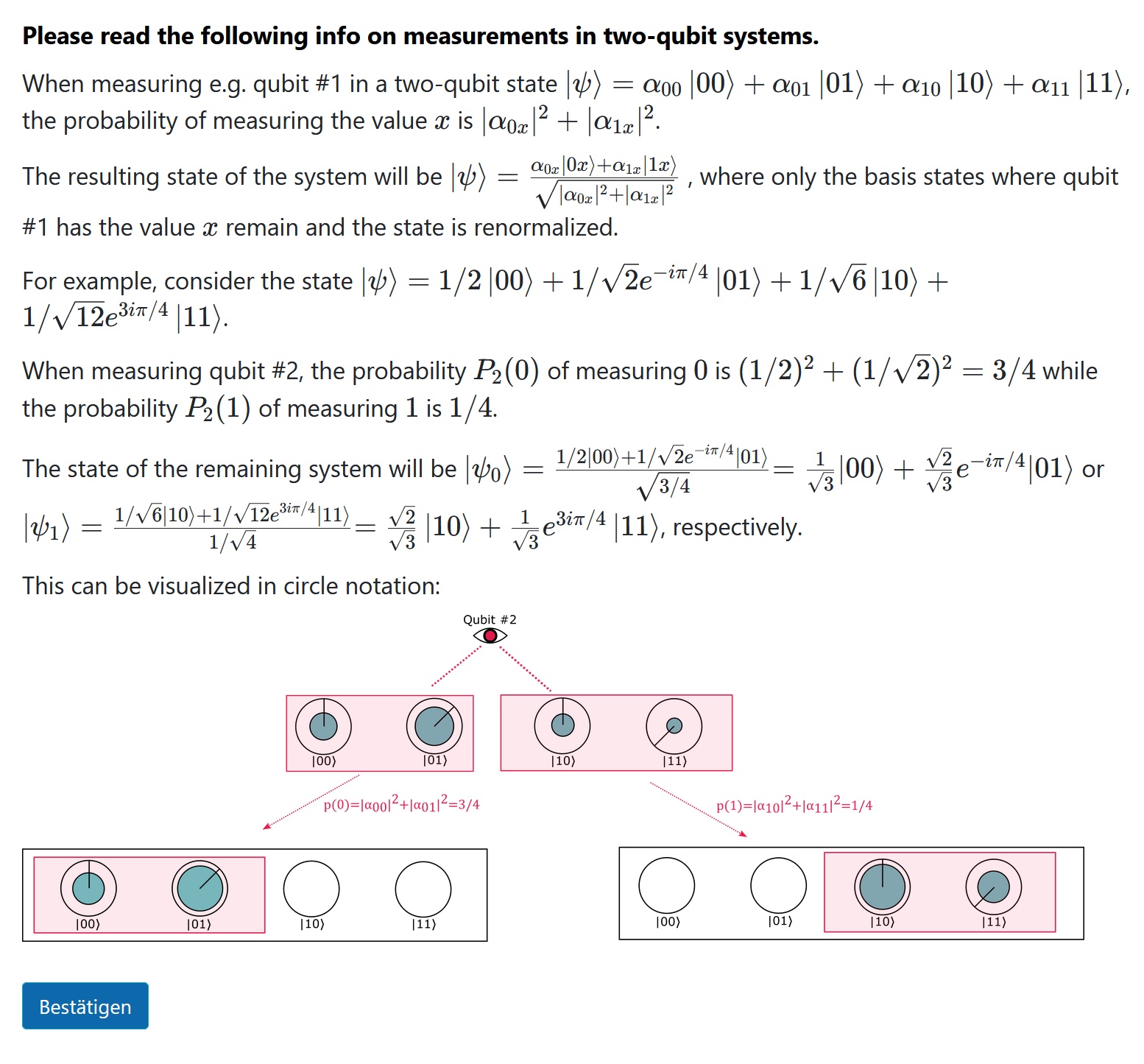}
    \caption{Introductory example slide for measurements in two-qubit systems visualized in \acl{dn}.}
    \label{fig:example_meas}
\end{figure*}

Because a quantum state $\ket{\psi}=\sum_i c_i\ket{i}$ is normalized, i.e., $\sum_i |c_i|^2=1$, the sum of the areas of the inner circles equals $1$. Each area can be interpreted as the measurement probability of the corresponding basis state. After measurement, the state of the system collapses into a renormalized state corresponding to the measurement outcome. An example introductory slide for participants of the two-qubit \ac{dn} survey is shown in Figure \ref{fig:example_meas}

\paragraph{Separability and Entanglement}

\begin{figure*}[ht]
    \centering
    \includegraphics[width=0.9\linewidth]{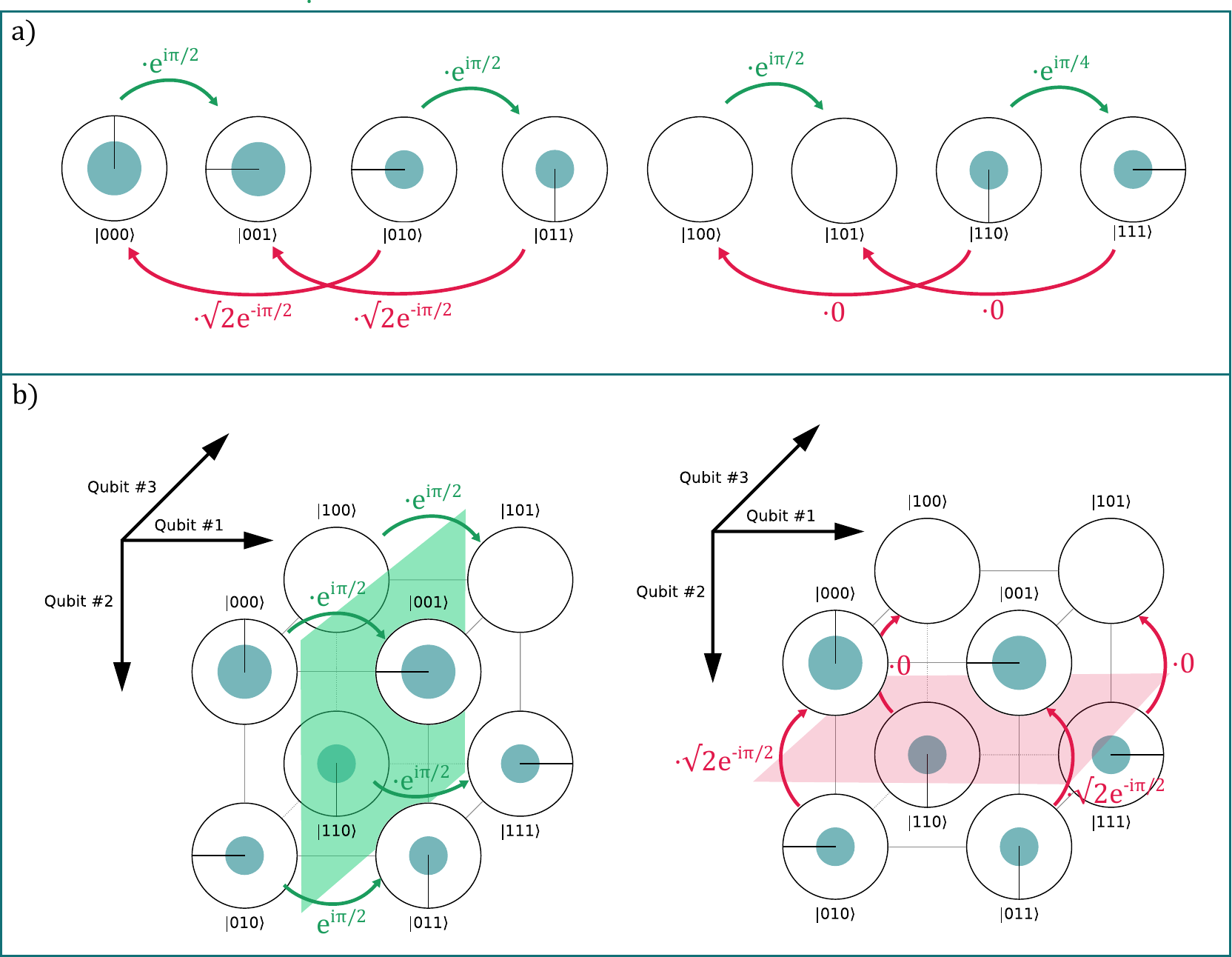}
    \caption{Partial separability of the state $\ket{\psi}_{\#3\#2\#1}=\frac{1}{2}\ket{000}+\frac{1}{2}e^{i\pi/2}\ket{001}+\frac{1}{2\sqrt{2}}e^{i\pi/2}\ket{010}-\frac{1}{2\sqrt{2}}\ket{011}-\frac{1}{2\sqrt{2}}\ket{110}+\frac{1}{2\sqrt{2}}e^{-i\pi/2}\ket{111}$~\cite{bley2024visualizing} in \acl{cn} and \acl{dcn}. a) Separability in standard circle notation in terms of qubit \#1 can be seen by the fact that there exists a ratio $r_1=e^{i\pi/4}$ (in green) such that for any coefficients $c_{xy0}$ and $c_{xy1}$, $c_{xy1}=r_1e^{i\pi/4}c_{xy0}$ is the same everywhere. However, as is shown in red, this is not the case for qubit \#2.  b) As depicted on the left hand side, the state is separable in terms of qubit \#1, because the state is symmetric apart from a complex factor in terms of qubit \#1. This is apparent by the green symmetry plane. For the other two qubits, there exists no such ratio, i.e., no such symmetry planes, as shown here for qubit \#2 with the red plane.}
    \label{fig:sep_ent}
\end{figure*}

In the computational basis, separability of pure states can be seen as symmetry -- apart from a complex factor -- with respect to each qubit of the system~\cite{bley2024visualizing}. To determine whether a system is separable, one must examine the ratios between coefficients along the axis of one qubit. If all ratios are equal to a constant $r$, the state is separable with respect to that qubit. In two-qubit systems, separability can be observed through symmetry along the axis of one qubit, apart from the ratio $r$. If this symmetry is absent, the system is entangled. In three-qubit systems, one can identify separability by symmetry with respect to the plane perpendicular to a qubit’s axis: a qubit is separable from the rest of the system if such a symmetry exists. A system is called partially separable if only a subset of qubits is separable from the rest. This is illustrated in Figure \ref{fig:sep_ent}. Alternatively, entanglement can be explained as a dependency of the resulting post-measurement state on the measurement outcome of the qubit in question. 

\subsection{Representational understanding}\label{sec:rep_und}

\begin{figure}[ht]
    \centering
    \includegraphics[width=\linewidth]{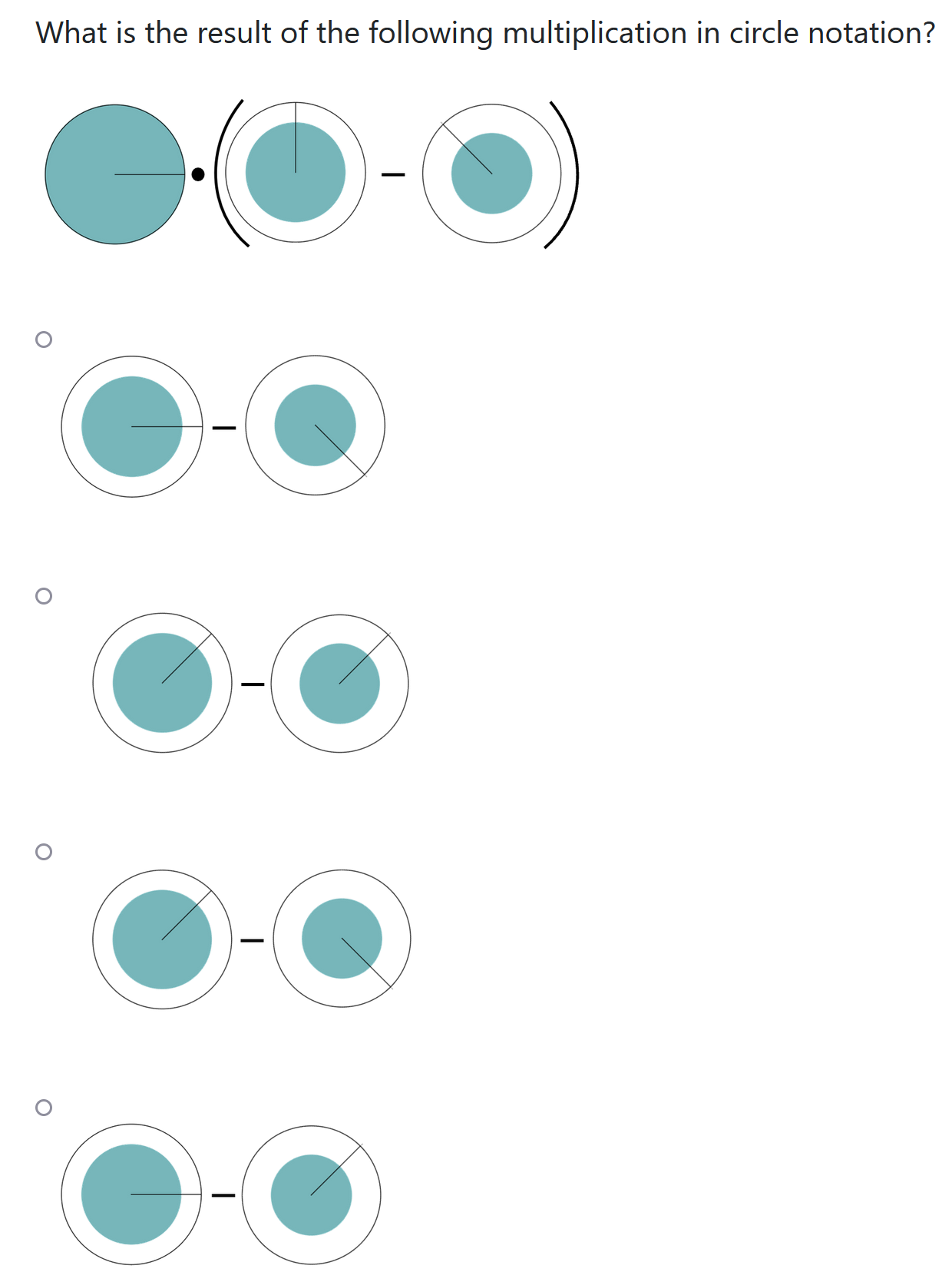}
    \caption{Example question for representational understanding in \acl{cn} for the calculation $e^{-i\pi/2}\cdot(\frac{\sqrt{2}}{\sqrt{3}}-\frac{1}{\sqrt{3}}e^{i\pi/4})=\frac{\sqrt{2}}{\sqrt{3}}e^{-i\pi/2}-\frac{1}{\sqrt{3}}e^{-i\pi/4}$. The correct answer is the bottom most. The distractors include variations in the phase of the complex numbers depicted by the circles.}
    \label{fig:rep_comp_example}
\end{figure}

We use the term \enquote{representational understanding} instead of \enquote{visual understanding} as described by Martina Rau~\cite{rau2017conditions}, because we are also interested in representational competence with the traditional mathematical-symbolic notation. In the one-qubit survey, participants were asked one question on the multiplication of complex numbers and one on the Euler formula (for example, translating between $e^{i\pi/4}$ and $1/\sqrt{2}+i/\sqrt{2}$), and similar questions using only \ac{cn}. For this, four different questions were designed, two of which each participant received in \ac{cn} and two of which each participant received in mathematical symbolism, assigned randomly. An example is shown in Figure \ref{fig:rep_comp_example}. These questions are limited in their applicability to basic arithmetic in \ac{dn} and \ac{cn} and do not capture spatial reasoning ability that is supposedly necessary for a benefit from \ac{dcn} in multi-qubit systems. There is a lack of alternatives to assess multi-qubit representational understanding which we did not address in this study due to time constraints on the side of the participants, as we did not want to exceed the 60 minutes threshold. Because many of the same students also participated in the other surveys and these questions are the most relevant for one-qubit systems, the representational competence questions of the one-qubit survey were not repeated in the other surveys. All representational understanding questions are provided in the supplementary material. The questions were designed to include possible misconceptions about \ac{cn}: for example, students might think that the gauge pointing to the right indicates a phase of 0, and that adding a positive phase means that the gauge turns clockwise.

Participants were asked to rate their confidence on each question using a Likert scale from 0 (random guess) to 4 (very sure). The responses were evaluated using a weighted scoring method as proposed in, e.g.,~\cite{Fehringer+2020+138+212} as having the benefit of reducing statistical error, but may be influenced by participant characteristics such as self-esteem. The extent to which participants use the visualization is a deciding factor in whether the visualization provides a benefit. We therefore expect the weighted measure to serve as a reliable moderator variable.

\subsection{Connectional understanding}\label{sec:con_und}

\begin{figure*}[ht]
    \centering
    \includegraphics[width=0.9\linewidth]{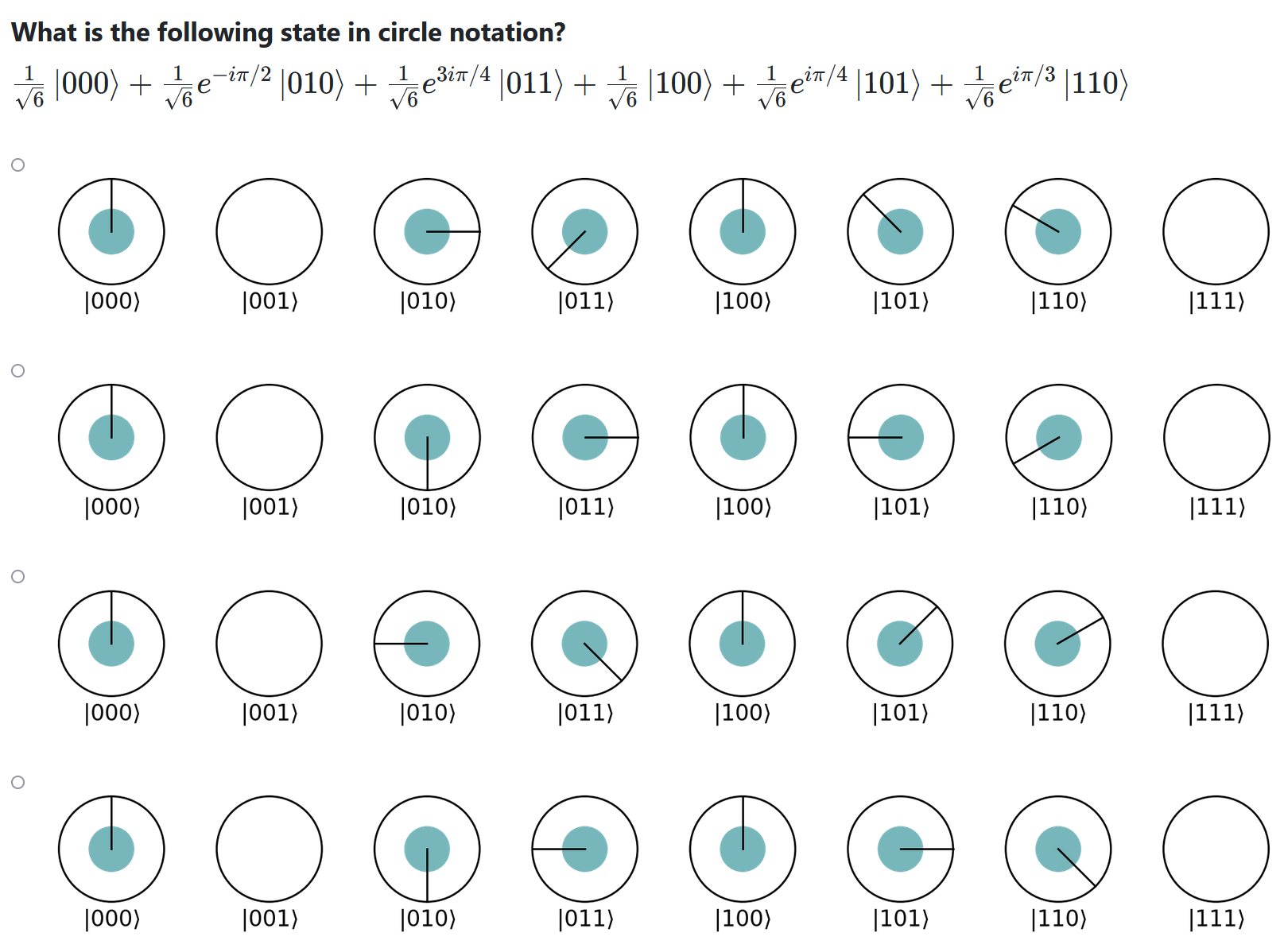}
    \caption{Example question for translating from \acl{dn} to \acl{cn}, where a focus lies in the understanding of the phase in \acl{cn}. The correct answer is the top most. The distractors are chosen to share features with the correct answer, with no answer standing out regarding a single feature.}
    \label{fig:example_translation}
\end{figure*}

We designed test items to measure participants' connectional understanding between \ac{dn} and the visualization. After the introductory slides, participants were asked four questions (six questions in the two-qubit survey) to translate between their assigned visualization and the \ac{dn}. Two (three in the two-qubit survey) of these questions required translating from \ac{dn} to visualization, and two (three) required translating back. One question addressed translation of phase, one of magnitude, and in the two-qubit survey, and in the two-qubit survey we added one question addressing both. An example task from the three-qubit survey is shown in Figure \ref{fig:example_translation}. All connectional understanding questions of the one-, two-, and three-qubit surveys are included in the supplementary material. 

The connectional understanding questions were designed to test participants’ ability to translate either phases or magnitudes between representations. The distractors were constructed so that no answer stood out regarding any specific feature -- or that all did -- ensuring that participants could not infer hints toward the correct answer without engaging with the question itself. In addition, the questions were intentionally diverse, illustrating different combinations of amplitudes.

\subsection{Quantum Operations \& entanglement}\label{sec:quantum_ops}

\begin{figure*}[ht]
    \centering
    \includegraphics[width=0.73\linewidth]{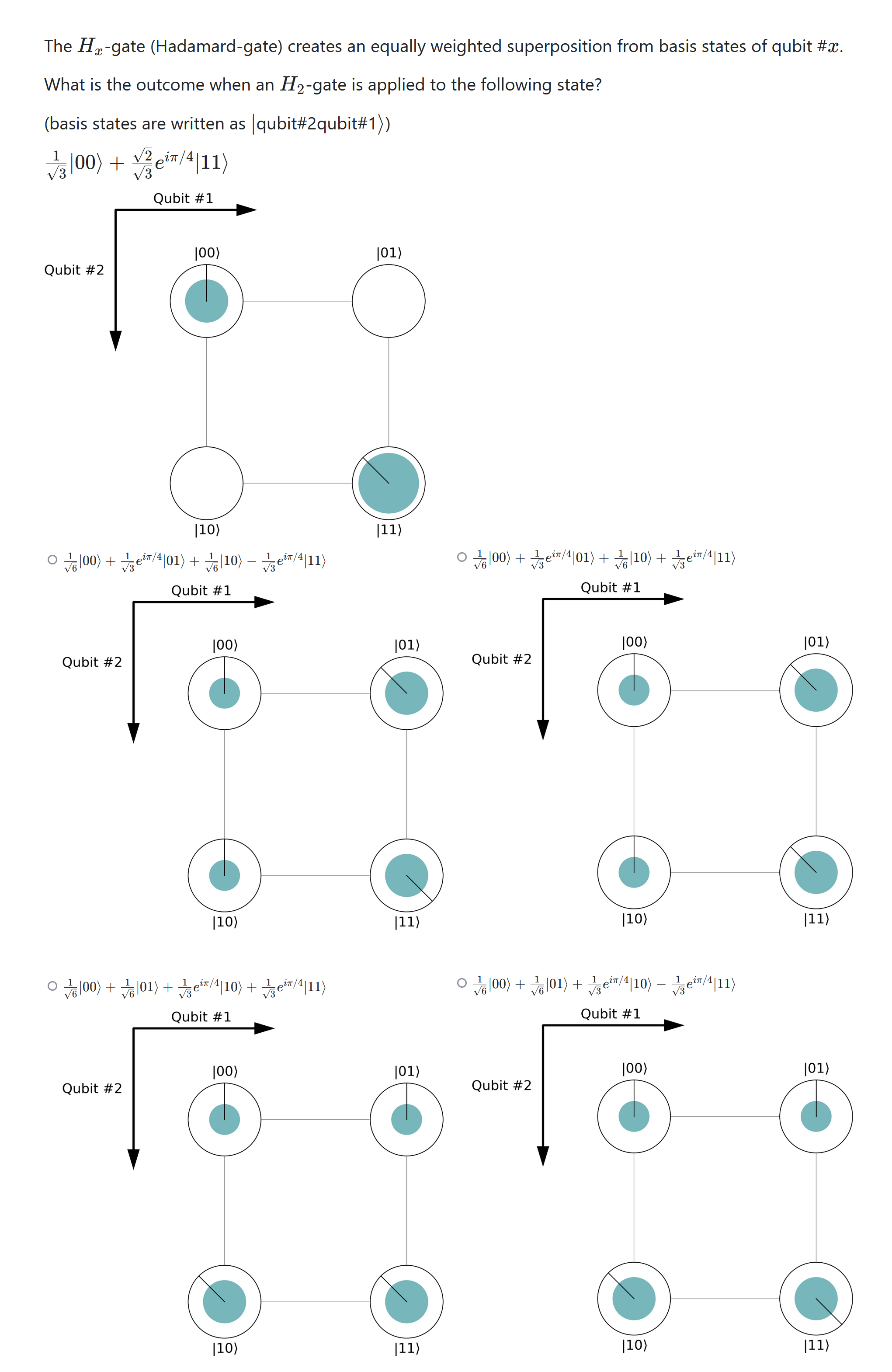}
    \caption{Example task of predicting the outcome of a Hadamard gate. The correct answer is the one in the top left. The distractors are chosen such that the Hadamard gate is applied to the wrong qubit (bottom right), and the two other distractors sharing features with the first and the correct answer.}
    \label{fig:example_task}
\end{figure*}

We tested the visualizations in one-, two-, and three-qubit systems. The X and Hadamard gates were tested for all system sizes, and the Z gate was included in the one- and three-qubit systems. In addition, questions about measurement probabilities and the resulting state after measurement were asked in all surveys. The phase gate was tested only in the one-qubit survey, while in the two- and three-qubit surveys the CNOT gate, (partial) separability, and (partial) entanglement were also examined. Figure \ref{fig:example_task} shows an example task on the effect of a Hadamard gate in a two-qubit system. A short explanation of how a given operation acts—or how entanglement or separability is defined—was provided to reduce the influence of prior knowledge, following the same reasoning as for the introductory slides. This also reduces the possibility of a false interpretation of the questions, in line with recommendations from~\cite{doi:10.3102/0034654317726529}.

\paragraph{Questions of the one-qubit survey}

In the one-qubit survey, we first asked about the equivalence of states that differ only by a global phase to test whether these phase differences are easier to recognize in the \ac{cn} representation. The distractors were designed to include confusion between shifting only one of the two phases and swapping only magnitudes. The X-gate questions were designed to include the confusion about which part of the amplitude to swap, i.e., one distractor was chosen such that only the magnitude is swapped and one such that only the phase is swapped. The last distractor was chosen as equalizing the magnitudes. For this question, participants needed to recognize the equivalence of two states differing only by a global phase ($\frac{1}{2}\ket{0}+\frac{\sqrt{3}}{2}e^{i\pi/2}\equiv \frac{1}{2}e^{-i\pi/2}\ket{0}+\frac{\sqrt{3}}{2}\ket{1}$). 

The H-gate questions in the one-qubit survey were designed to be more challenging, as they required participants to apply a combination of superposition creation and destruction. They were meant to involve competence in distributive multiplication of complex numbers; however, discussions with participants revealed that these particular questions were too complex. We therefore decided to ask only about situations where amplitudes were either created or destroyed (but not both) in the two- and three-qubit surveys. The correct answer to the part A question also differed by a global phase ($\frac{1}{\sqrt{2}}\ket{0}+\frac{1}{\sqrt{2}}e^{i\pi/2}\ket{1}\equiv \frac{1}{\sqrt{2}}e^{-i\pi/4}\ket{0}+\frac{1}{\sqrt{2}}e^{i\pi/4}\ket{1}$). These global phase differences might represent a disadvantage of \ac{cn} compared to the Bloch sphere, where global phase information is omitted by construction. The Bloch sphere also provides a unique strategy to solve such questions by imagining a rotation around the diagonal axis between the $x$- and the $z$-axis on the sphere.

The $Z$-gate question included distractors involving confusion of a $\pi$-phase shift with a $\pi/2$-phase shift, the misconception that Z-gates swap magnitudes, and a combination of both. For the phase-gate questions, we added distractors showing the phase shift in the wrong direction or doubled in either direction.

The measurement questions were designed to test whether the depiction of measurement probabilities by circle area aided participants’ reasoning. Distractors included interpreting the real part without squaring it, assuming no statement about measurement was possible, and assuming the measurement result was always 1. The second measurement question, on the resulting state after measurement, included distractors with non-normalized or not properly normalized states and unchanged states.

\paragraph{Quantum Operations in two- and three-qubit systems}

The two- and three-qubit surveys followed the same structure as the one-qubit survey, except that the phase gate and global-phase equivalence questions were omitted in favor of entanglement and separability. Distractors were included where the gate was applied to the wrong qubit or to both qubits (for CNOT, we included a SWAP gate as a distractor). Additional distractors were designed to share similar visual or numerical features, ensuring that no option stood out.

The Guess-the-Gates questions questions in the two-qubit survey asked which combination of gates had been applied and included all combinations of $H_iX_j$ or $H_iZ_j$ gates, where $i\neq j$. The corresponding questions in the three-qubit survey included the $H_iCNOT_{ij}$ or $H_jCNOT_{ji}$ operations in part A and measurement operations in part B.

The Measurement Probability questions were designed similarly to those in the one-qubit survey but referred to percentage differences in the two-qubit survey and multiples in the three-qubit survey. We expected that the visualized areas might provide hints for students less comfortable with fractional values. Distractors depicted normalized states where the measured qubit always appeared in state 0 or 1. The Measurement State questions were designed to include fewer misconceptions. One question asked participants to identify the correct phases after measurement; the correct answer was ($\varphi_1$, $\varphi_1+\pi/2$), while the distractors included ($\varphi_1+\pi$, $\varphi_1-\pi/2$), ($\varphi_1+\pi$, $\varphi_1+\pi/2$), and ($\varphi_1$, $\varphi_1-\pi/2$). In amplitude-based Measurement state questions, the amplitudes were swapped around. 

\paragraph{Entanglement and Separability}

For each of the two- and three-qubit surveys, we designed four questions on entanglement. In the two-qubit survey, two questions asked participants to identify separable states and two to identify entangled ones. One question focused on magnitude entanglement (with equal phases), and the other on phase entanglement (with equal amplitudes). In the magnitude entanglement questions, amplitudes were permuted so that only one state was entangled or separable. In the phase separability questions, distractors included different ratio configurations (see Figure \ref{fig:sep_ent}): $r_1 = e^{i\varphi}$, $r_2= e^{-i\varphi}$; $r_1 = e^{i\varphi}$, $r_2= e^{2i\varphi}$; and $r_1 = e^{i\varphi}$, $r_2= e^{-2i\varphi}$. Because there exist far more entangled states than separable ones~\cite{PhysRevA.65.064304}, the task \enquote{find the separable state} allowed for more distractor variations than \enquote{find the entangled state}. In the latter case, distractors were restricted to equal ratios, with the correct state chosen as $r_1 = e^{i\varphi}$, $r_2= e^{-i\varphi}$, and the others varying in phase direction and magnitude.

In the three-qubit survey, participants identified fully entangled or fully separable states, with partially entangled states used as distractors. To design the Entanglement questions, we started from a fully entangled state (phase or amplitude), and created distractors symmetric in regard to one qubit, making that qubit separable from the system. To design questions asking to find fully separable states, we started with a fully symmetric state as the correct answer and generated distractors by applying CNOT$_{xy}$ and CNOT$_{xz}$ for $y\neq z$, or by mixing amplitudes from multiple states. The amplitude questions only had four amplitudes that were non-zero, in the phase questions all amplitudes were non-zero. All magnitudes were kept equal to reduce task complexity. Ideally, the number of questions would be doubled to include all combinations of entanglement/separability and phase/amplitude with and without visualization. In this pilot study, however, each participant was shown all four types only in one condition (with or without visualization), not both.

\subsection{Cognitive load}\label{sec:cognitive_load}

Perceived cognitive load was measured after each part of the survey (A and B). The following Cognitive Load test items were adopted from~\cite{Klepsch2017} to measure \acf{icl}, \acf{gcl}, and \acf{ecl}: 

\begin{description}
    \item[ICL question 1] For the tasks of survey part A/B, many things needed to be kept in mind simultaneously.
    \item[ICL question 2] The tasks of survey part A/B were very complex.
    \item[GCL question 1] For the tasks of survey part A/B, I had to highly engage myself.
    \item[GCL question 2] For the tasks of survey part A/B, I had to think intensively what things meant.
    \item[ECL question 1] During the tasks of survey part A/B, it was exhausting to find the important information.
    \item[ECL question 2] The design of the tasks of survey part A/B was very inconvenient for learning. 
    \item[ECL question 3] During the tasks of survey part A/B, it was difficult to recognize and link the crucial information.
\end{description}

Responses were given on a Likert scale from 0 (completely wrong) to 6 (absolutely right).

We chose the naive version of the questionnaire from~\cite{Klepsch2017} to measure \ac{gcl}, as we concluded that these items fit our use case of problem-solving tasks more appropriately. However, given the limited effects observed for the naive \ac{gcl} items in~\cite{Klepsch2017}, the interpretation of \ac{gcl} should be viewed critically. The selected items target perceived mental effort -- specifically, cognitive engagement and intensity of thinking -- which are assumed to correlate with \ac{gcl}~\cite{deleeuw2008comparison}.

\subsection{Data analysis}

The small sample size of this pilot study does not allow for reliable statistical analysis. The sample is also heterogeneous, which is typical in \ac{qist} education, where participants’ academic backgrounds vary widely. To address these challenges, we combine insights from the quantitative results with findings from the qualitative part of the study. This approach allows us to evaluate the suitability of the test items for a heterogeneous target group and to identify potential causes of performance and cognitive load differences between visualization conditions.

From the quantitative data, we infer the appropriateness of question types by examining answer correctness, time taken, and cognitive load. Depending on whether the average correctness was above 84\%, between 51–84\%, or below 51\%, each question type in the main study was labelled as easy, moderate, or hard, respectively. We further distinguished between low and high variance, depending on the observed differences between and within groups with and without visualization. Larger differences indicate either stronger dependency on learner prerequisites, the presentation of visualization, or their interaction, and warrant further exploration. In this way, we identify promising directions for more targeted and robust future studies.

\section{Interview Summary and Discussion}

In order to gain a deeper understanding of the cognitive processes associated with answering the questions of this survey, we interviewed five students who had similar backgrounds and prior experiences to those of the main study cohort. In doing so, we also assess the quality of the questions, looking especially for false reasoning that leads to correct answers. Here, we thematically summarize the takeaways from the individual participants' thought processes with regard to the following questions: Do the students find answer options that correspond to their misconceptions? (How) are the visualizations utilized? What problems exist when solving tasks with the visualizations and do these problems correspond to problems in representational competence?

For an overview of the individual performances of the students in the tests, the scores of the five participants are shown in Table \ref{tab:interview_scores} and the perceived cognitive load is shown in Table \ref{tab:interview_cl}. 

\subsection{Representational competence}

Problems with the representational competence questions are summarized in Table \ref{tab:interv_rep_comp}. Although three of the five students achieved perfect scores in the symbolic and the visual representational competence, some problems still became apparent. Student no. 1 had no trouble with the symbolic form of complex numbers, but associated the left side of the circle with negative numbers. Student no. 2 had trouble with both the symbolic and the visual representation, twice reversing the direction of the angle of $e^{-i\pi/4}$ and associating the gauge in the circle with the polar form in the complex plane, describing the line going to the right as the number 1. Student no. 3 saw complex numbers as a representation of a quantum state as a superposition of the real part and the imaginary part and had problems with properties of exponents, while also having the misconception that positive angles are clockwise. Student no. 4 had no problems with the mathematical notation, but in the visual representation associated the right side of the circle with both negative real part and negative imaginary part of the complex number. Student no. 5 showed no difficulties.

\subsection{One-qubit Survey}\label{sec:interv_one_q}

The problems and strategies used in the one-qubit survey are shown in Tables \ref{tab:interv_one_qubit} (Z and phase gate), \ref{tab:interv_one_qubit_xz} (Global Phase and X gate), and Table \ref{tab:interv_one_qubit_h} and \ref{tab:interv_one_qubit_h_2} (H gate and Measurements) in the Appendix. Here, we summarize the key takeaways. In the Z and Phase gate questions, it became apparent that problems with the visualization included confusion about the gauge angle direction. Strategies that circle notation allows include rotating the line and solving exercises by approximation. Such visual strategies are also used in the Global Phase question, where student no. 1 had no strategy without the visualization. Student no. 2 solved the question with visualization, using the symbolic notation to check their answer, according to the \textit{constrain interpretation} function of the \acf{deft} framework. Most students were confused by the X gate question, leading them to disregard parts of the characteristics of the state in order to align the answers with their own ideas; only student no. 5 recognized that a global phase change should be added to arrive at the correct answer. The H gate question was similarly difficult for the students, with some giving up and others trying to make sense of it by the explanation of the H gate in the question, without success. Only student no. 4 solved this question correctly not by chance, and only with visualization, but with a mathematical strategy of matrix multiplication. No visual strategies were used to solve the Measurement Probability questions. This choice might be due to difficulty in approximating areas that has been observed in, e.g.,~\cite{Raidvee2020}.

Some students missed the fact that magnitudes must be squared to obtain measurement probabilities. Student no. 1 approximated the answer by squaring the numbers. Another identified the square root depiction as a hint that measurement probabilities should be squared. The students did not experience problems with the Measurement State questions. In summary, visual strategies only seemed relevant in the question involving manipulation of phases. In these questions, visualization complemented the mathematics by providing alternative strategies, in accordance with the \ac{deft} framework.

\subsection{Two-qubit Survey}\label{sec:interv_two_q}

In the two-qubit survey, the student were divided into two more groups: Students no. 1 and 2 were assigned to \acf{cn} and students no. 3, 4 and 5 to \acf{dcn}. The problems and strategies used in the one-qubit survey are shown in Tables \ref{tab:interv_two_qubit_x} (X and H gate), \ref{tab:interv_two_qubit_g} and \ref{tab:interv_two_qubit_g_2} (Guess-the-Gates and Measurements), and Table \ref{tab:interv_two_qubit_s} and \ref{tab:interv_two_qubit_s_2} (Separability and Entanglement) in the Appendix. In the X gate questions, the students assigned to \ac{cn} chose to swap the amplitudes of basis states in the standard way, while those assigned to \ac{dcn} adopted visual strategies. Only student no. 3 experienced difficulties in the question with visualization, as they swapped states where qubit \#2 is 1 (i.e., a CNOT$_{21}$), instead of applying an X$_2$ gate correctly. Here, the visual representation might have led to a reduction in the use of cognitive processes that are actually necessary to solve the question, as student no. 3 had answered the X gate question without visualization correctly and with correct reasoning.  

In the H gate questions, student no. 1 did not know the concept of the H gate collapsing superpositions, while student no. 3 and student no. 4 chose the wrong qubit to perform the H gate on. The other students adopted successful visual strategies, arguing with \ac{cn} or \ac{dcn}, but were equally successful without visualization. The Guess-the-Gates question was more difficult, but was still solved correctly by most of the students with and without visualization. In many cases a random guess or intuition was the reason for this, and the students were in general cognitively overloaded by the question or lacked a strategy for solving it. Only student no. 5 solved both Guess-the-Gates questions with a correct strategy, while student no. 4 solved the question correctly only without visualization by comparing special features of the coefficients. Student no. 2 solved the questions with visualization using a visual strategy, while relying on intuition without visualization. 

In the Measurement Probability questions, student no. 1 had conceptual difficulties regarding what a measurement of only one qubit of a two-qubit system represents. Student no. 2 calculated the probabilities, ignoring the visualization, while students no. 3 and 5 tried solving by approximation but lacked knowledge of a suitable strategy. Student no. 4 had mathematical difficulties. Like in the one-qubit survey, all students ignored the visualization in the Measurement Probability questions. This, again, hints towards the visualization not providing much benefit to the participants when area comparisons would be required for a visual task-solving strategy.

In the Separability and Entanglement questions, student no. 1 and 3 lacked conceptual understanding. They identified entanglement as a characteristic of \textit{what happens to one qubit also happens to another}, i.e., only definable by action, and picked answers at random or by intuition. Student no. 2 only lacked coherent strategies in the questions without visualization. They correctly identified separable states by a notion of \enquote{following the trend}, whereas they thought the question to find an entangled state was about finding a separable state. Student no. 4 and 5 adopted successful visual strategies, whereas student no. 4 imagined the circle notation in both questions where it was not presented, in order to solve the question. This strategy requires connectional understanding, and is an example of connectional understanding providing benefits even when the representation in question is not present. Student no. 5 remembered the rule $\alpha_{00}\alpha_{11}= \alpha_{01}\alpha_{10}$ for the coefficients of separable states to solve the question in symbolic notation. 

In summary, we see a variety of different strategies in two-qubit systems, both symbolic and visual. These are not only learner dependent, but also dependent on context. For example, the students did not use visual strategies to solve Measurement questions, but preferred such strategies for the Entanglement and Separability questions when they knew how to adopt them. The results point towards advantages of the visualization by offering strategies not available in the mathematical notation. At the end of the survey, student no. 1 stated: \enquote{At least with two qubits it's way easier to do the operations, it's just switching things around.} Student no. 5 stated: \enquote{I found the circle notation interesting, especially for entanglement and separability. I think it is easy to see whether they are separable or entangled and I did not know that it is possible to see it so easy and I also think it was sometimes hard for me to calculate things in my head.}, explaining a decrease in cognitive load with the visualization.

\subsection{Three-qubit Survey}\label{sec:interv_three_q}

Only two students (no. 2 and 4) participated in the three-qubit survey. We note that the voluntary participation of the two students in the three-qubit survey already involves a selection bias, as they were not turned off by the already quite complex one- and two-qubit surveys, such that these students could be seen as higher-performing students in general. The main interview results are shown in Tables \ref{tab:interv_three_qubit_x} (X and Z gate), \ref{tab:interv_three_qubit_h} (H gate), \ref{tab:interv_three_qubit_m} (Measurements), \ref{tab:interv_three_qubit_s} (Separability and Entanglement), and \ref{tab:interv_three_qubit_c} (CNOT and Guess-the-Gates) in the Appendix. In the X and Z gate questions, student no 4 adopted visual strategies where presented with visualization. Student no. 2 compared coefficients in the symbolic notation to find the correct answer to the X gate questions and had trouble with the Z gate question because they thought the Z gate would switch the phases instead of just changing the relative phase. By contrast, student no. 2 used a visual strategy to solve the H gate question. Here, student no. 4 observed the coefficients to arrive at an answer, even when presented with visualization, and had trouble identifying the \enquote{partner} states for the H gate to combine into one. 

In the Measurement Probability questions, both students calculated the answer with and without visualization. In the Measurement State question, student no. 2 lacked conceptual understanding of renormalization after measurement and indicated that the states merged to one upon being measured. By contrast, student no. 4 arrived at the correct answer by visual comparison of the phases. In the Separability and Entanglement questions, student no. 2 lacked a strategy on two questions. However, they picked the correct answer by intuition on the Separability question without visualization and adopted a visual strategy on the Entanglement question with visualization. Student no. 4 adopted purely visual strategies successfully, imagining \ac{dcn} in their head to solve also the questions where the visualization was not present. Both students adopted visual strategies also in the CNOT gate question when presented with visualization, swapping states along the axis of the qubit in question, but also managed to correctly solve the question without visualization. Both students had problems solving the Guess-the-Gates questions. Here, student no. 2 again had conceptual difficulties with measurements and was cognitively overloaded on the question without visualization. Student no. 4 applied the gates in the wrong order in the question with visualization. 

At the end of the three-qubit survey, we asked for key takeaways. The students responded with positive attitudes towards \ac{dcn}. Student 2 stated: \enquote{It was awful trying to do any switches and stuff in the non-DCN, the bra-ket notation. There was so much to manually think about in my head. In DCN I could just see it in front of me, it was comparatively easier so I didn't have to keep a lot in my head. It was absurdly better in the three-qubit system than in two. Three qubits needs DCN.} Student 4 said: \enquote{I found it humane that especially in those questions where the circle notation was not there anymore, that it didn't get out of hand, that at least the amplitudes were equal and only two phase shifts were there. The questions without circle notation were not too complex. Especially entanglement the circle notation helps a lot. Sometimes I use with Hadamard for example the matrix or the direct calculations, but all in all the circle notation helps in many things significantly. [translated from German]} We see students particularly benefiting from clarification of the role of a single qubit to the whole system, utilizing strategies along the axis of that qubit in the X, H and CNOT gate questions, while also utilizing visual strategies to compare coefficients to identify (partially) separable or (partially) entangled states. A requirement for the use of these strategies, however, is conceptual knowledge. 

In summary, we can see that \ac{dcn} can provide different strategies to solving these questions depending on context, previous knowledge, and representational competence of the students, in accordance with the \ac{deft} framework. The questions were suitable for the students, as the problem-solving strategies used were -- although very different -- generally reasonable, and problems mostly lay in the execution rather than in conceptual knowledge. We can see that connectional understanding can play an important role in solving these questions, even if \ac{dcn} is not present, as some students are able to imagine the representation in their heads. Interestingly, this could mean that connectional understanding reduces the difference in performance with and without visualization, suggesting that a measure of conceptual competence with visualization should be adopted in future studies in addition to measuring connectional understanding. Although the three-qubit survey had only two participants, further investigation in this realm may be fruitful.

\section{Quantitative Results \& Discussion}\label{sec:discussion}

In the following, we refer to the group \enquote{math-vis} as the group that was assigned questions without visualization in survey part A and questions with visualization in part B. Vice versa, the group \enquote{vis-math} was assigned questions with visualization in survey part A and questions without visualization in part B. The quantitative data are shown in Appendix \ref{sec:A_results}.

We discuss the results obtained regarding answer correctness and time taken for correct answers for the one-qubit (Section \ref{sec:disc_one}), two-qubit (Section \ref{sec:disc_two_CN} and \ref{sec:disc_two_DCN}), and three-qubit (Section \ref{sec:disc_three_CN} and \ref{sec:disc_three_DCN}) surveys with the intention of learning about the suitability of the questions and identifying indications of effects to explore in future studies. As described in~\cite{Lord_1952}, the difficulty of an ideal item with three distractors is 0.74. We label an item as difficult at or below an average correctness of 0.5, moderate between 0.51 and 0.84, and easy at or above 0.85~\cite{universityofwashingtonitemanalysis}. We also examine the variance of the results for each type of question with and without visualization and the differences between the two groups to separate the questions into high or low variance. We consider questions with moderate difficulty and high variance to be of the highest value for further exploration. We provide such an overview of the discussion in Section \ref{sec:disc_summary}. Table \ref{tab:dif_scores_times} summarizes the average connectional understanding, average scores, and average times for correctly solved questions. In addition, it shows the differences of these performance metrics (with visualization minus without visualization). Average \acf{icl}, \acf{gcl} and \acf{ecl} and the corresponding differences (with visualization minus without visualization) are shown in Table \ref{tab:dif_cl}. A more detailed analysis can be found in Appendix \ref{sec:A_results}.

\begin{table*}
\caption{Average scores (the maximum possible is 1 for 100\% correct) and time taken (in seconds) for correctly solved questions, average connectional understanding, and the difference of the performance metrics with visualization minus without in the different surveys.}
\begin{tabular}{@{}l|c|cc|cc@{}}
\toprule
\textbf{number of qubits} &  \textbf{one} & \multicolumn{2}{@{}c@{}}{\textbf{two}} & \multicolumn{2}{@{}c@{}}{\textbf{three}} \\
\cmidrule(r{0.75em}){1-2}\cmidrule(lr{0.75em}){3-4}\cmidrule(lr{0.75em}){5-6}
\textbf{visualization} &  \textbf{CN} & \textbf{CN} & \textbf{DCN} & \textbf{CN} & \textbf{DCN} \\
\midrule
\textbf{avg. conn. und.} & 0.81$\pm$0.23 & 0.78$\pm$0.21 & 0.7$\pm$0.23 & 0.85$\pm$0.17 & 0.89$\pm$0.12 \\ \hline
\textbf{score w/o vis.} & 0.67$\pm$0.24 & 0.59$\pm$0.25 & 0.58$\pm$0.29 & 0.62$\pm$0.26 & 0.73$\pm$0.28 \\
\textbf{score with vis.} & 0.7$\pm$0.24 & 0.61$\pm$0.23 & 0.54$\pm$0.31 & 0.68$\pm$0.25 & 0.67$\pm$0.24 \\
\textbf{score difference} & 0.03$\pm$0.25 & 0.01$\pm$0.14 & -0.04$\pm$0.3 & 0.06$\pm$0.1 & -0.06$\pm$0.13 \\ \hline
\textbf{time (s) w/o vis.} & 99$\pm$62 & 89$\pm$49 & 121$\pm$52 & 105$\pm$32 & 140$\pm$88 \\
\textbf{time (s) with vis.} & 87$\pm$49 & 114$\pm$73 & 117$\pm$75 & 110$\pm$53 & 121$\pm$53 \\
\textbf{time (s) difference} & -12$\pm$59 & 25$\pm$56 & -4$\pm$69 & 6$\pm$38 & -18$\pm$51 \\
\bottomrule
\end{tabular}
\footnotetext{Means and standard deviations are calculated taking the averages over the participants individual results.}
\label{tab:dif_scores_times}
\end{table*}

\begin{table*}
\caption{Average \acf{icl}, \acf{gcl} and \acf{ecl} with and without visualization, and the corresponding differences of with visualization minus without. The maximum possible cognitive load score is 6.}

\begin{tabular}{@{}l|c|cc|cc@{}}
\toprule
\textbf{number of qubits} &  \textbf{one} & \multicolumn{2}{@{}c@{}}{\textbf{two}} & \multicolumn{2}{@{}c@{}}{\textbf{three}} \\
\cmidrule(r{0.75em}){1-2}\cmidrule(lr{0.75em}){3-4}\cmidrule(lr{0.75em}){5-6}
\textbf{visualization} &  \textbf{CN} & \textbf{CN} & \textbf{DCN} & \textbf{CN} & \textbf{DCN} \\
\midrule
\textbf{ICL w/o vis.} & 3.73$\pm$1.22 & 3.92$\pm$1.04 & 4.69$\pm$1.55 & 4.62$\pm$0.99 & 4.57$\pm$1.29 \\
\textbf{ICL with vis.} & 3.3$\pm$1.27 & 4.0$\pm$1.13 & 4.0$\pm$1.46 & 4.4$\pm$1.11 & 3.5$\pm$1.8 \\
\textbf{ICL difference} & -0.42$\pm$1.32 & 0.18$\pm$1.11 & -0.55$\pm$1.19 & -0.25$\pm$1.2 & -0.83$\pm$1.77 \\ \hline
\textbf{GCL w/o vis.} & 4.2$\pm$1.39 & 4.73$\pm$0.86 & 4.84$\pm$1.15 & 4.89$\pm$1.37 & 4.86$\pm$1.36 \\
\textbf{GCL with vis.} & 3.91$\pm$1.21 & 4.27$\pm$0.86 & 4.73$\pm$1.0 & 4.4$\pm$1.36 & 3.57$\pm$1.59 \\
\textbf{GCL difference} & -0.25$\pm$1.41 & -0.5$\pm$0.92 & -0.11$\pm$0.92 & -0.44$\pm$1.26 & -1.29$\pm$1.83 \\ \hline
\textbf{ECL w/o vis.} & 2.67$\pm$1.35 & 2.48$\pm$0.82 & 3.98$\pm$1.34 & 3.05$\pm$1.17 & 3.21$\pm$1.46 \\
\textbf{ECL with vis.} & 2.4$\pm$1.39 & 2.59$\pm$0.82 & 3.23$\pm$1.62 & 2.75$\pm$1.33 & 1.86$\pm$1.38 \\
\textbf{ECL difference} & -0.3$\pm$1.72 & 0.11$\pm$1.26 & -0.75$\pm$1.66 & -0.3$\pm$1.33 & -1.36$\pm$2.03 \\
\bottomrule
\end{tabular}
\footnotetext{Means and standard deviations are calculated taking the averages over the participants individual results.}
\label{tab:dif_cl}
\end{table*} 

\subsection{One-qubit systems}\label{sec:disc_one}

In the first survey, participants were asked to find 

\begin{itemize}
    \item a state that is equal to another state apart from a global phase,
    \item the resulting state after application of an X gate (only in part A) or a Z gate (only in part B), phase gate, and a Hadamard (H) gate,
    \item the probability of finding a particular result upon measurement of a given state, and
    \item the resulting state after measurement.
\end{itemize}

The average scores were $0.67 \pm 0.24$ without visualization and $0.70 \pm 0.24$ with visualization, at average times of $104 \pm 74$ and $88 \pm 56$ seconds, respectively. In general, the questions seemed suitable for the participants, as the mean scores were mostly above 50\%, except for the X gate and Measurement Probability questions. The necessity to think about global phase equivalence could be a reason for the low performance in the X gate question, while many participants might have had no prior knowledge of or forgotten to square measurement probabilities. The average scores for single-qubit questions are shown in Table \ref{tab:1qubit_correctness} and the average times for correctly solved questions are shown in Table \ref{tab:1qubit_time}. Figure \ref{fig:score_2} shows the combined average score per question and Figure \ref{fig:time_2} shows the combined average time taken (for all answers, not only the correct ones).

\paragraph{Representational understanding} 

Participants were asked to calculate Euler formula identities and similar questions to translate between \acf{cn} and a complex number. In addition, they were asked to multiply the numbers written in \ac{dn} and \ac{cn} as shown in Figure \ref{fig:rep_comp_example}. Then they were asked to rate their confidence for the question from 0 (random guess), 1 (very unsure) to 4 (very sure). The correct answers were weighted (multiplied by $x/4$, where $x$ is the confidence). Detailed results are shown in Section \ref{sec:A_rep_comp}.

None of the participants scored higher with \ac{cn} than with \acf{dn} (weighted or unweighted), suggesting that the representational understanding of most participants of the visualization was lower than that of \ac{dn}. Presenting the visualization to students with high representational understanding with \ac{cn} but low representational understanding with the \ac{dn} would probably show greater benefits, but these students might be difficult to find.

All in all, we see that possible benefits of visualization are learner- and context-dependent and, if present, are probably not strongly dependent on the presence of the visualization, that is, we expect small effects if any. It is worth investigating further under which circumstances benefits occur. In one-qubit systems, the \ac{cn} is merely a visual representation of complex numbers, providing an aid, in theory, complementary to the \ac{dn}. In two- and three-qubit systems, \acf{dcn} is an extension of \ac{cn} that could provide additional benefits due to the introduction of dimensionality.

\paragraph{Connectional understanding and cognitive load}

The average connectional understanding was 0.81$\pm$0.23. Thirteen students achieved perfect scores and ten students achieved a score between 0.5 and 0.75. The students who did not score perfectly also scored lower in the rest of the survey (0.57$\pm$0.10 to 0.74$\pm$0.20 without visualization and 0.57$\pm$0.25 to 0.78$\pm$0.21 with visualization) and took less time on the correctly solved questions (82$\pm$74 to 114$\pm$71 seconds without visualization and 86$\pm$57 to 89$\pm$55 seconds with visualization) than the students who solved the connectional understanding questions perfectly. 
The average perceived \acf{ecl} was 2.4$\pm$1.39 with visualization and 2.67$\pm$1.35 without, average \acf{icl} was 3.3$\pm$1.27 with visualization and 3.73$\pm$1.22 without, and \acf{gcl} was on average, 3.91$\pm$1.21 with visualization and 4.2$\pm$1.39 without. In the difference of perceived \ac{ecl} of survey parts supported by visualization minus parts not supported by visualization, the numbers vary more (as can be seen in the high variance of 1.72 compared to 1.32 and 1.41 in \ac{icl} and \ac{gcl}, respectively; see Table \ref{tab:dif_scores_times}). This high variance could indicate a large dependency of \ac{ecl} reduction due to visualization on learner characteristics.

\subsection{Two-qubit systems}\label{sec:disc_two}

In both parts of the second survey, participants were split into two groups, one presented with \ac{cn} and one presented with \ac{dcn}, and were asked to find 

\begin{itemize}
    \item the resulting state after application of an X gate and a Hadamard (H) gate,
    \item the probability of finding a particular result upon measurement of a given state,
    \item a separable state and an entangled state (between non-separable, or non-entangled states, respectively), and
    \item a combination of two gates that were applied to transform one given state into another given state (\enquote{Guess-the-Gates}) questions).
\end{itemize}

The average scores for two-qubit questions are shown in Table \ref{tab:2qubit_correctness} and the average times for correctly solved questions are shown in Table \ref{tab:2qubit_time}. In the latter, the combined scores of part A and part B show only cases where participants managed to solve both questions, i.e., both questions on the same concept with and without visualization, correctly. Figure \ref{fig:score_2} shows the combined average score per question and Figure \ref{fig:time_2} shows the combined average time taken (for all answers, not only the correct ones).

\subsubsection{Two-qubit systems with Circle Notation}\label{sec:disc_two_CN}

In the two-qubit survey with \ac{cn}, the average scores were $0.59 \pm 0.25$ without visualization and $0.61 \pm 0.23$ with visualization, at average times of $82 \pm 47$ and $113 \pm 76$ seconds, respectively. In both survey parts, we see group math-vis scoring higher than group vis-math on all questions except the Separability and Entanglement questions, where group vis-math scored higher. This could be explained by learner characteristics, such as the ability to identify entanglement by visual means or using mathematical strategies. The H gate question shows a large difference in score between the two groups, where group math-vis scored 0.86 without visualization and group vis-math 0.17. This could hint toward the Hadamard gate question performance being especially sensitive to learner prerequisites.

When presented with the visualization, participants took longer when solving questions correctly, at $113 \pm 76$ seconds versus $82 \pm 47$ seconds. This could be the result of alternative strategies to solve the question that the visualization provides, which bring the question's difficulty closer to the level of the participant's ability, according to the distance-difficulty hypothesis~\cite{THISSEN1983179,ferrando2007item}. 

\paragraph{Connectional understanding and cognitive load}

Similarly to the one-qubit survey, participants scored $0.85 \pm 0.14$ in translating between \ac{cn} and \ac{dn}. Again, we see that participants who scored perfectly when solving these questions also scored higher in the rest of the survey ($0.88 \pm 0.14$ without and $0.83 \pm 0.17$ with visualization) than participants without a perfect score ($0.45 \pm 0.15$ without visualization and $0.50 \pm 0.16$ with visualization).

The group of participants with perfect connectional understanding scores also took more \enquote{reasonable} amounts of time to correctly solve questions, i.e., the variance in time taken was lower (100$\pm$39 seconds without visualization and 120$\pm$40 with visualization) than in the group without a perfect connectional understanding score (68$\pm$47 seconds without visualization and 107$\pm$91 with visualization). Presumably, students that did not know the answer and chose to take a guess took a shorter amount of time, while students that tried hard to find an answer in tasks where this would not be necessary for students with greater prior knowledge will take a long time. Therefore, a higher variance in the time taken for task solving could be expected for lower-performing students~\cite{THISSEN1983179,ferrando2007item}.

The average \ac{icl} was $3.92 \pm 1.04$ without and $4.0 \pm 1.13$ with visualization, \ac{gcl} ($4.73 \pm 0.86$ without and $4.27 \pm 0.86$ with visualization, and \ac{ecl} was $2.48 \pm 0.82$ without and $2.59 \pm 0.82$ with visualization. It seems that, on average, the questions succeeded in offering a high degree of intrinsic difficulty, as indicated by the comparatively high \ac{icl}.

\subsubsection{Two-qubit systems with Dimensional Circle Notation}\label{sec:disc_two_DCN}

In the two-qubit survey with \ac{dcn}, the average scores were $0.58 \pm 0.29$ without visualization and $0.54 \pm 0.31$ with visualization, at average times of $121 \pm 58$ and $139 \pm 99$ seconds, respectively. In all question types except the X gate, one group performed better than the other group in both part A and part B (regardless of whether visualization is presented). This consistency can possibly be explained by learner prerequisites predicting performance better than the visualization condition. In addition, this could reflect context dependency of student competence, i.e., competence in one quantum process does not necessarily transfer to others. The differences between the scores of the groups are particularly apparent in the H gate and Entanglement questions in favor of group vis-math (averaging over the question types: 0.56 vs 0.14, and 0.81 vs 0.43, respectively) and the Guess-the-Gates questions in favor of group math-vis (0.44 vs 0.1). 

The H gate and Guess-the-Gates questions seemed quite difficult, as the average scores in these questions were below 0.4. Feedback from some students after the test suggested that the Guess-the-Gates questions were particularly challenging and mentally demanding.

Considering the time spent on correctly solved questions, we see outliers of above 6 minutes in the group vis-math on the Z gate A, H gate A, and H gate questions of part B. Participants took, on average, a similar amount of time when presented with visualization or without (121$\pm$52 seconds without visualization vs 117$\pm$75 seconds without).

\paragraph{Connectional understanding and cognitive load}

Average connectional understanding of participants in the two-qubit \ac{dcn} survey was $0.7 \pm 0.23$. No one in the \ac{dcn} group correctly answered all the connectional understanding questions. The average scores in this survey were also lower than in all other surveys, at $0.58 \pm 0.29$ without visualization and $0.54 \pm 0.31$ with visualization.

For participants of the \ac{dcn} group, \ac{icl} and \ac{ecl} were higher than for participants of the \ac{cn} group when solving questions without visualization, at $4.7 \pm 1.6$ and $4.0\pm1.3$, respectively, with \ac{ecl} higher than in all other surveys. We see indications of reductions in \ac{icl} and \ac{ecl} when participants are presented with \ac{dcn} of $-0.6 \pm 1.2$ and $-0.8 \pm 1.7$. We do not see a similar reduction in \ac{gcl} as we do in the \ac{cn} group. The cognitive load results indicate potential reductions in \ac{icl} and \ac{ecl} when students are presented with \ac{dcn} specifically, as opposed to \ac{cn}. If students adopt new strategies to solve questions using \ac{dcn} rather than \acl{dn}, a reduction in \ac{icl} could be theoretically explained if these strategies make the tasks easier to solve. 

\subsection{Three qubits}\label{sec:results_three}

In the third survey, participants were asked to find: 

\begin{itemize}
    \item the resulting state after application of an X gate, a Z gate, and a Hadamard (H) gate,
    \item the probability of finding a particular result upon measurement of a given state,
    \item the resulting state after a measurement,
    \item the resulting state after application of a CNOT gate,
    \item a fully separable state and a fully entangled state (between partially separable/entangled states), and
    \item a combination of two gates that were applied to transform one given state into another given state (``Guess-the-Gates'' questions).
\end{itemize}

The average scores for the three-qubit questions are shown in Table \ref{tab:3qubit_correctness} and the average times for correctly solved questions are shown in Table \ref{tab:3qubit_time}. In the latter, the combined score of part A and part B show only cases where participants managed to solve both questions, i.e., two questions of the same type, correctly. Figure \ref{fig:score_3} shows the combined average score per question and Figure \ref{fig:time_3} shows the combined average time taken (for all answers, not just the correct ones). Cases where a participant took more than 10 minutes for one question were excluded.

\subsubsection{Three-qubit systems with Circle Notation}\label{sec:disc_three_CN}

In general, the participants scored marginally higher when presented with \ac{cn} ($0.68 \pm 0.25$) than when presented with the questions without visualization ($0.62 \pm 0.28$). The overall score reflects that the questions were suitable for the participants in terms of difficulty. The results of the H gate and Guess-the-Gates questions could indicate a benefit of presenting the visualization in these contexts, as both groups scored higher when visualization was presented (0.9 vs 0.6 for the H gate question and 0.6 vs 0.1 for the Guess-the-Gates question). 
In the entanglement question, group math-vis scored higher (0.6 without visualization and 0.4 with visualization) than group vis-math (0.2 with and without visualization) in both parts of the survey. These results could indicate differences in context-dependent competence of the two groups, similar to what we observed in the two-qubit survey, in the identification of entangled states. In some cases, students seemed to benefit from the visualization, whereas on average there is no difference. 

The results warrant further investigation in the regime of the more complex H gate, Guess-the-Gates, Separability, and Entanglement questions to find the reasons for the increases in performance with visualization for some participant groups and large deviations. However, due to the especial complexity of the Guess-the-Gates questions, it should probably not be included in future studies involving students not as well-versed in \ac{qist}. To benefit from visualization in the context of separability and entanglement, students may need a thorough introduction to the strategy to identify separability and entanglement (see Figure \ref{fig:sep_ent}). We received feedback from some participants that they did not comprehend the process of identifying entangled states before or during the study. Perhaps, the topic is too advanced as only a subtopic of a study like this and should be introduced more thoroughly in a study with only this context.

\paragraph{Connectional understanding and cognitive load}

In the three-qubit survey, participants presented with \ac{cn} scored on average $0.85 \pm 0.17$ on the connectional understanding questions, similar to the results of the one- and two-qubit \ac{cn} surveys. Participants who scored perfectly in connectional understanding also scored comparatively high in the rest of the survey at $0.80 \pm 0.16$ without visualization and $0.84 \pm  0.17$ with visualization, compared to $0.44 \pm 0.21$ and $0.51 \pm 0.19$ for the other participants. However, we do not see this connectional understanding translate into an improvement of score when visualization is presented.

When participants were presented with \ac{cn}, we observe indications of a reduction in \ac{icl}, \ac{gcl} and \ac{ecl} of $-0.25\pm1.28$, $-0.44\pm1.33$ and $-0.3\pm1.4$, respectively. These results justify further investigation into the circumstances under which a reduction in cognitive load occurs through presentation of \ac{cn}. 

\subsubsection{Three-qubit systems with Dimensional Circle Notation}\label{sec:disc_three_DCN}

Group math-vis consists of two students and group vis-math of three participants. Hence, the resolution on the average score value per question is coarse. Due to the large impact a single person's performance can have on the results, it is difficult to find reliable indications of effects. Future studies could focus more on the impact of \ac{dcn} in the domain of three qubits in particular. If there is any benefit of visualizing quantum states representing qubits as axes in space, one would expect the benefit to be the largest here, because the difference between the dimensional approach and standard \ac{cn} increases with system size. Similarly to the results for answer correctness, it would be unreasonable to make statements about time taken for correct answers due to the small number of participants.

\paragraph{Connectional understanding and cognitive load}

The impact of the results of one participant seems too large to make any firm assumptions about the connectional understanding and cognitive load in the three-qubit \ac{dcn} context. 

\section{Summary of discussion \& Limitations}\label{sec:disc_summary}

In the qualitative part of the study, we observed that students adopt a multitude of different strategies when solving questions with visualization. These strategies are learner- and context-dependent, as predicted by the \acf{deft} framework. They lie in the \textit{constrain interpretation} and \textit{complementing} domains, using symbolic notation to perform more accurate calculations or to check answers, and also utilizing visualization especially to determine actions where phase changes are involved, where gates act along the axis of qubits, or to identify (partially) separable or (partially) entangled states. This lays the foundation for expecting performance and cognitive load differences when students solve questions with visualization as compared to when they do not.  However, we also observe hints of context dependency: for example, visualization seems to be used less when the question requires comparison of areas, as in the Measurement Probability questions. In general, we find the suitability of the questions for a heterogeneous target group, as these questions were neither too difficult nor too easy for the five students. The group was recruited from our quantum computing lecture and was relatively new to the field, and as such can be seen as representative, although the size of the group means that the number of different strategies that could be identified is limited. 

Considering the suitability of the questions in terms of performance alone and the variance between the two groups with or without visualization, we summarize the results of this study in Table \ref{tab:question_overview}, where we assigned questions a label of easy (average score above 0.84), moderate (between 0.51 and 0.84) or hard (below 0.51), and labeled half the questions with differences between groups and/or whether visualization was presented above the median as high-variance, and below the median as low-variance. We consider questions of moderate difficulty and questions with higher variance to be of value for further investigation, as it is unclear whether this variance is due to learner characteristics or the presentation of the visualization. Overall, questions of moderate difficulty and high variance in at least two different surveys include the H and the Z gate (three-qubit systems), CNOT gate (two-qubit \acf{cn} and three-qubit \acf{dcn}), and Entanglement (two qubits) questions. However, in both three-qubit surveys, one participant group answered the Z-gate question perfectly, indicating possible ceiling effects in that question type in particular. 

In general, we do not see indications of a large change in performance when participants are presented with a visualization (see Table \ref{tab:dif_scores_times}), although there are indications of context-dependent performance increases with visualization that require further investigation. These increases are apparent in the Global Phase Equivalency question of the one-qubit survey, the Z gate, the Measurement Probability, and the Separability question of the two-qubit \ac{cn} survey, the H gate question of the two-qubit \ac{dcn} survey, and the H gate and Guess-the-Gates questions of the three-qubit \ac{cn} survey. However, context-dependent learner characteristics seem to have a greater impact on performance than the presence or absence of a visualization. This could be due to students with better background knowledge already knowing how to solve the questions regardless of whether visualization is present or not, while students with less background knowledge might not know how to utilize the visualization. Some groups seem to perform better than other groups specifically in certain types of questions, as we see in the two-qubit \ac{cn} survey in the X gate, the H gate, Separability and Entanglement questions, all questions in the two-qubit \ac{dcn} survey, and in the Z gate, Measurement Probability, Entanglement, CNOT gate, and Guess-the-Gates questions in the three-qubit \ac{cn} survey, further highlighting the dependency of possible effects on the interaction between context and task solvers.

\begin{table*}[ht]
\centering
\begin{tabular*}{\linewidth}{@{\extracolsep\fill}l|cc|cc|cc|cc|cc}
\toprule
\textbf{Question Type} & \multicolumn{2}{@{}c@{}}{\textbf{1-Qubit CN}} & \multicolumn{2}{@{}c@{}}{\textbf{2-Qubit CN}} & \multicolumn{2}{@{}c@{}}{\textbf{2-Qubit DCN}} & \multicolumn{2}{@{}c@{}}{\textbf{3-Qubit CN}} & \multicolumn{2}{@{}c@{}}{\textbf{3-Qubit DCN}} \\
\midrule
& \textbf{dif.} & \textbf{var.} & \textbf{dif.}&  \textbf{var.} & \textbf{dif.} & \textbf{var.}& \textbf{dif.}&\textbf{var.}&\textbf{dif.}&\textbf{var.} \\
\cmidrule(r{0.75em}){2-3}\cmidrule(lr{0.75em}){4-5}\cmidrule(lr{0.75em}){6-7}\cmidrule(lr{0.75em}){8-9}\cmidrule(lr{0.75em}){10-11}
\textbf{Global Phase Equiv.}   & \textbf{m} & \textbf{hv}        & -        & -        & -       & - &- &- &- & -\\
\textbf{X Gate}                & h & lv &\textbf{m} & \textbf{hv}& e &lv &e &lv & e & hv \\
\textbf{Z Gate }               & m & - & - & - &  -& -&\textbf{m}&\textbf{hv}&\textbf{m}&\textbf{hv} \\
\textbf{H Gate}                & m & lv & e & hv &h  &hv&\textbf{m}&\textbf{hv}&\textbf{m}& \textbf{hv}\\
\textbf{Phase Gate}            & m & lv & - & - & - & - & - & - & - &-  \\
\textbf{Measurement Prob.}     & h & hv  & h&lv  &h  &hv &\textbf{m}&\textbf{hv}&e & hv \\
\textbf{Measurement State}     & m & hv & - & - & - &-&m&lv&\textbf{m}& \textbf{hv}\\
\textbf{CNOT Gate}             & -  & - & - &- & - &-&\textbf{m} &\textbf{hv} &\textbf{m}& \textbf{hv}\\
\textbf{Guess-the-Gates}       & - & - & h& lv&h &hv& h& hv& h&  hv\\
\textbf{Separability}          & - & - &  \textbf{m} & \textbf{hv}  & m & lv&h&lv&\textbf{m} &\textbf{hv}  \\
\textbf{Entanglement}          & - & - & \textbf{m}  & \textbf{hv}  & \textbf{m}&\textbf{hv}&h&hv&\textbf{m} & \textbf{hv} \\
\bottomrule
\end{tabular*}
\caption{Overview of question types across survey contexts (one-, two-, and three-qubits in \acf{cn} and \acf{dcn}). Each cell lists a difficulty level (hard (h) for scores below 0.51, moderate (m) for scores between 0.51 and 0.84, and easy (e) for scores above 0.84~\cite{universityofwashingtonitemanalysis}) and a variance indicator (low variance (lv) for a standard deviation of the four group averages (with or w/o vis. and group math-vis or group vis-math) below the median standard deviation of 0.13, and high variance (hv) for standard deviation at or above 0.13). Questions with the m/h combination are highlighted, as these questions might be of most interest in future studies.}
\label{tab:question_overview}
\end{table*} 

In order to find indicators of complexity and suitability of questions, we considered the time taken for all answers, not just correct ones (see Figure \ref{fig:time_1}, Figure \ref{fig:time_2}, and Figure \ref{fig:time_3} for the one-, two-, and three-qubit surveys, respectively). During the one-qubit survey, the Global Phase and the H gate questions took the participants the longest time. We would omit the H gate and X gate questions, as they were associated with unnecessary confusion due to incorporation of the global phase equivalency. During the two-qubit survey, the Guess-the-Gates and the Measurement Probability questions took participants in the \ac{cn} survey the longest, and the Guess-the-Gates and H gate questions for participants of the \ac{dcn} survey (see Figure \ref{fig:time_2}). In the three-qubit survey, participants presented with both \ac{cn} and \ac{dcn} took the longest in the Guess-the-Gates and H gate questions (see Figure \ref{fig:time_3}), in the former of which some participant groups also performed quite poorly. Long times could in some cases be due to greater complexity and in some cases to the distance-difficulty hypothesis~\cite{THISSEN1983179,ferrando2007item}, which would, in contrast, state that these questions were quite suitable for the participants in terms of level of ability, making the performance in these questions an important indicator. Due to these results, we would omit the Guess-the-Gates questions in two- and three-qubit systems in future studies due to their high difficulty, as the interview results also suggest.

We also measured the cognitive load imposed on the students during the surveys. The generally high \acf{icl} and average scores between 0.5 and 0.7 indicate that the difficulty of the problems was adequate overall. We see reductions in cognitive load when students are presented with visualization in both \ac{cn} and \ac{dcn} surveys. One possibility is that the amount of space \ac{dcn} takes up compared to \ac{cn} leads to increased \acf{ecl}, due to participants not being able to see all answers at once (which means page scrolling becomes necessary), resulting in a possible split-attention effect~\cite{chandler1992split}. On the other hand, the lower connectional understanding of participants in the two-qubit \ac{dcn} survey could also be a reason for the higher \ac{ecl}. To summarize the results in regard to \textbf{RQ1}, on average, the test items were suitable, even for a heterogeneous target group, with a few exceptions.

Most of the participants were recruited from our quantum computing lecture. While it is an optional Master's lecture in information science, students from other fields also select it as an elective subject. We expect and have experienced a higher reluctance towards mathematical notation among students from other fields than Physics or Mathematics, especially if they were not introduced to \ac{dn} before. When answering the question of whether they have been introduced to \ac{dn} before, we received the feedback that some students thought that this referred to pre-lecture knowledge. We have also received positive feedback regarding the visualizations from students newer to the field, while more advanced physics students often do not see a benefit in the visualizations. A possible expertise-reversal effect remains to be investigated. The A-B survey design should mitigate previous knowledge as a relevant factor, if the questions are neither too easy nor too difficult. For future studies with such heterogeneous groups, eye tracking or a think-aloud study design might allow us to gain a better understanding of the underlying thought processes. In studies with more participants, the analysis can be performed by separating the groups according to experience with \ac{dn}. 

In this study, we have seen how performance and cognitive load can be context- and learner-dependent. The results give an overview of the cognitive load with and without visualization, but context-specific assumptions cannot be made with these results. It can also considered that measurements of cognitive load after multiple questions could be biased toward the most recently answered questions, which would be the Measurement questions in the one-qubit survey and the Guess-the-Gates questions in the two- and three-qubit surveys. Performance on these questions is probably very dependent on previous knowledge. This, combined with the high expected heterogeneity, motivates separate analyses of the results for different types of questions in future studies. 

The test instruments presented in this study are purely designed to measure the benefits of visualization in learning basic quantum operations and entanglement. They cover a broad range of quantum gates in one-, two-, and three-qubit systems, but do not address more complicated processes such as whole quantum algorithms or larger qubit-systems, and they leave out real-world applications like different qubit implementations, discussions about consequences of quantum information theory like the no-cloning theorem, and decoherence, and omit discussion about various applications.

The measures of representational competence presented in this study are limited. The questions measure only a part of representational competence -- representational and connectional understanding. Other factors for representational competence would include representational and connectional fluency and meta-representational competencies, considering the symbolic \acf{dn} as a representation and therefore extending the framework in~\cite{rau2017conditions}. In the future, additional ideas for questions that provide a more general metric should be generated. For example, meta-representational competence -- choosing between \ac{dn}, \ac{cn}, or \ac{dcn} to most efficiently solve different questions -- could be investigated using eye tracking. Representational understanding is only measured in the first survey and includes only understanding of the phase in \ac{cn}. In future studies, this should be extended to finding a measure of representational understanding of amplitudes as well, e.g., by asking students to compare numbers and sizes of circles. This type of understanding was only included in the connectional understanding questions. 

Due to small sample sizes (especially in the three-qubit \ac{dcn} survey) and large fluctuations (possibly due to large deviations of learner prerequisites), we can only identify indications of positive effects of presenting a visualization regarding performance and cognitive load. The circumstances and contexts in which these possible effects occur should be investigated more thoroughly. This study covered a broad range of different contexts, and, therefore, compromises had to be made with regard to the number of questions for each question type. To answer \textbf{RQ3}, reevaluation of the test instruments can be based on fine-tuning to make the questions in parts A and B of the surveys more similar. This can be done with respect to questions asking about, e.g., the state after measurement or entanglement, ensuring that all participants are presented with amplitude-type and phase-type questions in both parts of the survey. This would either mean an increase in the number of questions or a narrowing of question types. Due to the already large number and high complexity of the questions, we recommend the second option. %

\section{Conclusion \& Outlook}\label{sec:conclusion}

In general, the interviews, as well as the performance and cognitive load results, indicate that most of the questions were suitable for the participants. However, there were some notable exceptions. The X gate and Measurement State questions were possibly too easy, except when a global phase was also involved, as in the X gate question of the one-qubit survey. The H gate, Measurement Probability, Guess-the-Gates, Separability, and Entanglement questions were too difficult for some participant groups. Considering the overall time taken, the number of questions appears feasible for the participants. The A-B-crossover design allowed for within-subject as well as between-subject comparisons, while mitigating multiple testing effects due to randomization of the order (A-B or B-A) and we therefore see it as suitable to answer the overarching questions (see Section \ref{sec:res_ques}).

The results of this study open up a wide range of questions to investigate in future studies. Some students performed better and experienced reduced cognitive load when presented with visualization, while others did not. Therefore, it is important to investigate further under which circumstances exactly students benefit from visualizations of qubit systems, and which strategies are actually useful in this regard. In particular, contexts within the three-qubit \acf{dcn} setting should be investigated more thoroughly. Here, \ac{dcn} differs the most from the other two representations used, since the number of steps necessary to transform \acf{cn} to \ac{dcn} increases with the number of qubits. If, in fact, educators wish to teach students about systems of multiple qubits, future studies should also be restricted to multi-qubit systems. If the goal is to teach only one-qubit systems, visualizations like the Bloch sphere or vector notation could also be included in the studies. The Bloch sphere could alleviate issues with the one-qubit X and H gate questions that the students encountered during this study.

The variance in results imply a substantial learner and context dependency. Therefore, future studies should control for these characteristics in a more detailed way to draw conclusions about the potential benefits of visualizations, as also proposed in~\cite{Schewior2024}. This could also alleviate issues with question difficulty -- for example, with the Hadamard and Entanglement/Separability questions -- where proper strategies to solve these questions could be emphasized more thoroughly in introduction design. In order to support the understanding of cognitive processes involved in problem solving with visualization, the use of eye tracking techniques could be a valuable tool~\cite{PhysRevPhysEducRes.18.013102}. With eye tracking, it would be possible to observe which strategies the participants adopted to solve the questions and which strategies were used by more successful participants. This could result in further design guidelines for educators. It could also be investigated whether there are other benefits of visualization, such as a shorter fixation duration on incorrect answers when a visualization is presented~\cite{LINDNER201791}.

Representational competence as a learner characteristic is theoretically the deciding factor for whether a visualization supports learning~\cite{rau2017conditions}. In addition, we saw one student adopting a strategy of imagining the \ac{dcn} visualization in their head which necessitates a considerable amount of representational competence. This suggests that high connectional understanding could lead to smaller differences in performance for those students that adopt such strategies. Therefore, in future studies, it will be necessary to measure representational competence not only in terms of ability to translate between representations, but also in terms of conceptual ability to reason within one representation. As we have seen from the qualitative results of this study, this reasoning is very context-dependent, as the choice of representation and strategies differs per student depending on the question type. In following studies, representational understanding should also be measured utilizing the same contexts as those that are tackled.

Many students answered the representational questions perfectly, and the number of different representational competence scores was not high. Therefore, it should be considered to measure representational competence more thoroughly in future studies. This could be done, for example, by increasing the number of questions focusing on connectional understanding, and finding other ways to measure representational competence that also include visual/conceptual understanding of \ac{cn} or \ac{dcn} and quantum operations in these visualizations. Future studies could further investigate the competence of students when presented with only a visualization and how this representational competence impacts the possible benefits of presenting a visualization in addition to \ac{dn}. Perhaps also, mental-rotation ability can help explain possible beneficial effects of \ac{dcn}~\cite{Taylor2023}.

In general, we saw indications of reductions in perceived cognitive load by presenting \ac{cn} or \ac{dcn}. Due to the consistency with which this occurs, it is possible that there are beneficial effects regarding a reduction of cognitive load when students are presented with a visualization. A decrease in \acl{ecl} should, in theory, lead to improved learning outcomes when learning with visualization, provided this decrease coincides with an increase in \acl{gcl}~\cite{vanMerrienboer2005Cognitive,gerjets2009scientific}. If we also find a decrease in \acl{icl}, it could be that the questions become inherently easier when asked and solved in \ac{cn} or \ac{dcn}, which would mean they provide beneficial complementary strategies according to the \ac{deft} framework. Another point to consider is that the cognitive load of learners could be context-dependent. For example, for some question types, an effect on cognitive load from presentation of a visualization may differ from other question types. In addition, learning with visualization could also improve the competency of participants when solving questions in uncomplemented \ac{dn}. Investigating further whether there is a benefit for students who are presented with visualization, and under which circumstances and in which contexts this occurs, could therefore be of great interest. 
Studies with higher numbers of participants should separate by experience with \ac{dn} or quantum physics in general, as we expect higher reluctance towards mathematical notation for those who are not as familiar with it. 

Going beyond static multiple external representations, conveying processes in visualizations using video sequences and audio could be worth studying. Furthermore, investigating learning outcomes by including interactivity in the teaching process -- for example, via gamification -- could be an alternative path forward. There already exists a multitude of quantum games~\cite{Seskir_2022} and incorporating \ac{cn} or \ac{dcn} into these games is a possibility. Also, the possible beneficial effects of other visualization techniques and their context- and learner-dependency remain to be investigated in multi-qubit systems. 

In conclusion, there remains a large space to explore the benefits of visualizations in \acl{qist} education. In this work, we show that it is worth exploring the benefits of visualizations of quantum states due to a multitude of additional strategies that the visualizations enable. We presented test instruments to investigate the circumstances under which visualizations can support students in the field of \ac{qist} and first results that narrow down promising contexts for future studies. In order to counteract ceiling effects, we identify possible adjustments in the measurement of representational competence. The other test instruments cover a broad range of contexts and can be adapted easily for further research, also with other visualizations. However, we recommend more focused studies that use eye tracking and are much more restricted in question types. Restricting the question types will decrease the influence of previous knowledge on the variance of the results. Promising contexts include entanglement, the CNOT gate, and the Hadamard gate in multi-qubit systems, which are also some of the most important concepts of quantum computing. We expect future studies structured in this way to result in valuable takeaways for educators and consider this path forward to be highly relevant to the quantum ecosystem.

\begin{acknowledgments}
We thank the research initiative Quantum Computing for Artificial Intelligence at the RPTU Kaiserslautern-Landau for covering the study expenses.
E.R., J.B., P.L., M. K.-E., and A.W. acknowledge support by the German Federal Ministry of Education and Research (BMBF) in the QuanTUK project under grant number 13N15995. E.R. and A.W. acknowledge support by the European Union via the Interreg Oberrhein programme in the project Quantum Valley Oberrhein (UpQuantVal).
This work was supported by LMUexcellent, funded by the Federal Ministry of Education and Research (BMBF) and the Free State of Bavaria under the Excellence Strategy of the Federal Government and the Länder.
\end{acknowledgments}

All procedures performed in the study were in accordance with the ethical standards of the national research committee and the Declaration of Helsinki of 1964 and its subsequent amendments. The study involved data collection in three online surveys that took about 45-60 minutes each and interviews with some of the students. Participation in this study was voluntary and anonymous, and under informed consent. The data collected included participants’ age, gender, field of study, and highest educational achievement. Personal data of participants is kept confidential and used solely for research purposes.

\appendix

\setlength{\parindent}{0pt}

\section{Qualitative results}

In this section, data collected from the five students of the qualitative part of the study is shown; in section \ref{sec:interv_overall} the performance scores and cognitive load, in section \ref{sec:rep_comp_interv} the problems with the representational competence questions, and in section \ref{sec:one_qubit_interv}, \ref{sec:two_qubit_interv}, \ref{sec:three_qubit_interv}, their problems and problem-solving strategies in the main parts of the one-, two-, and three-qubit survey, respectively.

\subsection{Overall Scores}\label{sec:interv_overall}

Table \ref{tab:interview_scores} shows the average performance scores of the students that took part in the qualitative study. Table \ref{tab:interview_cl} shows the perceived cognitive load.

\begin{table*}
\caption{Average scores for the students that partook in the qualitative study. Scores in brackets are weighted.}
\begin{tabularx}{\linewidth}{@{}l|XXX|XX|XXX|XXX@{}}
\toprule
\textbf{no. of qubits} & \multicolumn{5}{@{}c@{}}{\textbf{one-qubit}} & \multicolumn{3}{@{}c@{}}{\textbf{two-qubit}} & \multicolumn{3}{@{}c@{}}{\textbf{three-qubit}} \\
\cmidrule(r{0.75em}){2-6}\cmidrule(lr{0.75em}){7-9}\cmidrule(lr{0.75em}){10-12}
 & \multicolumn{3}{@{}c@{}}{\textbf{rep. comp.}} & \multicolumn{2}{@{}c@{}}{\textbf{score}} & \textbf{rep. comp.} & \multicolumn{2}{@{}c@{}}{\textbf{score}} & \textbf{rep. comp.} & \multicolumn{2}{@{}c@{}}{\textbf{score}}\\
\cmidrule(r{0.75em}){1-1}
\cmidrule(r{0.75em}){2-4}\cmidrule(lr{0.75em}){5-6}\cmidrule(lr{0.75em}){7-7}\cmidrule(lr{0.75em}){8-9}\cmidrule(lr{0.75em}){10-10}\cmidrule(lr{0.75em}){11-12}
\textbf{student no.} &  \textbf{symbolic} & \textbf{visual} &  \textbf{conn. und.} &  \textbf{w/o vis.} &  \textbf{with vis.} & \textbf{conn. und.} &  \textbf{w/o vis.} &  \textbf{with vis.} & \textbf{conn. und.} &  \textbf{w/o vis.} &  \textbf{with vis.} \\
\midrule
\textbf{1} & 1(0.63) & 0.5(0.25) & 0.5 & 0.67 & 0.33 & 0.5 &0.33& 0.33 & - & - & - \\ 
\textbf{2} & 0.5(0.5) & 0.5(0.5) & 0.25 & 0.83 & 0.83 & 0.33 &0.8 & 0.8 & 0.75 & 0.67 & 0.67 \\
\textbf{3} & 0.75(0.5) & 0(0) & 0.5 & 0.33 & 0.5 & 0.33 &0.5&0.2&-&-&- \\
\textbf{4} & 1(0.88) & 0.75(0.63) & 0.75 & 0.67 & 1 &0.67 &0.8&0.6&1 &1 & 0.9 \\
\textbf{5} & 1(1) & 1(0.88)& 1 & 1 & 0.83 & 0.83 & 1 & 0.8 &-&-&- \\
\bottomrule
\end{tabularx}
\label{tab:interview_scores}
\end{table*}

\begin{table*}
\caption{Perceived Cognitive Load (\acl{icl}, \acl{gcl}, \acl{ecl}, respectively) of the students that partook in the qualitative study.}
\begin{tabular}{@{}l|c|c|c|c|c|c@{}}
\toprule
\textbf{no. of qubits} & \multicolumn{2}{@{}c@{}}{\textbf{one-qubit}} & \multicolumn{2}{@{}c@{}}{\textbf{two-qubit}} & \multicolumn{2}{@{}c@{}}{\textbf{three-qubit}} \\
\cmidrule(r{0.75em}){1-1}\cmidrule(r{0.75em}){2-3}\cmidrule(lr{0.75em}){4-5}\cmidrule(lr{0.75em}){6-7}
\textbf{student no.} &    \textbf{w/o vis.} &  \textbf{with vis.}   &  \textbf{w/o vis.} &  \textbf{with vis.}   &  \textbf{w/o vis.} &  \textbf{with vis.} \\
\midrule
\textbf{1} & 2, 6, 1 & 1, 5, 0 &4, 1, 1.5 & 6, 6, 1.5 & - & - \\
\textbf{2} & 1, 3, 2 & 1, 4, 3 & 1, 4, 1.5 &3, 4, 3 & 4, 5, 4  & 3 , 4, 1.5 \\
\textbf{3} & 5, 4, 3.5 & 3, 4, 2 &6, 6, 5.5 & 5, 5, 2.5 &-&- \\
\textbf{4} & 3, 3, 1.5 & 5, 5, 1.5 & 5, 5, 3 & 5, 6, 3 & 6, 6, 2& 5, 4, 0 \\
\textbf{5} & 5, 6, 2.5 & 2, 3, 0.5 & 4, 4, 4 & 2, 4, 0.5 &-&- \\ 
\bottomrule
\end{tabular}
\label{tab:interview_cl}
\end{table*} 

\subsection{Representational competence}\label{sec:rep_comp_interv}

Table \ref{tab:interv_rep_comp} shows the problems that emerged in the representational competence parts of the survey of those students that took part in the qualitative study. 

\begin{sidewaystable*}
\caption{Representational competence problems of the students that partook in the qualitative study.}
\begin{tabularx}{\textwidth}{lXX|XX|XX}
\toprule
\textbf{question type:} & \multicolumn{2}{@{}c@{}}{\textbf{complex numbers}} & \multicolumn{2}{@{}c@{}}{\textbf{circle notation}} & \multicolumn{2}{@{}c@{}}{\textbf{translation}} \\
\cmidrule(r{0.75em}){1-1}\cmidrule(r{0.75em}){2-3}\cmidrule(lr{0.75em}){4-5}\cmidrule(lr{0.75em}){6-7}
\textbf{student no.} &    \textbf{prob./strat.} &  \textbf{example}   &  \textbf{prob./strat.} &  \textbf{example}   &  \textbf{prob./strat.} &  \textbf{example} \\
1&complex numbers can't be real & "It has to be either $-i$ or $+i$ because a complex number has to have an imaginary part to it." & +1 corresponds to line to the right \newline \newline  association of left side with negative values \newline \newline  \newline  & "It's going to the right side so that means positive" \newline "$-\frac{1}{\sqrt{2}}-\frac{i}{\sqrt{2}}$ because [...] the angle is completely on the y-axis so when it goes on the left side it is completely negative"  & views the 60$^\circ$ angle as a 45$^\circ$ angle \newline  \newline  \newline positive angles go clockwise \newline wrong denominator comparison & "This is the half of 90$^\circ$ so that is 45$^\circ$ but it makes zero sense if it is 45." \newline "This is +45$^\circ$ so it goes on the right side."\newline "$1/\sqrt{2}$ has to be smaller than $1/\sqrt{6}$ and $1/\sqrt{3}$" 
\\ \cmidrule(lr{0.75em}){1-7}
2 & $e^{-i\pi/2}=1$ \newline $e^{i\pi/4}=\frac{1}{\sqrt{2}}-\frac{1}{\sqrt{2}}i$ & "it's 1. Somewhat sure." \newline "It doesn't have a minus so it's in the opposite direction so the angle is in the opposite so it gets a minus in the complex plane." & unsure whether polar coordinates or circle notation convention 
& "If it's the qubit notation it would be measured from here, if it's normal circle it's measured from here, but it says \textit{normal circle}." \textit{selects correct answer}   &  left is the negative side & "$e^{i\pi/4}$ is on the right hand side so it's this" (selects circle corresponding to $\frac{1}{\sqrt{2}}\ket{0}+\frac{1}{\sqrt{2}}e^{-i\pi/4}\ket{1}$)."\\\cmidrule(lr{0.75em}){1-7}
3 & the number $\frac{1}{\sqrt{2}}+\frac{1}{\sqrt{2}}i$ is a quantum state \newline \newline \newline \newline \newline \newline
problems with power laws &   "It looks like here the probabilities are in question like how probable it is for this state to be measured, like if it's a one or a zero." \newline 
"$e^{i\pi/6}\cdot e^{-i\pi/3}=e^{-i\pi/18}$, because the $1/3$ goes into the $1/6$ and the minus into the plus." & -i corresponds to line to the left \newline 
one circle representing one state that the qubit is in \newline\newline \newline \newline \newline \newline \newline negative and positive makes negative &  "left is I assume the negative side of the axis" \newline \textit{When asked about one single full circle representing} $e^{-i\pi/4}$ "How am I supposed to pick a complex number here? because based on what I know this circle notation represents which state a qubit is in."\newline "Because this is a negative complex number and this is $\pi/2$, the answer is negative" & positive angles are clockwise & $e^{i\pi/4}$ is equal to a circle corresponding to $e^{-i\pi/4}$ \\ \cmidrule(lr{0.75em}){1-7}
4 & - & -& circle corresponding to $e^{-i\pi/4}$ is equal to $-\frac{1}{\sqrt{2}}-\frac{i}{\sqrt{2}}$ & "I have the double minus, for double plus I go left around and for double minus I go right around." &- &-  \\ \cmidrule(lr{0.75em}){1-7}
5 &- &- &- &- &- &  -\\ 
\bottomrule
\end{tabularx}
\label{tab:interv_rep_comp}
\end{sidewaystable*}

\subsection{One-qubit survey}\label{sec:one_qubit_interv}

Table \ref{tab:interv_one_qubit} shows the problems and problem-solving strategies in the Z gate and Phase gate questions of the one-qubit survey of the students that took part in the qualitative study. Further, Table \ref{tab:interv_one_qubit_xz} shows the problems and problem-solving strategies in the Global Phase and X gate questions of the one-qubit survey, and tables \ref{tab:interv_one_qubit_h} and \ref{tab:interv_one_qubit_h_2} show those of the H gate and measurement questions. 

\begin{sidewaystable*}
\caption{Identified problems and strategies in the one-qubit survey. Indices are connected to examples and quotations in Section \ref{sec:interv_one_q}.}
\begin{tabularx}{\textwidth}{lXX|XX|XX|XX}
\toprule
\textbf{question type:} & \multicolumn{4}{@{}c@{}}{\textbf{Z gate}} & \multicolumn{4}{@{}c@{}}{\textbf{Phase gate}} \\
\cmidrule(r{0.75em}){1-1}\cmidrule(r{0.75em}){2-5}\cmidrule(lr{0.75em}){6-9}
 condition: & \multicolumn{2}{@{}c@{}}{\textbf{with vis.}} & \multicolumn{2}{@{}c@{}}{\textbf{w/o vis.}} & \multicolumn{2}{@{}c@{}}{\textbf{with vis.}} & \multicolumn{2}{@{}c@{}}{\textbf{w/o vis.}} \\
\cmidrule(r{0.75em}){1-1}\cmidrule(r{0.75em}){2-3}\cmidrule(lr{0.75em}){4-5}\cmidrule(lr{0.75em}){6-7}\cmidrule(lr{0.75em}){8-9}
\textbf{student no.} &    \textbf{prob./strat.} &  \textbf{example}   &  \textbf{prob./strat.} &  \textbf{example}   &  \textbf{prob./strat.} &  \textbf{example} &  \textbf{prob./strat.} &  \textbf{example} \\
\textbf{1} & - & - & calculation by addition & "I just did the maths, if $\varphi$ is positive, you arrive at $-2\pi/3+\pi=\pi/3$" &positive angle means counterclockwise  & "I know the $P(\pi/3)$ changes the state by eighty degrees, so that's the only reason I picked this option." & correct strategy: adding the angles & "Simply because the phase change should be $-3\pi/4$ and the math says it's $-7/20\pi$." \\
\textbf{2} & - & - & easily solved & "It's substracting by, no it adds." &relative phase = global phase & "It's applied to both, so it's gonna stay the same [state]." & easily solved & "just adding $-3\pi/4$ and $2\pi/5$ is $-7\pi/20$" \\
\textbf{3} & confusing Z with NOT gate & "this one [$\frac{\sqrt{3}}{2}\ket{0}+\frac{1}{2}e^{2\pi/3}\ket{1}$], because the Z gate swaps the values" & - & - & approximation & "if you do the calculations $-3\pi/4$ and $2\pi/5$, $-3$ by $2$ and $4$ by $5$, $-7\pi/20$ should be the closest" & multiplying phases instead of adding & "the $-$ is there because of the $-$ in $e^{-i\pi/5}$  \\
\textbf{4} &easily solved & "we just flip the phase of $\ket{1}$ by $\pi$"& - & - & circle for approximation, maths for checking answer & "It flips the phase by almost $\pi$ in the $y$ direction, and this circle seems like it fits, but I computed it in my head and yeah it fits." & easily solved& "$-3\pi/15+5\pi/15$ is $2\pi/15$" \\
\textbf{5} & visual strategy & "just rotate this here by $\pi$"  &-&-& visual strategy & "I think I would rotate this somewhere here because this would be a rotation of $-\pi$ and then $3/4$ times." & easily solved & "Here I add $\pi/3$ to $-\pi/5$ and if I calculated correctly it should be $2/15$." \\
\bottomrule
\end{tabularx}
\label{tab:interv_one_qubit}
\end{sidewaystable*}

\begin{sidewaystable*}
\caption{Identified problems and strategies in the one-qubit survey. Indices are connected to examples and quotations in Section \ref{sec:interv_one_q}.}
\begin{tabularx}{\textwidth}{lXX|XX|XX|XX}
\toprule
\textbf{question type:} & \multicolumn{4}{@{}c@{}}{\textbf{Global Phase}} & \multicolumn{4}{@{}c@{}}{\textbf{X gate}} \\
\cmidrule(r{0.75em}){1-1}\cmidrule(r{0.75em}){2-5}\cmidrule(lr{0.75em}){6-9}
 condition: & \multicolumn{2}{@{}c@{}}{\textbf{with vis.}} & \multicolumn{2}{@{}c@{}}{\textbf{w/o vis.}} & \multicolumn{2}{@{}c@{}}{\textbf{with vis.}} & \multicolumn{2}{@{}c@{}}{\textbf{w/o vis.}} \\
\cmidrule(r{0.75em}){1-1}\cmidrule(r{0.75em}){2-3}\cmidrule(lr{0.75em}){4-5}\cmidrule(lr{0.75em}){6-7}\cmidrule(lr{0.75em}){8-9}
\textbf{student no.} &    \textbf{prob./strat.} &  \textbf{example}   &  \textbf{prob./strat.} &  \textbf{example}   &  \textbf{prob./strat.} &  \textbf{example} &  \textbf{prob./strat.} &  \textbf{example} \\
\textbf{1} & equal angle change to both lines in the circle & "This is something my professor taught me very well. Global Phase means equal change to both circles." & strategy to solve unknown & "don't know the logic on how to solve this without having some kind of phase given." &  amplitude blindness & "I know the NOT gate reverses everything, so I first looked at the amplitudes [...]." \textit{picks option where the amplitudes are not swapped.}& -  & -  \\
\textbf{2} &rotation in circles, mathematics to check &"So just these two are rotated the same. This one. It's rotated by $\pi/10$." & checks denominators first, correct calculation & "the denoimators are all correct. This one. Just adding $-2\pi/5$." & confusion due to non-observance of global phase & "I initially thought it also changed the phase but apparently it does not change the phase and just swapping the bits directly?" & &  \\
\textbf{3} &  multiples phases instead of adding & "this one because $-2\pi/5$ times $2\pi/7$ equals $-4\pi/35$"  & chooses answer with reasoning independent of question & "This one [$\frac{1}{\sqrt{2}}e^{i\pi/5}+\frac{3}{\sqrt{2}}e^{-i\pi/5}$], $\pi/5$ gets applied to both states" & -  & - & disregards phase & The X gate basically shifts the values" \textit{points at $\sqrt{3}/2$ and $1/2$} \\
\textbf{4} & symbolic strategy & "$2/7$ is $10/35$ and $-2/5$ is $-14/35$ so if I apply $-14$ to $10$ I get $-4$." &starting difficulties &"that [$\frac{1}{2}e^{i\pi/10}\ket{0}+\frac{\sqrt{3}}{2}e^{-i\pi/10}\ket{1}$] doesn't look like a global phase to me"&-&-& confusion due to non-observance of global phase & "so in my opinion everything is wrong because in my head I just switch $\ket{0}$ and $\ket{1}$ but nothing matches."  \\
\textbf{5} & visual strategy & "I think I would rotate this somewhere here because this would be a rotation of $-\pi$." & factoring out & "here you can factor out the $e^{i\pi/10}$, then you would have $e^{-2i\pi/10}$ which is $1/5$."& - & - & remembering global phase changes  &  "NOT gate flips the state of the qubits so I would I expect same state of 1 here at the 0 again so I think maybe the global phase has changed. If I would factor $e^{-i\pi/2}$ out then I would have $e^{i\pi/2}$ and it's the first answer." \\
\bottomrule
\end{tabularx}
\label{tab:interv_one_qubit_xz}
\end{sidewaystable*}

\begin{sidewaystable*}
\caption{Identified problems and strategies in the H gate and Measurements questions of the one-qubit survey (students 1 to 3). For a summary, see Section \ref{sec:interv_one_q}.}
\begin{tabularx}{\textwidth}{lXX|XX|XX|XX}
\toprule
\textbf{question type:} & \multicolumn{4}{@{}c@{}}{\textbf{H gate}} & \multicolumn{4}{@{}c@{}}{\textbf{Measurements}} \\
\cmidrule(r{0.75em}){1-1}\cmidrule(r{0.75em}){2-5}\cmidrule(lr{0.75em}){6-9}
\textbf{condition:} & \multicolumn{2}{@{}c@{}}{\textbf{with vis.}} & \multicolumn{2}{@{}c@{}}{\textbf{w/o vis.}} & \multicolumn{2}{@{}c@{}}{\textbf{with vis.}} & \multicolumn{2}{@{}c@{}}{\textbf{w/o vis.}} \\
\cmidrule(r{0.75em}){1-1}\cmidrule(r{0.75em}){2-3}\cmidrule(lr{0.75em}){4-5}\cmidrule(lr{0.75em}){6-7}\cmidrule(lr{0.75em}){8-9}
\textbf{student no.} &    \textbf{prob./strat.} &  \textbf{example}   &  \textbf{prob./strat.} &  \textbf{example}   &  \textbf{prob./strat.} &  \textbf{example} &  \textbf{prob./strat.} &  \textbf{example} \\
\textbf{1} & unidentified problem with H-gate on $\frac{1}{\sqrt{2}}\ket{0}+\frac{1}{\sqrt{2}}e^{i\pi/4}\ket{1}$ & "Why waste time" & unidentified problem with H-gate on $\frac{1}{\sqrt{2}}\ket{0}+\frac{1}{\sqrt{2}}e^{i\pi/2}\ket{1}$  & "I don't get the question" & absolute values = measurement probabilities & "By just looking at the numbers, 0.45 and 0.9" \textit{picks option that measuring 1 is twice as likely as measuring 0} & approximation & "0.5 is 0.025 and I would consider this at 0.9 squared is 0.081. [...] 3 times 0.025 is 0.075 I think that kind of makes sense." \\
\textbf{2} & $H(\frac{1}{\sqrt{2}}\ket{0}+\frac{1}{\sqrt{2}}e^{i\pi/2}\ket{1})=\ket{1}$ & "I find it weird the $\ket{1}$ is missing. [...] I am not sure what's going on here I thought the H gate  collapses the value." & knows correct strategy in principle, but too difficult to execute & "What happened here? Is it supposed to be this way? Ah the phase is gonna affect it. $1/\sqrt{2} + 1/\sqrt{2}\cdot e^{i\pi/4}$, so $\sqrt{2}$ goes out, and $1/\sqrt{2} - 1/\sqrt{2}\cdot e^{i\pi/4}$. But what is that? I have no clue." & derives probabilities from square roots in state depiction &"it's supposed to be in the root format so it's squared I think." & easily solved & "measuring 0 is three times as likely as measuring 1." \\
\textbf{3} & Hadamard gate creates an equally weighted superposition from any state & "The Hadamard gate is supposed to create an equally weighted superposition. None of them is equally weighted so I am picking a random one." & easily solved & "It's on the y axis"  & not squaring the amplitudes & "if you take out the $\sqrt{3}$ I think the chance of measuring 0 is $\sqrt{3}$ times measuring 1" & not squaring the amplitudes & "because 0.9 is twice 0.45, measuring 1 is twice as likely as measuring 0" \\
\bottomrule
\end{tabularx}
\label{tab:interv_one_qubit_h}
\end{sidewaystable*}

\begin{sidewaystable*}
\caption{Identified problems and strategies in the H gate and Measurements questions of the one-qubit survey (students 4 and 5). For a summary, see Section \ref{sec:interv_one_q}.}
\begin{tabularx}{\textwidth}{lXX|XX|XX|XX}
\toprule
\textbf{question type:} & \multicolumn{4}{@{}c@{}}{\textbf{H gate}} & \multicolumn{4}{@{}c@{}}{\textbf{Measurements}} \\
\cmidrule(r{0.75em}){1-1}\cmidrule(r{0.75em}){2-5}\cmidrule(lr{0.75em}){6-9}
\textbf{condition:} & \multicolumn{2}{@{}c@{}}{\textbf{with vis.}} & \multicolumn{2}{@{}c@{}}{\textbf{w/o vis.}} & \multicolumn{2}{@{}c@{}}{\textbf{with vis.}} & \multicolumn{2}{@{}c@{}}{\textbf{w/o vis.}} \\
\cmidrule(r{0.75em}){1-1}\cmidrule(r{0.75em}){2-3}\cmidrule(lr{0.75em}){4-5}\cmidrule(lr{0.75em}){6-7}\cmidrule(lr{0.75em}){8-9}
\textbf{student no.} &    \textbf{prob./strat.} &  \textbf{example}   &  \textbf{prob./strat.} &  \textbf{example}   &  \textbf{prob./strat.} &  \textbf{example} &  \textbf{prob./strat.} &  \textbf{example} \\
\textbf{4} & matrix multiplication & "In my head I just applied the matrix of the Hadamard gate to this state and I had $1+e^{i\pi/4}$ and $1-e^{i\pi/4}$ and I flipped so the $\ket{0}$ has no phase and this is then $-\pi/2$" & lack of strategy & "I can't really explain that. Intuition."& mathematical strategy & "The probabilities are 3/4 and 1/4 so 0 is three times more likely than 1"& easily solved by calculation & "because if I take the absolute values squared it's 20\% vs. 80\% and 80\% is four times 20\%" \\
\textbf{5} & uncertain & "when I try to apply the H gate to each qubit individually then add the amplitudes. But I'm not sure." \textit{selects correct answer} & cannot compute & "If I apply the H gate to each qubit then I get the equal superposition of $\ket{0}$ and $\ket{1}$ and here $\ket{0}$ and $-\ket{1}$. I can't compute it in my head." & easily solved by calculation & "Measuring 0 is three times as likely as measuring 1 so I get 3/4 and 1/4"  & comparing in root notation & "$1/\sqrt{5}$ and $2/\sqrt{5}$ makes 4/5 so it's four times as likely." \\
\bottomrule
\end{tabularx}
\label{tab:interv_one_qubit_h_2}
\end{sidewaystable*}

\subsection{Two-qubit survey}\label{sec:two_qubit_interv}

Table \ref{tab:interv_two_qubit_x} shows the problems and problem-solving strategies in the X gate and H gate questions of the two-qubit survey of the students that took part in the qualitative study. Further, tables \ref{tab:interv_two_qubit_g} and \ref{tab:interv_two_qubit_g_2} show the problems and problem-solving strategies in the Guess-the-Gates and Measurement questions, and tables \ref{tab:interv_two_qubit_s} and \ref{tab:interv_two_qubit_s_2} show those of the Separability and Entanglement questions. 

\begin{sidewaystable*}
\caption{Identified problems and strategies in the X and H gate questions of the two-qubit survey. For a summary, see section \ref{sec:interv_two_q}.}
\begin{tabularx}{\textwidth}{lXX|XX|XX|XX}
\toprule
\textbf{question type:} & \multicolumn{4}{@{}c@{}}{\textbf{X gate}} & \multicolumn{4}{@{}c@{}}{\textbf{H gate}} \\
\cmidrule(r{0.75em}){1-1}\cmidrule(r{0.75em}){2-5}\cmidrule(lr{0.75em}){6-9}
\textbf{condition:} & \multicolumn{2}{@{}c@{}}{\textbf{with vis.}} & \multicolumn{2}{@{}c@{}}{\textbf{w/o vis.}} & \multicolumn{2}{@{}c@{}}{\textbf{with vis.}} & \multicolumn{2}{@{}c@{}}{\textbf{w/o vis.}} \\
\cmidrule(r{0.75em}){1-1}\cmidrule(r{0.75em}){2-3}\cmidrule(lr{0.75em}){4-5}\cmidrule(lr{0.75em}){6-7}\cmidrule(lr{0.75em}){8-9}
\textbf{student no.} &    \textbf{prob./strat.} &  \textbf{example}   &  \textbf{prob./strat.} &  \textbf{example}   &  \textbf{prob./strat.} &  \textbf{example} &  \textbf{prob./strat.} &  \textbf{example} \\
\textbf{1 (CN)} & interchanging coefficients next to each other & "They interchange, so what was $1/3$ will become $1/\sqrt{3}e^{i\pi/4}$ and this will become $1/3$ and similar in the second half." & X$_2$ means two swaps & "the X$_2$ gate is different from the X$_1$ gate, it changes either with the second to last or the last one. Then if there is a NOT gate these two must change as well." & Hadamard gate always creates equal superpositions & "I was thinking equal probability would go to all. But that's not happening. Probability is being divided between the first two and the last two and I have no reason for that. So I am going to pick a random answer." & lack of strategy & "I'm going with this one." \\
\textbf{2 (CN)} & easily solved by coefficient comparison & "X$_2$ so it's switching these two [$\ket{00}$ and $\ket{10}$]" & easily solved by coefficient comparison & "The ones were the first bit is changing are swapped." & visual strategy & "First one so next to each other. First one is state $\ket{1}$ because it's negative and second has 0 relative [phase] so it's in the $\ket{0}$." & splitting states  & "This [$1/\sqrt{3}\ket{00}$] would be split in this and this [$1/\sqrt{6}\ket{00}$ and $1/\sqrt{6}\ket{01}$] and here since we are starting from 1 this would get a negative." \\
\textbf{3 (DCN)} & mistaking X$_2$ gate for CNOT$_{21}$ & "When qubit \#2 is one these two should should switch so that happens over here." &  easily solved by coefficient comparison & "this one because these two shift and these two shift values."  & visual reasoning, mistaking H$_2$ for H$_1$ gate & "Along the $x$ axis the values have an equally weighted superposition." & lack of strategy & "It applies when the first qubit is 1. So that's why, here the first qubit is 1." \textit{selects $\frac{1}{\sqrt{2}}\ket{00}-\frac{1}{\sqrt{2}}\ket{10}$} \\
\textbf{4 (DCN)} & visual strategy & "X$_1$ as in the introduction just flips the circles around the axis of qubit \#1." & easily solved by coefficient comparison & "So the first number is the second qubit. So this [first 0 in $\ket{00}$] is flipped to 1." & wrong qubit &  "$\ket{10}$ should get this $e^{i\pi/4}$."  & canceling out of amplitudes & "$\ket{00}$ gets erased and $\ket{11}$ gets erased so I think its [$\frac{1}{\sqrt{2}}\ket{00}-\frac{1}{\sqrt{2}\ket{10}}$] \\
\textbf{5 (DCN)} & visual strategy & "should be this answer because here these colors are swapped [colored inner circles]." & easily solved by coefficient comparison  &  "I have to flip the qubit \#2, the $\ket{00}$ would then correspond to the $\ket{10}$ state and we have the same coefficients." &  visual strategy &  "Here we distribute the probability from the $\ket{00}$ state to the $\ket{10}$ state and not change the angle, [...] and here to the $\ket{01}$ state and rotate the angle where the qubit is 1." &  apply reverse gate  &  "If I apply the H$_1$ gate to this state because it is it's own inverse, I should end up in this state."  \\
\bottomrule
\end{tabularx}
\label{tab:interv_two_qubit_x}
\end{sidewaystable*}

\begin{sidewaystable*}
\caption{Identified problems and strategies in the Guess-the-Gates and Measurements questions of the two-qubit survey (students 1-3). For a summary, see section \ref{sec:interv_two_q}.}
\begin{tabularx}{\textwidth}{lXX|XX|XX|XX}
\toprule
\textbf{question type:} & \multicolumn{4}{@{}c@{}}{\textbf{Guess-the-Gates}} & \multicolumn{4}{@{}c@{}}{\textbf{Measurements}} \\
\cmidrule(r{0.75em}){1-1}\cmidrule(r{0.75em}){2-5}\cmidrule(lr{0.75em}){6-9}
\textbf{condition:} & \multicolumn{2}{@{}c@{}}{\textbf{with vis.}} & \multicolumn{2}{@{}c@{}}{\textbf{w/o vis.}} & \multicolumn{2}{@{}c@{}}{\textbf{with vis.}} & \multicolumn{2}{@{}c@{}}{\textbf{w/o vis.}} \\
\cmidrule(r{0.75em}){1-1}\cmidrule(r{0.75em}){2-3}\cmidrule(lr{0.75em}){4-5}\cmidrule(lr{0.75em}){6-7}\cmidrule(lr{0.75em}){8-9}
\textbf{student no.} &    \textbf{prob./strat.} &  \textbf{example}   &  \textbf{prob./strat.} &  \textbf{example}   &  \textbf{prob./strat.} &  \textbf{example} &  \textbf{prob./strat.} &  \textbf{example} \\
\textbf{1 (CN)} & - & - & -  & - & a circle as a qubit, measurement result 0 corresponds to measuring 00 & "If it was just two qubits I would just square it and make an educated guess, but now the 1 probability is being divided between four circles." "It would make sense if I just remove this clutter and only look at this [00] and [11]." & lack of knowledge where qubit \#1 can be found in the Dirac notation & "I don't understand the question. Both these bits have 01 and 10. I can't tell the difference." \\
\textbf{2 (CN)} & visual strategy & "Hadamard on the second one. [...] and phase shift. " & intuition &"I feel like here its phase was shifted and it was a NOT" \textit{picks correct answer $H_2Z_1$}& calculations in symbolic notation & "$1/3 +2/9$ is $5/9$,  $1/9+1/3$ is $4/9$. So it's $5/4$" & calculating & "$(\frac{1}{2\sqrt{2}})^2$ plus $(1/2)^2$ is $3/8$ and on $P(1)$ it was 5/8" \\
\textbf{3 (DCN)} & Z gate does a $\pi/2$ phase shift & "$1/\sqrt{2}$ was given to both sides for an equal superposition. And then there was a phase shift applied by $\pi/2$." &  disregard for gate indices, correct solution & "A Hadamard gate was applied because it delegated superpositions equally. And it's a not gate because the states have switched and I am assuming it switched on the $x$ axis again." & approximation, looking at wrong values & "Just looking at the values, measuring 1 is $\sqrt{2}/3$ or something it's $1/\sqrt{3}$ for 0" & approximation, no squaring, wrong qubit &  "0 has a probability of $1/2\sqrt{2}$ plus $\sqrt{3}/2\sqrt{2}$ and 1 has a probability of $1/2$ plus $1/2$" \\
\bottomrule
\end{tabularx}
\label{tab:interv_two_qubit_g}
\end{sidewaystable*}

\begin{sidewaystable*}
\caption{Identified problems and strategies in the Guess-the-Gates and Measurements questions of the two-qubit survey (students 4 and 5). For a summary, see section \ref{sec:interv_two_q}.}
\begin{tabularx}{\textwidth}{lXX|XX|XX|XX}
\toprule
\textbf{question type:} & \multicolumn{4}{@{}c@{}}{\textbf{Guess-the-Gates}} & \multicolumn{4}{@{}c@{}}{\textbf{Measurements}} \\
\cmidrule(r{0.75em}){1-1}\cmidrule(r{0.75em}){2-5}\cmidrule(lr{0.75em}){6-9}
\textbf{condition:} & \multicolumn{2}{@{}c@{}}{\textbf{with vis.}} & \multicolumn{2}{@{}c@{}}{\textbf{w/o vis.}} & \multicolumn{2}{@{}c@{}}{\textbf{with vis.}} & \multicolumn{2}{@{}c@{}}{\textbf{w/o vis.}} \\
\cmidrule(r{0.75em}){1-1}\cmidrule(r{0.75em}){2-3}\cmidrule(lr{0.75em}){4-5}\cmidrule(lr{0.75em}){6-7}\cmidrule(lr{0.75em}){8-9}
\textbf{student no.} &    \textbf{prob./strat.} &  \textbf{example}   &  \textbf{prob./strat.} &  \textbf{example}   &  \textbf{prob./strat.} &  \textbf{example} &  \textbf{prob./strat.} &  \textbf{example} \\
\textbf{4 (DCN)} & narrow down with patterns, but problem computing & "I don't think that an X gate was applied because we have this $1/\sqrt{6}\ket{00}$ which we get by applying a Hadamard to the first or the second qubit. I also have the impression somewhere was a Z gate. I don't know. First intuition was H$_1$Z$_2$." & coefficient comparison & "I can't get rid of this $\sqrt{3}$ in this state $\ket{00}$ so there has to be an X gate somewhere after applying the Hadamard. So then maybe a flip of the second one." & issue with percentages & "This is $1/8+1/4=3/8$ and this should add up to $5/8$. So what is $5/8$ over $3/8$? So I think measuring 1 is 33\% more likely than measuring 0, because 66\% seems too much." & incorrect squaring & "Here I have $1/3$ and here I have $2/6$ so they add up to $4/6$. I don't remember the formula [...] but if the probabilities are doubled then 50\% more is more likely than 25\% more." \\
\textbf{5 (DCN)} & visual strategy & "Z$_1$ would rotate this angle to the other side and then H$_2$ would distribute the amplitudes around the qubit \#2 axis." &  focussing on one basis state in the superposition & "If I flip the second qubit $\ket{01}$ becomes $\ket{11}$ and then the H gate would give this expression then." &  use of insufficient small amount of information & "I think it's 33\% more likely because of this $\sqrt{3}$ here but I'm not sure."  & approximation & "I'm not completely sure but we have only $\sqrt{2}$ more so I think it's only 25\% more likely [to measure 0 than 1]. \\
\bottomrule
\end{tabularx}
\label{tab:interv_two_qubit_g_2}
\end{sidewaystable*}

\begin{sidewaystable*}
\caption{Identified problems and strategies in the Separability and Entanglement questions of the two-qubit survey (students 1 and 2). For a summary, see section \ref{sec:interv_two_q}.}
\begin{tabularx}{\textwidth}{lXX|XX|XX|XX}
\toprule
\textbf{question type:} & \multicolumn{4}{@{}c@{}}{\textbf{Separability}} & \multicolumn{4}{@{}c@{}}{\textbf{Entanglement}} \\
\cmidrule(r{0.75em}){1-1}\cmidrule(r{0.75em}){2-5}\cmidrule(lr{0.75em}){6-9}
\textbf{condition:} & \multicolumn{2}{@{}c@{}}{\textbf{with vis.}} & \multicolumn{2}{@{}c@{}}{\textbf{w/o vis.}} & \multicolumn{2}{@{}c@{}}{\textbf{with vis.}} & \multicolumn{2}{@{}c@{}}{\textbf{w/o vis.}} \\
\cmidrule(r{0.75em}){1-1}\cmidrule(r{0.75em}){2-3}\cmidrule(lr{0.75em}){4-5}\cmidrule(lr{0.75em}){6-7}\cmidrule(lr{0.75em}){8-9}
\textbf{student no.} &    \textbf{prob./strat.} &  \textbf{example}   &  \textbf{prob./strat.} &  \textbf{example}   &  \textbf{prob./strat.} &  \textbf{example} &  \textbf{prob./strat.} &  \textbf{example} \\
\textbf{1 (CN)} & lack of strategy, separability as a process rather than a property & "If I was given the input and output of two qubits, and ask a question regarding is that separable or not, that would have been a far easier problem to solve. This I cannot fathom how to solve." & lack of strategy, separability as a process rather than a property &"As I explained previously. If there was some change made to the qubits and shown what the results was I would have been able to tell." & entanglement means that action on one also is action on the other, four basis states = four qubits & 
"How do I tell if they're entangled if no action is being performed on them? [...] Here I'm just given like four qubits with already predefined phases. This is something I cannot comprehend." &  lack of strategy, entanglement means that action on one also is action on the other  & "I understand the concept of entanglement [as]: if you make an action on one qubit is has an impact on the other qubit." \\
\textbf{2 (CN)} & visual strategy & "[going through answer possibilities] Not following the trend. Not following the trend. This is separable because it is following the trend, its a shift on this and a similar shift on this." & lack of strategy & "I don't know how it's written in the bracket notation, I'm going to pick the third [correct] one, no reason" & checking for separability instead of entanglement & "This ratio [$1/\sqrt{2}$ to $1/2$ and $1/\sqrt{6}$ to $1/2\sqrt{3}$ in circles] seems better than this one [$1/\sqrt{2}$ to $1/\sqrt{6}$ and $1/2$ to $1/2\sqrt{3}$]." & incorrect phase shift application & "If I do a $-1/3$ on this [$-\pi/6$] it's gonna change, then it won't be separable." \\
\bottomrule
\end{tabularx}
\label{tab:interv_two_qubit_s}
\end{sidewaystable*}

\begin{sidewaystable*}
\caption{Identified problems and strategies in the Separability and Entanglement questions of the two-qubit survey (students 3-5). For a summary, see section \ref{sec:interv_two_q}.}
\begin{tabularx}{\textwidth}{lXX|XX|XX|XX}
\toprule
\textbf{question type:} & \multicolumn{4}{@{}c@{}}{\textbf{Separability}} & \multicolumn{4}{@{}c@{}}{\textbf{Entanglement}} \\
\cmidrule(r{0.75em}){1-1}\cmidrule(r{0.75em}){2-5}\cmidrule(lr{0.75em}){6-9}
\textbf{condition:} & \multicolumn{2}{@{}c@{}}{\textbf{with vis.}} & \multicolumn{2}{@{}c@{}}{\textbf{w/o vis.}} & \multicolumn{2}{@{}c@{}}{\textbf{with vis.}} & \multicolumn{2}{@{}c@{}}{\textbf{w/o vis.}} \\
\cmidrule(r{0.75em}){1-1}\cmidrule(r{0.75em}){2-3}\cmidrule(lr{0.75em}){4-5}\cmidrule(lr{0.75em}){6-7}\cmidrule(lr{0.75em}){8-9}
\textbf{student no.} &    \textbf{prob./strat.} &  \textbf{example}   &  \textbf{prob./strat.} &  \textbf{example}   &  \textbf{prob./strat.} &  \textbf{example} &  \textbf{prob./strat.} &  \textbf{example} \\
\textbf{3 (DCN)} & lack of strategy & "I am probably not getting the idea of the separable states correctly. It feels like if I calculate the first bit somehow this will also impact the second bit." & entanglement = dependency & "I cannot determine how they will not depend on eachother. It feels like they are all entangled. No matter how I calculate, measurement on the first one is going to impact qubit two as well, because their states are together." & intuition & "I don't know how am I supposed to determine if I calculate this whether I get a separable state or not. Both the separable and entangled states are still a vague concept to me. [...] Feels like these two states might be entangled based on the coefficients." & lack of strategy & "All of them look entangled. I probably don't understand the topic that well." \\
\textbf{4 (DCN)} & visual strategy: amplitude comparisons & "This looks very nice maybe. Because from this to this and from this to this. The others don't make sense." & imagining circle notation and phase shift comparison & "I just imagined the circle notation. If I have $-\pi/2$ and going $\pi/3$ to the left direction I should end up in $-\pi/6$." & visual strategy: phase shifts &  "Here you need a negative shift and here you need a positive shift."  & imagining circle notation and amplitude comparison & "I again imagined the circle notation. I just computed the probabilities of the left side and tried to find a factor to get these on the right side. And it worked for these three and not for this." \\
\textbf{5 (DCN)} & visual strategy & "here we could multiply the amplitudes by the same factor from the left to the right." & using the rule $\alpha_{00}\alpha_{11}=\alpha_{01}\alpha_{10}$  & "If I multiply $1/2$ with the factor of this then I get $1/4e^{-i\pi/6}$ and if I multiply these I get the same." & visual strategy & "we would rotate from the left to the right the angles in this direction and here in the other direction and it would not be separable." & using the rule $\alpha_{00}\alpha_{11}=\alpha_{01}\alpha_{10}$ & "If I multiply these two numbers I do not get the same as when I multiply these two." \\
\bottomrule
\end{tabularx}
\label{tab:interv_two_qubit_s_2}
\end{sidewaystable*}

\subsection{Three-qubit survey}\label{sec:three_qubit_interv}

Table \ref{tab:interv_three_qubit_x} shows the problems and problem-solving strategies in the X gate and Z gate questions of the three-qubit survey of the students that took part in the qualitative study. Table \ref{tab:interv_three_qubit_h} shows those of the Hadamard gate question and Table \ref{tab:interv_three_qubit_m} those of the Measurement questions.  Further, Table \ref{tab:interv_three_qubit_s} shows the problems and problem-solving strategies in the Separability and Entanglement questions, and Table \ref{tab:interv_three_qubit_c} shows that of the CNOT and Guess-the-Gates questions. 

\begin{sidewaystable*}
\caption{Identified problems and strategies in the X and Z gate questions of the three-qubit survey. For a summary, see section \ref{sec:interv_three_q}.}
\begin{tabularx}{\textwidth}{lXX|XX|XX|XX}
\toprule
\textbf{question type:} & \multicolumn{4}{@{}c@{}}{\textbf{X gate}} & \multicolumn{4}{@{}c@{}}{\textbf{Z gate}}
\\
\cmidrule(r{0.75em}){1-1}\cmidrule(r{0.75em}){2-5}\cmidrule(lr{0.75em}){6-9}
\textbf{condition:} & \multicolumn{2}{@{}c@{}}{\textbf{with vis.}} & \multicolumn{2}{@{}c@{}}{\textbf{w/o vis.}} & \multicolumn{2}{@{}c@{}}{\textbf{with vis.}} & \multicolumn{2}{@{}c@{}}{\textbf{w/o vis.}} \\
\cmidrule(r{0.75em}){1-1}\cmidrule(r{0.75em}){2-3}\cmidrule(lr{0.75em}){4-5}\cmidrule(lr{0.75em}){6-7}\cmidrule(lr{0.75em}){8-9}
\textbf{student no.} &    \textbf{prob./strat.} &  \textbf{example}   &  \textbf{prob./strat.} &  \textbf{example}   &  \textbf{prob./strat.} &  \textbf{example} &  \textbf{prob./strat.} &  \textbf{example} \\
\textbf{2 (DCN)} & coefficient comparison & "$\ket{000}$ is $1/4$. This should be $1/\sqrt{2}$" & key feature search & "There is no $\ket{100}$. So this one is the right one." \textit{checks the other terms}. & Z gate switches phases instead of a $\pi$ relative phase shift & "$\ket{101}$ and $\ket{111}$ should just stay the same. Does it not switch? The option I picked [$\frac{1}{\sqrt{6}}e^{2i\pi/3}\ket{001}+\frac{1}{\sqrt{6}}\ket{010}+\frac{1}{\sqrt{6}}e^{3i\pi/4}-\frac{1}{\sqrt{6}}\ket{101}+\frac{1}{\sqrt{6}}e^{i\pi/2}\ket{110}-\frac{1}{\sqrt{6}}\ket{111}$] had this and this [$\ket{001}$, $\ket{011}$, and $\ket{101},\ket{111}$] both on the opposite side because they had similar phases." & intuition, Z gate switches phases instead of a $\pi$ relative phase shift & "This one seems more correct. I don't know. This one [$\ket{000}$] and this one $\ket{001}$ should have swapped the phase but I don't see that anywhere." \\ \cmidrule(lr{0.75em}){1-9}
\textbf{4 (DCN)} & visual strategy: action in direction of axis & "The X$_3$ gate swaps this axis. So $\ket{000}$ should get 0 probability and the probabilities of the third qubit is 1 so from left to right it gets bigger so I go with this." & coefficient comparison  & "The X$_1$ gate flips for example this $\ket{000}$ with this $\ket{001}$ so the amplitudes for this state is the only one where this matches." & visual strategy: phase flips & "If the phase of the first qubit is flipped than I have to look at these four and this is the only one where all are flipped." & phase shift examination & "we add $\pi$ if the second qubit is 1. We get the $-$ here [$\ket{010}$] we get from $-\pi/4$ to $3\pi/4$, from $\pi/2$ to $-\pi/2$ that's also correct and here is just the minus. " \\
\bottomrule
\end{tabularx}
\label{tab:interv_three_qubit_x}
\end{sidewaystable*}

\begin{table*}
\caption{Identified problems and strategies in the H gate questions of the three-qubit survey. For a summary, see Section \ref{sec:interv_three_q}.}
\begin{tabularx}{\textwidth}{lXX|XX}
\toprule
\textbf{condition:} & \multicolumn{2}{@{}c@{}}{\textbf{with vis.}} & \multicolumn{2}{@{}c@{}}{\textbf{w/o vis.}} \\
\cmidrule(r{0.75em}){1-1}\cmidrule(r{0.75em}){2-3}\cmidrule(lr{0.75em}){4-5}
\textbf{student no.} &    \textbf{prob./strat.} &  \textbf{example}   &  \textbf{prob./strat.} &  \textbf{example} \\
\textbf{2 (DCN)} & visual strategy along qubit axis & because superposition in this [$\ket{000}$] and superposition in this [$\ket{011}$], these two are reduced [$\ket{001}$, $\ket{101}$] and these two are reduced [$\ket{010}$, $\ket{110}$] & key feature search followed by answer validation & "Because it is on the second bit, this one [$\ket{000}$] and this one [$\ket{010}$] come together. This one doesn't have anything, so it makes another one and same over here." \\ \cmidrule(lr{0.75em}){1-5}
\textbf{4 (DCN)} &  coefficient analysis & "If I apply H on the second, the amplitude of 0 just be $\sqrt{2}/3$, so this or this. $\ket{100}$ has a positive sign. If I apply H here [$\sqrt{2}/3\ket{110}$] I get a positive sign. Only 1 gets a negative sign."& problem identifying "partner" state & "Hadamard to third I have only one times $\ket{000}$ after that with amplitude $1/\sqrt{12}$ so $1/2\sqrt{3}$ so [one of] these two. This [$\ket{001}$] produces $-1/\sqrt{12}$ or not? And this also. Here [$1/\sqrt{6}\ket{011}$] I would get $1/\sqrt{12}$ and that [$\ket{111}$] is the only possible appearance. \\
\bottomrule
\end{tabularx}
\label{tab:interv_three_qubit_h}
\end{table*}

\begin{sidewaystable*}
\caption{Identified problems and strategies in the Measurement probabilities and measurement state questions of the three-qubit survey. For a summary, see section \ref{sec:interv_three_q}.}
\begin{tabularx}{\textwidth}{lXX|XX|XX|XX}
\toprule
\textbf{question type:} & \multicolumn{4}{@{}c@{}}{\textbf{Measurement probability}} & \multicolumn{4}{@{}c@{}}{\textbf{Measurement State}}\\
\cmidrule(r{0.75em}){1-1}\cmidrule(r{0.75em}){2-5}\cmidrule(lr{0.75em}){6-9}
\textbf{condition:} & \multicolumn{2}{@{}c@{}}{\textbf{with vis.}} & \multicolumn{2}{@{}c@{}}{\textbf{w/o vis.}} & \multicolumn{2}{@{}c@{}}{\textbf{with vis.}} & \multicolumn{2}{@{}c@{}}{\textbf{w/o vis.}} \\
\cmidrule(r{0.75em}){1-1}\cmidrule(r{0.75em}){2-3}\cmidrule(lr{0.75em}){4-5}\cmidrule(lr{0.75em}){6-7}\cmidrule(lr{0.75em}){8-9}
\textbf{student no.} &    \textbf{prob./strat.} &  \textbf{example}   &  \textbf{prob./strat.} &  \textbf{example}   &  \textbf{prob./strat.} &  \textbf{example} &  \textbf{prob./strat.} &  \textbf{example} \\
\textbf{2 (DCN)} & calculations & "0 is this $1/\sqrt{6}\ket{000}$ and this $1/\sqrt{6}\ket{100}$ so $1/3$ and 1 $2/3$. Measuring 1 is twice as likely as measuring 0." & calculations & "From taking the probabilities where this one [qubit \#2] is 0, it came $1/3$ and this one [where qubit \#2 is 1] was $2/3$, so 1 is two times as likely as 0." & intuition, misconception: basis states are merged and not erased & "This one seems better. It's not combining, I'm confused." & lack of understanding of normalization & "I don't know how it's normalized." \textit{picks a random answer}\\ \cmidrule(lr{0.75em}){1-9}
\textbf{4 (DCN)} & coefficient calculations & "The probability of getting 0 for the second qubit should be 1/3 and for getting 1 it should be 2/3 so measuring 1 is twice as likely as measuring 0." & coefficient calculations & "So the probability of 0 should be $1/3$ and for 1 $2/3$ so 1 is twice as likely as 0." & visual strategy: phase comparison & "In my mind the phases should not change so maybe there is a global phase change somewhere. $\pi/2$, $\pi/2$, $\pi/2$, this should be correct."  & probability calculations to get normalization constant & "We have to divide the probabilities I think it was the square root of $1/2$ so $1/4$. No it's $1/\sqrt{2}$, so the amplitude over here [$\ket{001}$] should be $1/\sqrt{6}$, that's fine. And here [$\ket{100}$] it should be $1/\sqrt{2}$ I guess. I can check the last one $1/\sqrt{3}$ yeah, should be correct." \\
\bottomrule
\end{tabularx}
\label{tab:interv_three_qubit_m}
\end{sidewaystable*}

\begin{sidewaystable*}
\caption{Identified problems and strategies in the Separability and Entanglement questions of the three-qubit survey. For a summary, see section \ref{sec:interv_three_q}.}
\begin{tabularx}{\textwidth}{lXX|XX|XX|XX}
\toprule
\textbf{question type:} & \multicolumn{4}{@{}c@{}}{\textbf{Separability}} & \multicolumn{4}{@{}c@{}}{\textbf{Entanglement}}\\
\cmidrule(r{0.75em}){1-1}\cmidrule(r{0.75em}){2-5}\cmidrule(lr{0.75em}){6-9}
\textbf{condition:} & \multicolumn{2}{@{}c@{}}{\textbf{with vis.}} & \multicolumn{2}{@{}c@{}}{\textbf{w/o vis.}} & \multicolumn{2}{@{}c@{}}{\textbf{with vis.}} & \multicolumn{2}{@{}c@{}}{\textbf{w/o vis.}} \\
\cmidrule(r{0.75em}){1-1}\cmidrule(r{0.75em}){2-3}\cmidrule(lr{0.75em}){4-5}\cmidrule(lr{0.75em}){6-7}\cmidrule(lr{0.75em}){8-9}
\textbf{student no.} &    \textbf{prob./strat.} &  \textbf{example}   &  \textbf{prob./strat.} &  \textbf{example}   &  \textbf{prob./strat.} &  \textbf{example} &  \textbf{prob./strat.} &  \textbf{example} \\
\textbf{2 (DCN)} & lack of strategy & "I don't really know what fully separable means. One of the first three. They're all similar. " & lack of strategy, picks correct answer by intuition &"I don't know this one either. This one [$\frac{1}{2\sqrt{2}}\ket{000}+\frac{1}{2\sqrt{2}}\ket{001}+\frac{1}{2\sqrt{2}}e^{i\pi/2}\ket{010}+\frac{1}{2\sqrt{2}}e^{i\pi/2}\ket{011}+\frac{1}{2\sqrt{2}}\ket{100}+\frac{1}{2\sqrt{2}}\ket{101}+\frac{1}{2\sqrt{2}}e^{i\pi/2}\ket{110}+\frac{1}{2\sqrt{2}}e^{i\pi/2}\ket{111}$] has common values in a weird way." & visual strategy & "This one [$\frac{1}{2\sqrt{2}}\ket{000}+\frac{1}{2\sqrt{2}}\ket{001}+\frac{1}{2\sqrt{2}}\ket{010}-\frac{1}{2\sqrt{2}}\ket{011}+\frac{1}{2\sqrt{2}}\ket{100}-\frac{1}{2\sqrt{2}}\ket{101}+\frac{1}{2\sqrt{2}}\ket{110}+\frac{1}{2\sqrt{2}}\ket{111}$] is different in all directions." &  lack of strategy & "Nothing specifically, I don't know which it could be." \\
\textbf{4 (DCN)} & visual strategy: phase comparison along planes & This is $\pi/2$, this should be $\pi/2$, so no. [next answer possibility] For the second qubit this is $\pi/2$ and this is $-\pi/2$, so no. This is $\pi/2$, this is 1, and this is 1, seems fully separable." & imagining visualization & "Should be this one when I think about the circle notation then they all should, ah yeah the second qubit is always 1. So is factor 0 allowed? Okay let me check the others again. It does not work for this. This does also not work." & visual strategy: comparison along planes &"Here if I draw the plane for the first qubit it's just 1. This plane does not work. This plane does not work. Here the plane for the third qubit has factor one. I think this is the fully entangled one. There is no plane that has a symmetry." & imagining visualization  & "So I imagined the circle representation in my head and checked it for the first one where only the $\ket{011}$ and $\ket{101}$ have a negative sign and they are direct neighbors in this area for the third qubit and then I looked at the other states, both states have a minus sign and they only differ in one qubit. If I draw a plane again I think they would be separable and this state differs in two different qubits which means I cannot draw a plane there." \\
\bottomrule
\end{tabularx}
\label{tab:interv_three_qubit_s}
\end{sidewaystable*}

\begin{sidewaystable*}
\caption{Identified problems and strategies in the CNOT gate and Guess-the-Gates questions of the three-qubit survey. For a summary, see section \ref{sec:interv_three_q}.}
\begin{tabularx}{\textwidth}{lXX|XX|XX|XX}
\toprule
\textbf{question type:} & \multicolumn{4}{@{}c@{}}{\textbf{CNOT gate}} & \multicolumn{4}{@{}c@{}}{\textbf{Guess-the-Gates}} \\
\cmidrule(r{0.75em}){1-1}\cmidrule(r{0.75em}){2-5}\cmidrule(lr{0.75em}){6-9}
\textbf{condition:} & \multicolumn{2}{@{}c@{}}{\textbf{with vis.}} & \multicolumn{2}{@{}c@{}}{\textbf{w/o vis.}} & \multicolumn{2}{@{}c@{}}{\textbf{with vis.}} & \multicolumn{2}{@{}c@{}}{\textbf{w/o vis.}} \\
\cmidrule(r{0.75em}){1-1}\cmidrule(r{0.75em}){2-3}\cmidrule(lr{0.75em}){4-5}\cmidrule(lr{0.75em}){6-7}\cmidrule(lr{0.75em}){8-9}
\textbf{student no.} &    \textbf{prob./strat.} &  \textbf{example}   &  \textbf{prob./strat.} &  \textbf{example}   &  \textbf{prob./strat.} &  \textbf{example} &  \textbf{prob./strat.} &  \textbf{example} \\
\textbf{2 (DCN)} & visual strategy & "switch in the 2 direction this and this [$\ket{100}$ and $\ket{110}$] and this and this [$\ket{101}$ and $\ket{111}$]. This one seems correct."  & key feature search followed by answer validation  & "These two [$\ket{001}$ and $\ket{101}$] switch. This one [$\ket{111}$] with that [$\ket{011}$]." & X gate correctly identified, doesn't know how measurements work & "3 would be this [moves cursor along axis \#3], 2 would be this [moves cursor along axis \#2]. Could be [answer option] 2 or 3 [X$_3$M$_2$ or X$_3$M$_1$]. I don't know what the difference would be."  & cognitive overload & "This is complicated. I don't know how it is working on this [$1/\sqrt{3}\ket{110}+\frac{1}{\sqrt{6}}e^{i\pi/2}\ket{111}$]. I think this one [H$_1$CNOT$_{21}$] seems more correct. I did it on H$_1$ and it was correct so far. After that I am just confused." \\
\textbf{4 (DCN)} & visual strategy & "If the first qubit is 1 then the third gets flipped so if I have here 1 I flip on axis 3 so I flip this and this [$\ket{001}$ and $\ket{101}$] and this and this [$\ket{011}$ and $\ket{111}$]."& bit flip calculations & "Nothing should happen with this state $\ket{001}$ so one of these two. So we flip if the third is 1. So it should be this one." & gates applied in wrong order & "It's a Hadamard on the first one, because I don't get another $\ket{000}$ here [in the rest of the state]. And now a CNOT. Okay I have everything somehow [no amplitude is 0]. If I flip the first if the second is 1 then this $\ket{001}$ state would do nothing but the $\ket{001}$ state has a phase. So I guess I flip the second if the first is 0. Then it's this [H$_1$CNOT$_{12}$]." & key feature search & "So the second qubit is measured, because they are all 1. But then which one is flipped? This [$1/\sqrt{6}e^{-\pi/2}\ket{010}$] I only have here [$1/\sqrt{6}e^{-\pi/2}\ket{110}$], so the third qubit got flipped. " \\
\bottomrule
\end{tabularx}
\label{tab:interv_three_qubit_c}
\end{sidewaystable*}

\section{Quantitative Survey results}\label{sec:A_results}

The quantitative data supporting the findings of this article is openly available~\cite{bley_2024_14051040}.

\begin{figure*}
    \centering
    \includegraphics[width=0.8\linewidth]{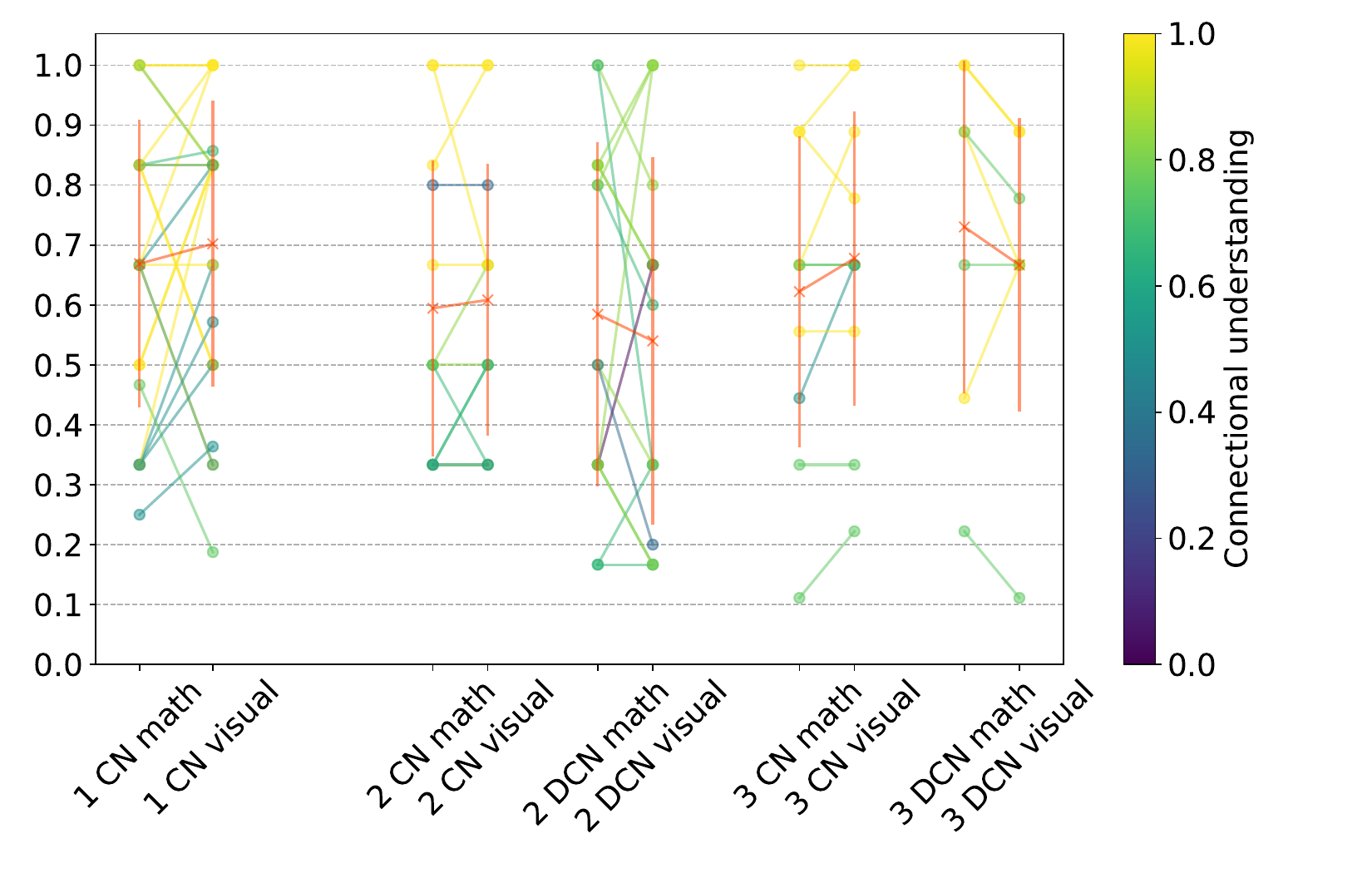}
    \caption{Average scores for different numbers of qubits. Two points connected by a line represent an individual participant, colored according to the connectional understanding score of the participant in that survey. Mean and standard error are shown in orange.}
    \label{fig:score}
\end{figure*}

\begin{figure*}
    \centering
    \includegraphics[width=0.8\linewidth]{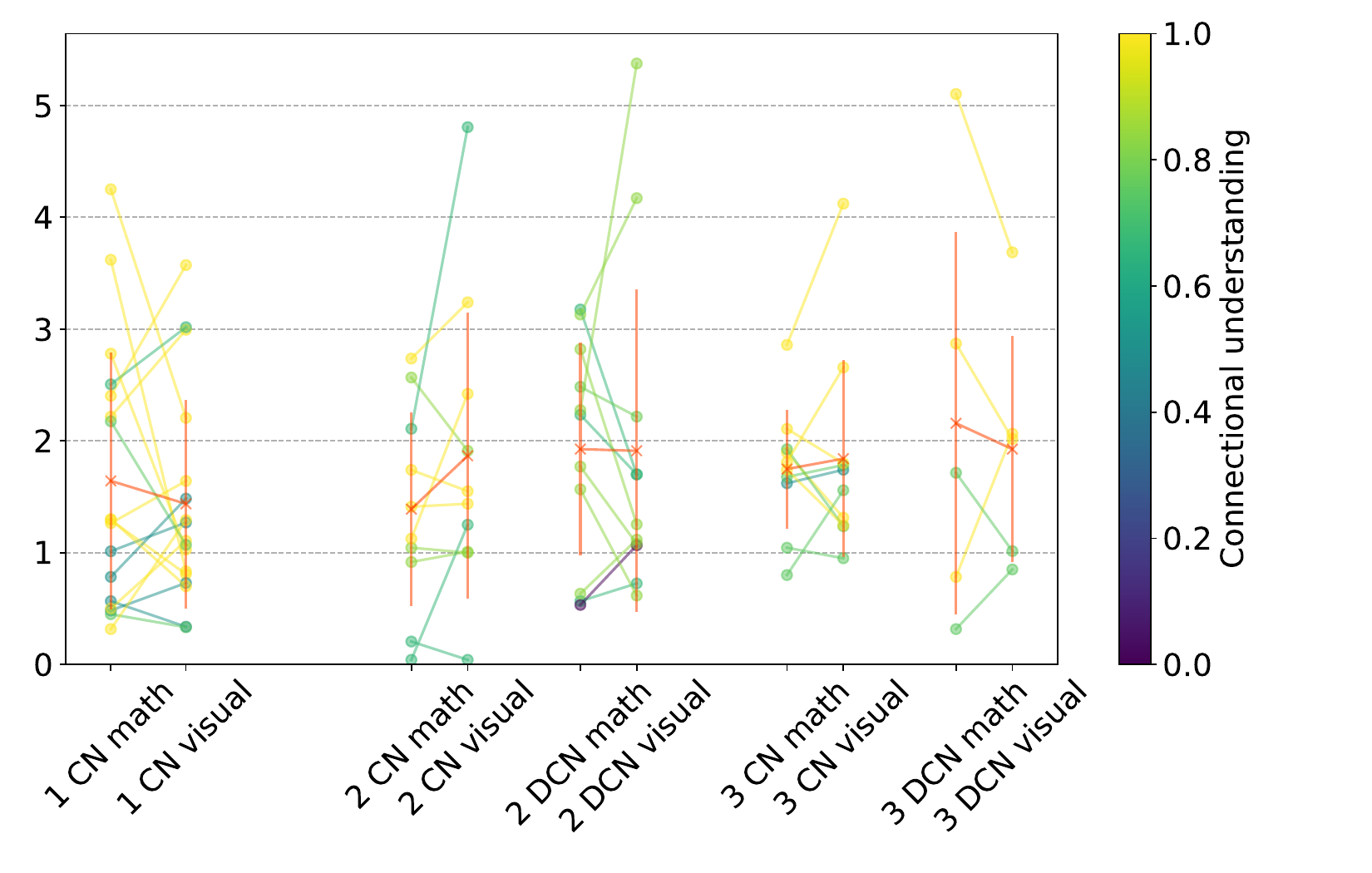}
    \caption{Average time taken per participant for correctly solved questions and connectional understanding.  Two points connected by a line represent an individual participant, colored according to the connectional understanding score of the participant in that survey. Mean and standard error are shown in orange. Times above 20 minutes were excluded.}
    \label{fig:time}
\end{figure*}

\begin{figure*}
    \centering
    \includegraphics[width=0.8\linewidth]{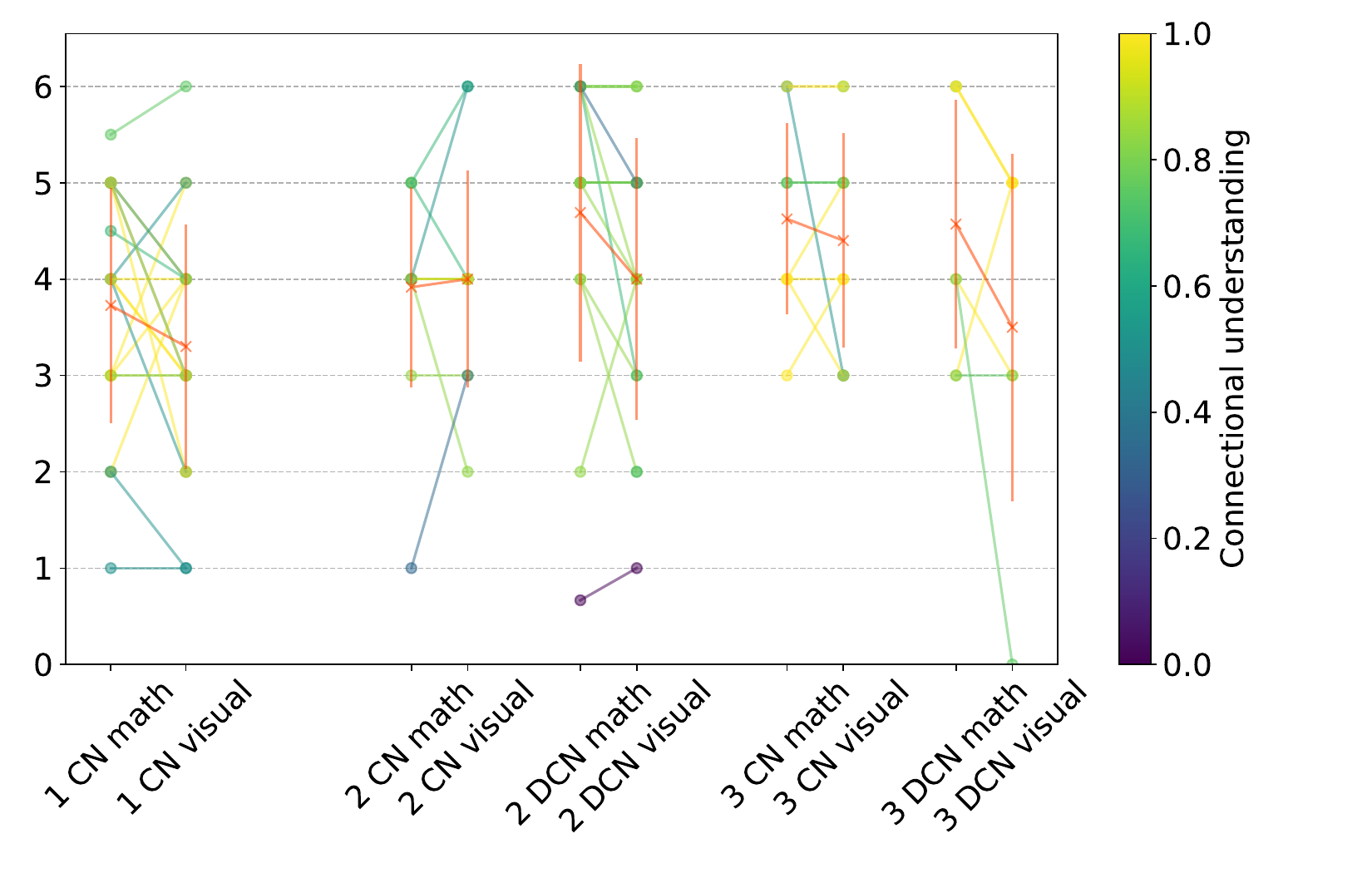}
    \caption{\acl{icl} by number of qubits, without visualization (math) and with visualization. Two points connected by a line represent single participants, colored according to the connectional understanding score of the participant in that survey. Mean and standard error are shown in orange.}
    \label{fig:icl}
\end{figure*}

\begin{figure*}
    \centering
    \includegraphics[width=0.8\linewidth]{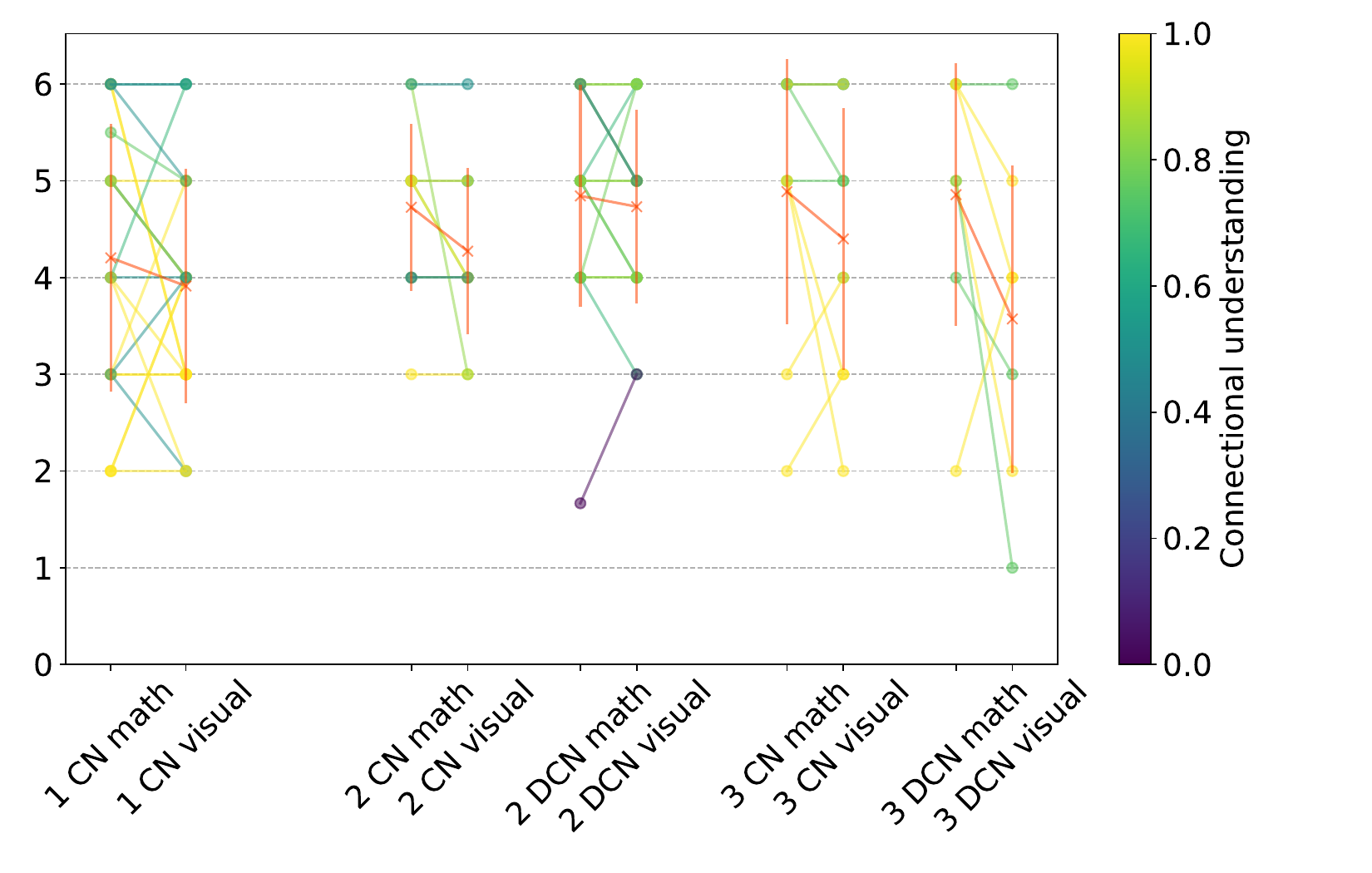}
    \caption{\acl{gcl} by number of qubits, without visualization (math) and with visualization. Two points connected by a line represent single participants, colored according to the connectional understanding score of the participant in that survey. Mean and standard error are shown in orange.}
    \label{fig:gcl}
\end{figure*}

\begin{figure*}
    \centering
    \includegraphics[width=0.8\linewidth]{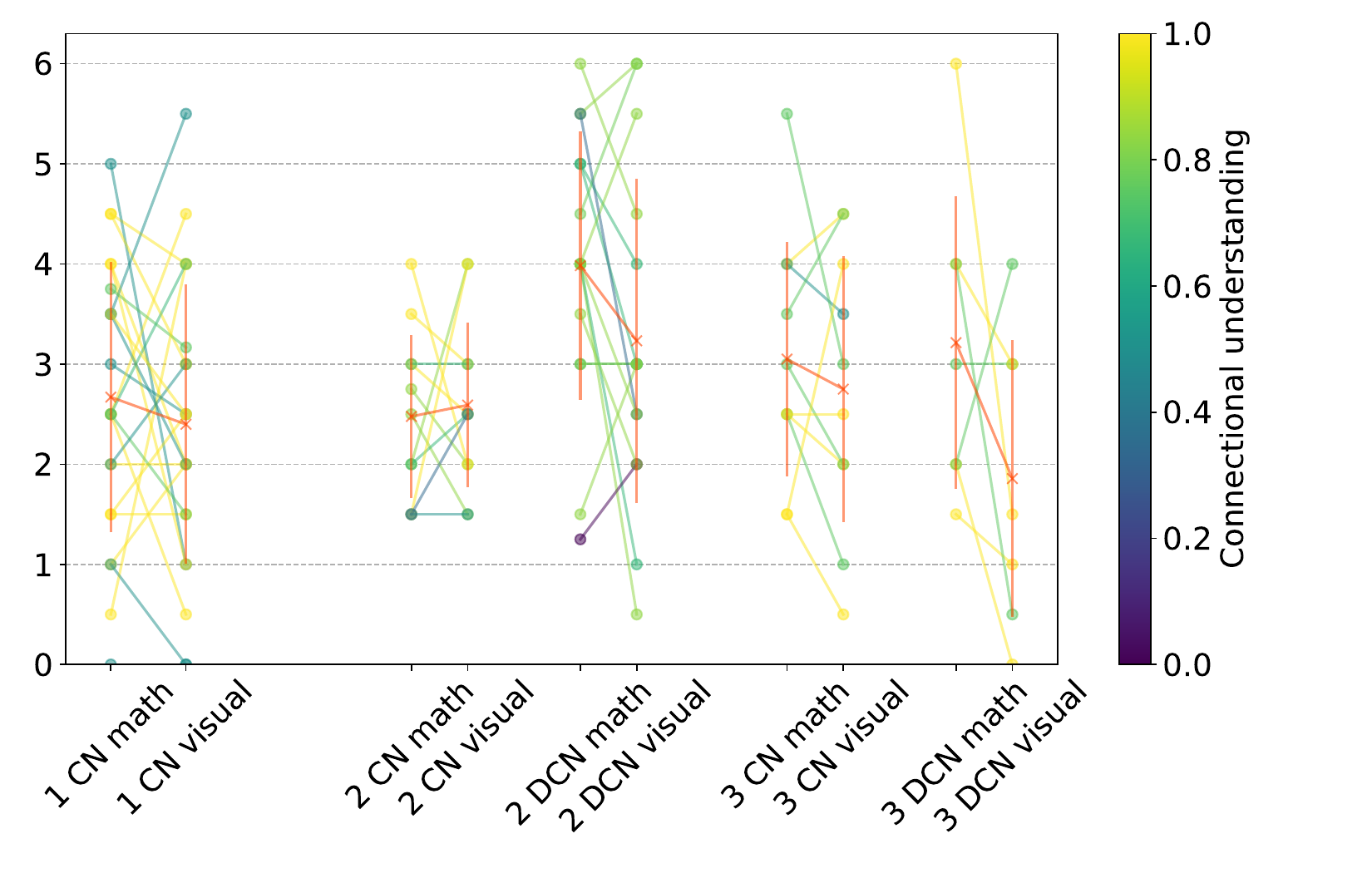}
    \caption{\acl{ecl} by number of qubits, without visualization (math) and with visualization. Two points connected by a line represent single participants, colored according to the connectional understanding score of the participant in that survey. Mean and standard error are shown in orange.}
    \label{fig:ecl}
\end{figure*}

Table \ref{tab:dif_scores_times} summarizes the average connectional understanding, average scores, and average times for correctly solved questions. In addition, it shows the differences of these performance metrics with visualization minus without visualization. Average \acf{icl}, \acf{gcl} and \acf{ecl} and the corresponding differences (between with visualization and without visualization) are shown in Table \ref{tab:dif_cl}. Figure \ref{fig:score} shows the average score per participant, and Figure \ref{fig:time} the average time taken for correctly solved questions. The influence of connectional understanding on \ac{icl}, \ac{gcl} and \ac{ecl} as well as the average cognitive load is shown in figures \ref{fig:icl}, \ref{fig:gcl} and \ref{fig:ecl} for tasks with one, two, and three qubits.

In the following, we show detailed results for single-qubit (Section \ref{sec:A_results_single}), two-qubit (Section \ref{sec:A_results_two}) and three-qubit studies (Section \ref{sec:A_results_three}). We describe the average correctness per question with and without visualization and survey parts A and B and the combined score of part A and B. Then, we do the same for average time taken for correct answers, considering for the combination of parts A and B only cases where both questions of each topic, with and without visualization, were answered correctly.

\subsection{One-qubit survey}\label{sec:A_results_single}

\begin{figure*}
    \centering
    \includegraphics[width=\linewidth]{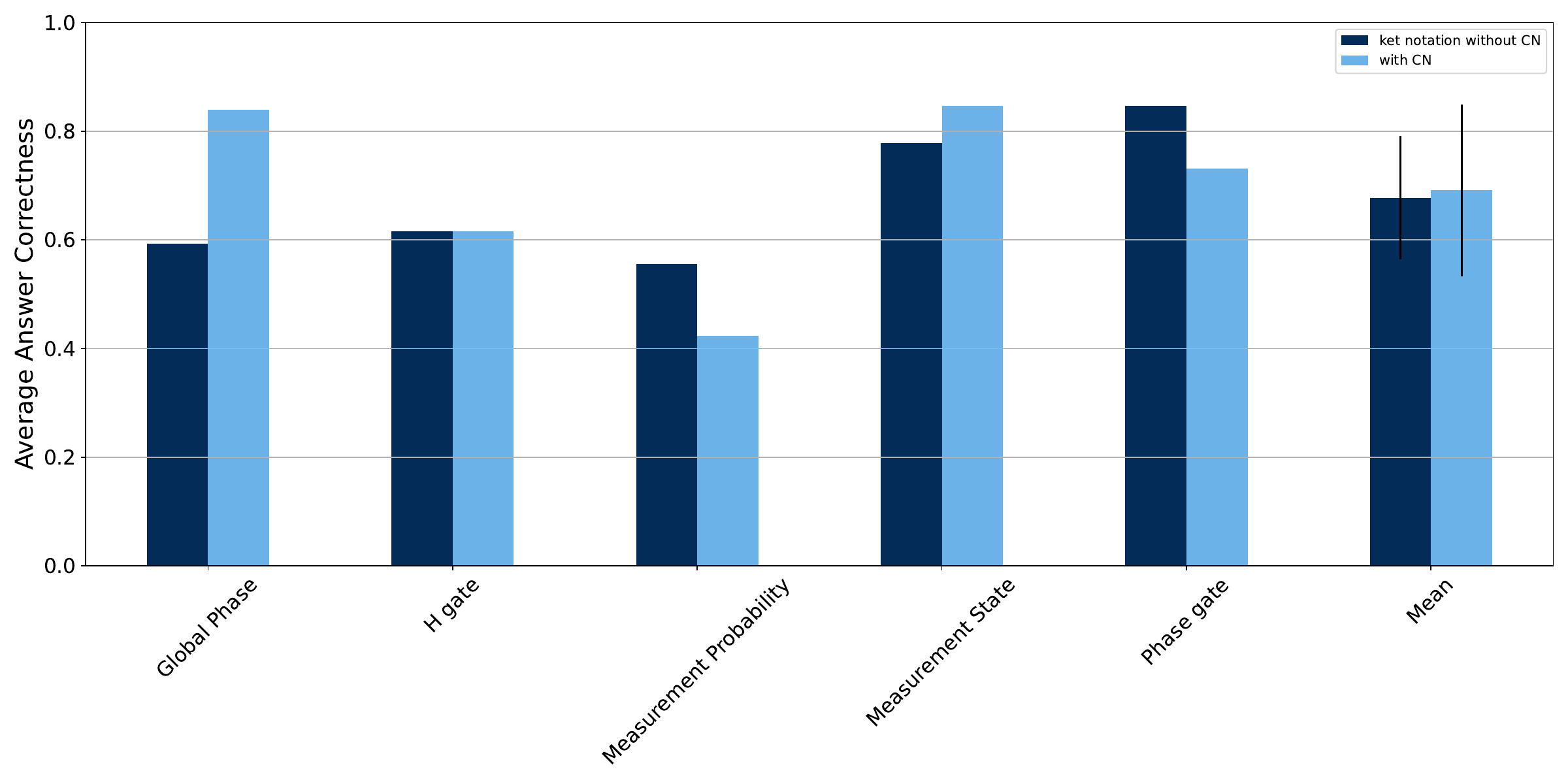}
    \caption{Average score for each question in the one-qubit survey, for questions were participants were presented with \acf{cn} (dark blue) and questions were the participants weren't presented with \ac{cn} (light blue). (N=23).}
    \label{fig:score_1}
\end{figure*}

\begin{figure*}
    \centering
    \includegraphics[width=\linewidth]{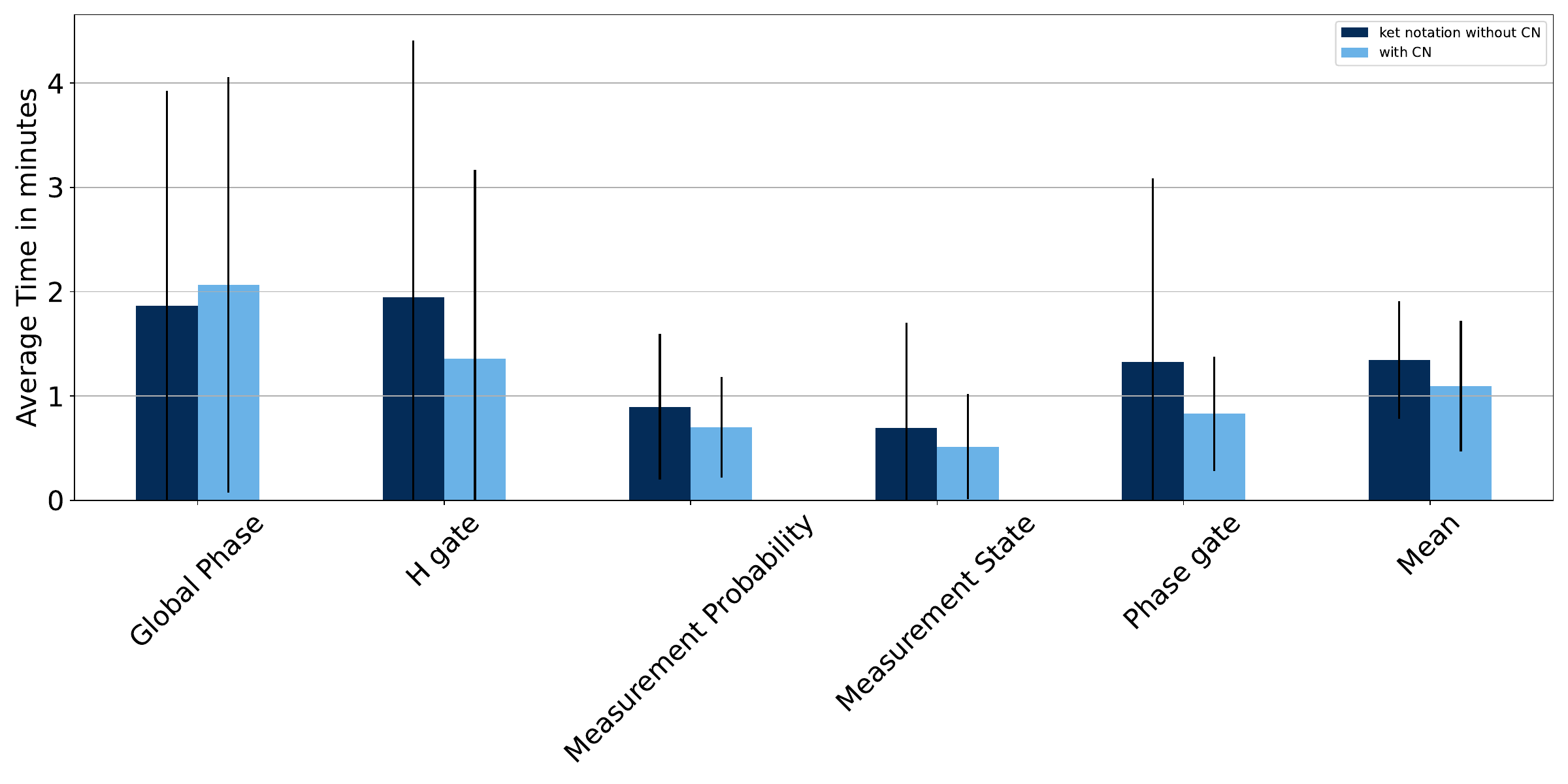}
    \caption{Average time taken for correct answers to each question in minutes in the one-qubit survey, for questions were participants were presented with \acf{cn} (dark blue) and questions were the participants weren't presented with \ac{cn} (light blue). (N=18).}
    \label{fig:time_1}
\end{figure*}

The average scores for single-qubit questions are shown in Table \ref{tab:1qubit_correctness} and average times for correctly solved questions in Table \ref{tab:1qubit_time}. In the latter, the combined score of part A and part B show only cases where participants managed to solve both questions, i.e. both questions on the same concept with and without visualization, correctly. Figure \ref{fig:score_1} shows the combined average score per question and Figure \ref{fig:time_1} shows the combined average time taken (for all answers, not just the correct ones). Cases in which a participant took more than 20 minutes for one question were excluded.

During part A, group math-vis solved questions without visualization with an average score of $0.74 \pm 0.14$ and an average time of $1.60 \pm 0.64$ minutes for correctly solved questions, while group vis-math answered the same questions with visualization with an average score of $0.66 \pm 0.28$ and an average time of $2.28 \pm 1.06$ minutes. During part B, the average scores were 0.57$\pm$0.17 (group vis-math) and $0.6 \pm 0.11$ (group math-vis), respectively, at average times for correctly solved questions of $1.57 \pm 0.63$ and $1.78 \pm 1.17$ minutes. When combining parts A and B of the survey, the average scores were $0.66 \pm 0.12$ without visualization and $0.63 \pm 0.12$ with visualization, at $1.55 \pm 0.51$ and $2.07 \pm 1.1$ minutes.

In survey part A, group math-vis scored higher than group vis-math on the global phase, X gate, phase gate, H gate, Measurement Prob., and Measurement State questions with an average score of $0.67 \pm 0.13$ to $0.5 \pm 0.24$ on these questions and did not score lower on any questions. Group math-vis took longer on correct answers to the X gate, phase gate, H gate, and Measurement State questions with an average of $151 \pm 74$ to $56 \pm 22$ seconds while taking less time on the Global Phase Equiv. and Measurement Prob. questions with $84 \pm 4$ to $75 \pm 1$ seconds.

In part B of the survey, group math-vis scored higher than group vis-math on the Global Phase Equiv., Z gate, H gate, Measurement Prob., and Measurement State questions, with an average of $0.84 \pm 0.13$ to $0.62 \pm 0.14$. Group vis-math scored higher on the phase gate question, with 0.88 to 0.78. Group math-vis took more time on the global phase, Z gate, and Measurement State questions with  $178 \pm 120$ to $140 \pm 85$ seconds and group vis-math took longer on phase gate, H gate and Measurement Prob. questions with $77 \pm 23$ to $63 \pm 19$ seconds.

When presented with visualization, group math-vis scored higher on the global phase, H gate, Measurement Prob., and Measurement State questions with average scores of $0.86 \pm 0.14$ to $0.67 \pm 0.14$. When presented questions with visualization, group math-vis took longer on the Global Phase Equiv. question with 177 to 77 seconds and took a shorter amount of time on the phase gate, H gate, Measurement Prob., and Measurement State questions with $104 \pm 43$ to $56 \pm 21$ seconds than when the group was not presented with a visualization. Group vis-math scored higher when presented without visualization on the global phase, phase gate, H gate, and Measurement Prob. question with scores of $0.66 \pm 0.18$ to $0.5 \pm 0.23$. Group vis-math took longer on the Measurement Prob. and Measurement State questions when presented with visualization with $57 \pm 23$ to $37 \pm 15$ seconds and took less time on the global phase, phase gate, and H gate questions with $118 \pm 43$ to $63 \pm 20$ seconds.

\begin{table*}
\caption{Average scores for one-qubit questions, with and without visualization (N=17).}\label{tab:1qubit_correctness}
\begin{tabular*}{\linewidth}{@{\extracolsep\fill}lcccccc}
\toprule%
\textbf{survey part:} &  \multicolumn{2}{@{}c@{}}{\textbf{A}} & \multicolumn{2}{@{}c@{}}{\textbf{B}} & \multicolumn{2}{@{}c@{}}{\textbf{combined}} \\
\cmidrule(r{0.75em}){1-1}\cmidrule(lr{0.75em}){2-3}\cmidrule(lr{0.75em}){4-5}\cmidrule(lr{0.75em}){6-7}
\textbf{question} & \textbf{w/o vis.\footnotemark[1]} & \textbf{with vis.\footnotemark[2]} & \textbf{w/o vis.\footnotemark[2]} & \textbf{with vis.\footnotemark[1]} & \textbf{w/o vis.\footnotemark[3]} & \textbf{with vis.\footnotemark[3]} \\
\midrule%
\textbf{Global Phase} & 0.62 & 0.67 & 0.57 & 1.0 & 0.59 & 0.84 \\
\textbf{X gate} & 0.46 & 0.29 & - & - & - & - \\
\textbf{Z gate} & - & - & 0.54 & 0.75 & - & - \\
\textbf{Phase gate} & 0.77 & 0.64 & 0.92 & 0.83 & 0.85 & 0.73 \\
\textbf{H gate} & 0.69 & 0.57 & 0.54 & 0.67 & 0.62 & 0.62 \\
\textbf{Measurement Prob.} & 0.62 & 0.21 & 0.5 & 0.67 & 0.56 & 0.42 \\
\textbf{Measurement State} & 0.92 & 0.71 & 0.64 & 1.0 & 0.78 & 0.85 \\
\midrule%
\textbf{Mean} & 0.68$\pm$0.14 & 0.52$\pm$0.19 & 0.62$\pm$0.14 & 0.82$\pm$0.14 & 0.68$\pm$0.11 & 0.69$\pm$0.16 \\
\botrule
\end{tabular*}
\footnotetext[0]{Note: Standard deviation is used for confidence intervals.}
\footnotetext[1]{N = 12}
\footnotetext[2]{N = 11}
\footnotetext[3]{N = 23}
\end{table*}

\begin{table*}
\caption{Average time taken for correct answers to one-qubit questions, with and without visualization in minutes (N=17).}\label{tab:1qubit_time}
\begin{tabular*}{\linewidth}{@{\extracolsep\fill}lcccccccccccc}
\toprule%
\textbf{survey part:} &  \multicolumn{2}{@{}c@{}}{\textbf{A}} & \multicolumn{2}{@{}c@{}}{\textbf{B}} & \multicolumn{2}{@{}c@{}}{\textbf{combined}} \\
\cmidrule(r{0.75em}){1-1}\cmidrule(lr{0.75em}){2-3}\cmidrule(lr{0.75em}){4-5}\cmidrule(lr{0.75em}){6-7}
\textbf{question} & \textbf{w/o vis.\footnotemark[1]} & \textbf{with vis.\footnotemark[2]} & \textbf{w/o vis.\footnotemark[2]} & \textbf{with vis.\footnotemark[1]} & \textbf{w/o vis.\footnotemark[3]} & \textbf{with vis.\footnotemark[3]} \\
\midrule%
\textbf{Global Phase Equiv.}& 1.28$\pm$0.63(6) & 1.48$\pm$1.57(5) & 2.94$\pm$2.99(6) & 2.95$\pm$2.2(10) & 1.63$\pm$1.08(10) & 2.26$\pm$2.27(10) \\
\textbf{X gate} & 4.31$\pm$2.61(5) & 1.52$\pm$0.92(3) & - & - & - & - \\
\textbf{Z gate} & - & - & 3.69$\pm$5.56(5) & 5.43$\pm$4.11(7) & - & - \\
\textbf{phase gate} & 1.65$\pm$2.49(8) & 1.01$\pm$0.44(7) & 1.19$\pm$1.05(9) & 0.87$\pm$0.42(7) & 1.47$\pm$1.99(11) & 0.93$\pm$0.46(11) \\
\textbf{H gate} & 2.93$\pm$2.9(7) & 0.68$\pm$0.41(6) & 1.78$\pm$2.74(6) & 1.5$\pm$0.83(7) & 2.13$\pm$2.59(10) & 0.84$\pm$0.4(10) \\
\textbf{Measurement Prob.} & 1.23$\pm$1.0(6) & 1.33(1) & 0.86$\pm$0.59(4) & 0.81$\pm$0.58(6) & 1.13$\pm$0.96(6) & 0.91$\pm$0.61(6) \\
\textbf{Measurement State} & 1.14$\pm$1.39(9) & 0.55$\pm$0.24(7) & 0.38$\pm$0.17(6) & 0.54$\pm$0.68(9) & 0.83$\pm$1.14(14) & 0.55$\pm$0.55(14) \\
\midrule%
\textbf{Mean} & 2.09$\pm$1.27 & 1.09$\pm$0.41 & 1.81$\pm$1.28 & 2.01$\pm$1.88 & 1.44$\pm$0.49 & 1.1$\pm$0.67 \\
\botrule
\end{tabular*}
\footnotetext[0]{Note: The number of participants N that solved the question correctly and are therefore included in the average is shown in brackets. Standard deviations are shown where N $>1$. The confidence interval of the mean is the mean of the standard deviations.}
\footnotetext[1]{N = 9}
\footnotetext[2]{N = 8}
\footnotetext[3]{N = 17}
\end{table*} 

\subsubsection{Representational competence}\label{sec:A_rep_comp}

Participants were asked to calculate Euler formula identities and similar questions to translate between \ac{cn} to a complex number. In addition, they were asked to multiply numbers written in \ac{dn} and \ac{cn} as shown in Figure \ref{fig:rep_comp_example}. Then they were asked to rate their confidence in the question from 0 (random guess), 1 (very unsure) to 4 (very sure). The correct answers were weighted (multiplied by $x/4$, where $x$ is the confidence). The weighted (unweighted) average symbolic representational understanding was $0.76 \pm 0.25$ ($0.87 \pm 0.23$). The weighted (unweighted) average \ac{cn} representational understanding was $0.47 \pm 0.34$ ($0.69 \pm 0.32$). The average weighted (unweighted) difference of \ac{cn} representational understanding minus mathematics representational understanding was $-0.29 \pm 0.23$ ($-0.18 \pm0.19$). 

\begin{figure*}
    \centering
    \includegraphics[width=0.8\linewidth]{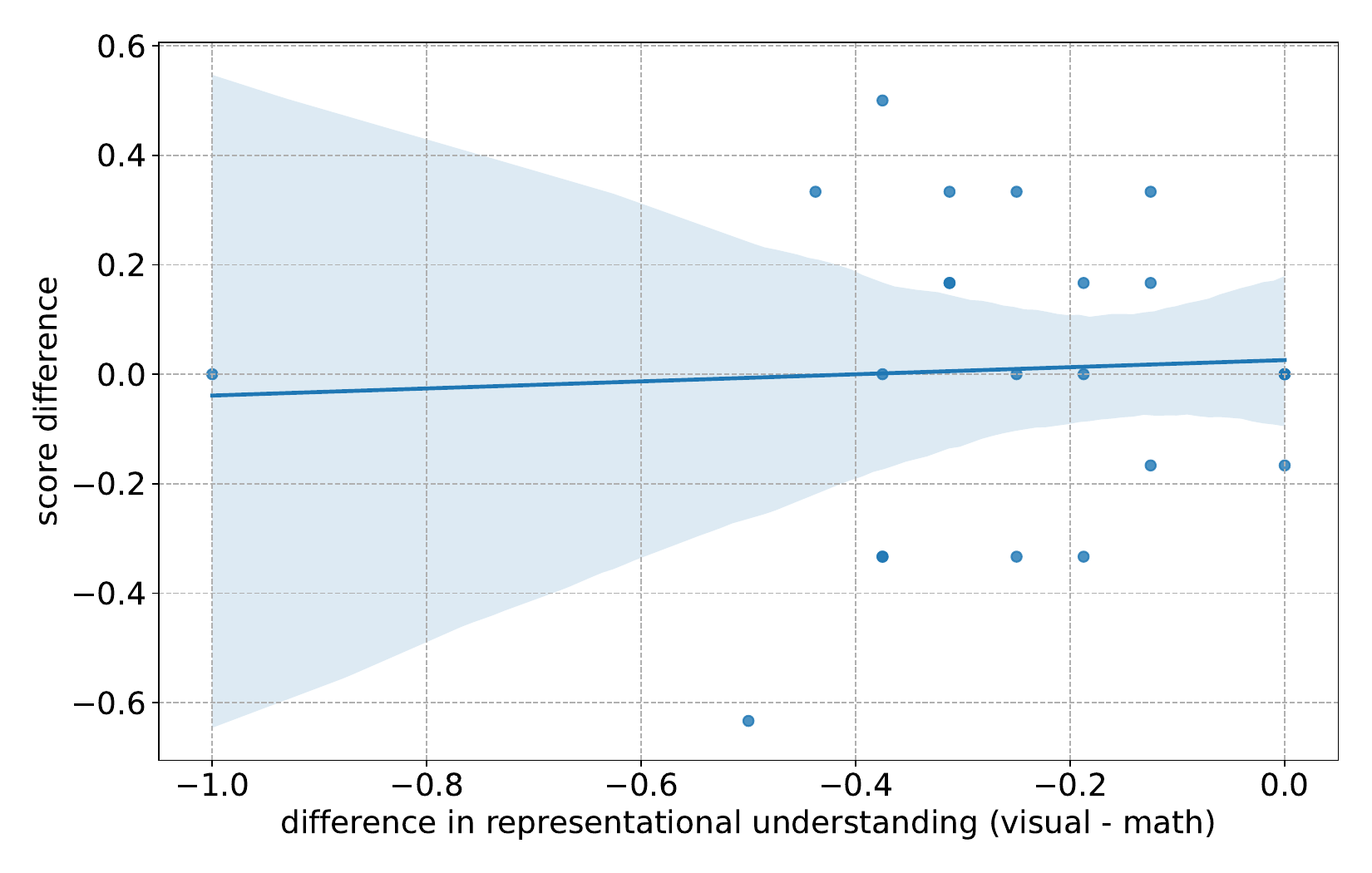}
    \caption{Difference of average score (with \ac{cn} minus without \ac{cn}) plotted against difference of representational understanding (with \ac{cn} minus without \ac{cn}) in the one-qubit survey. The linear regression has slope $b=0.04$, intercept $a=0.01$, Pearson $r=0.03$ at $p=0.91$, standard error $\sigma=0.32$. The shaded areas represents a 95\% confidence interval.}
    \label{fig:rep_comp_score}
\end{figure*}

\begin{figure*}
\centering
    \includegraphics[width=0.9\linewidth]{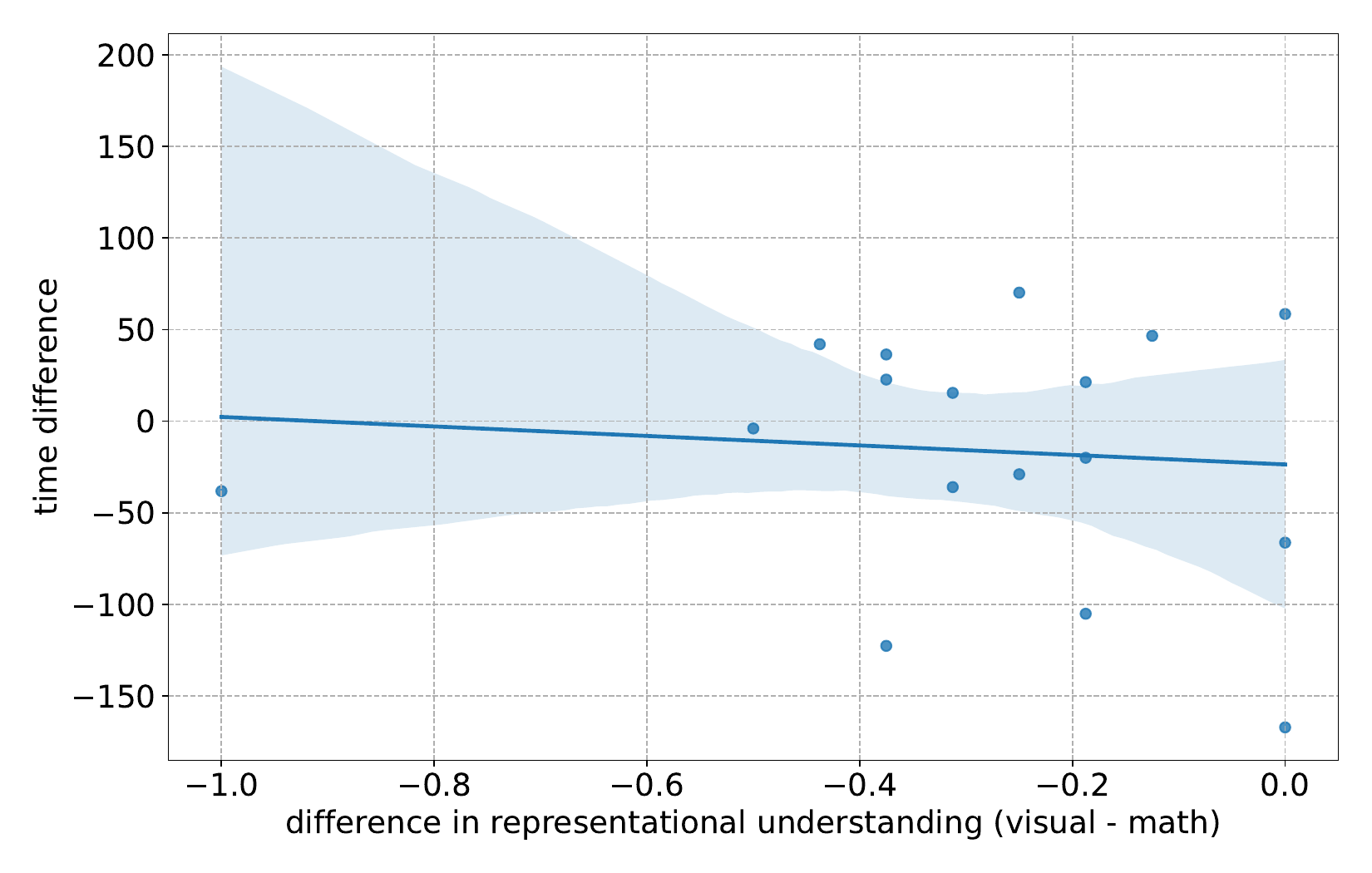}
    \caption{Average time taken for correct answers (with \ac{cn} minus symbolic notation) plotted against difference of representational understanding (with \ac{cn} minus without \ac{cn}) in the one-qubit survey. The linear regression has slope $b=-26$, intercept $a=-24$, Pearson $r=-0.09$ at $p=0.73$ and standard error $\sigma=73$. The shaded areas represents a 95\% confidence interval.}
    \label{fig:rep_comp_time}
\end{figure*}

\begin{figure*}
    \centering
    \includegraphics[width=0.9\linewidth]{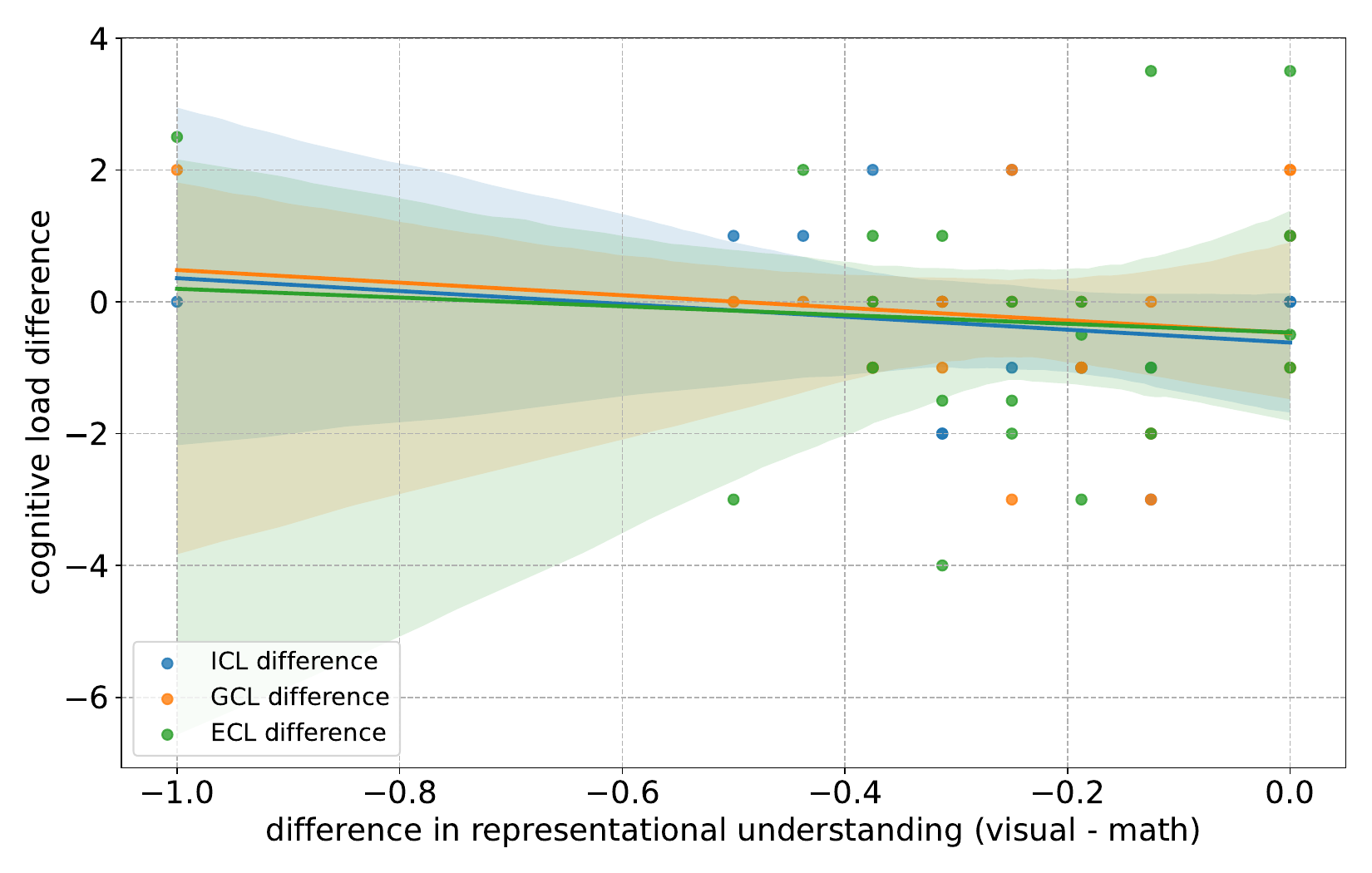}
    \caption{Perceived intrinsic, germane and \ac{ecl} difference (with \ac{cn} minus without \ac{cn}) and their dependency on the representational understanding difference (again, with \ac{cn} minus symbolic notation) in the one-qubit survey. Shown are also the linear regressions between the three variables and the representational understanding difference with slope $b$ and intercept $a$. Pearson's $r$, $p$ value and standard error $\sigma$ are as follows. \ac{icl}: $b=-0.68$, $a=-0.27$, $r=-0.16$, $p=0.59$, $\sigma=1.2$ \ac{gcl}: $b=-0.83$, $a=-0.36$, $-0.16$, $p=0.56$, $\sigma=1.4$ \ac{ecl}: $b=-0.87$, $a=-0.37$, $r=-0.09$, $p=0.73$, $\sigma=2.5$. The shaded areas represent 95\% confidence intervals.}
    \label{fig:rep_comp_CL_summary}
\end{figure*}

Figure \ref{fig:rep_comp_score} shows, for each participant in the first survey, the difference in scores (with visualization minus without visualization) and Figure \ref{fig:rep_comp_time} the difference in time, both plotted against the difference in representational understanding between visualization and math. Figure \ref{fig:rep_comp_CL_summary} shows the dependency of the difference in intrinsic, germane, and \ac{ecl} on the representational understanding difference. 

\subsubsection{Connectional understanding and cognitive load}

On average connectional understanding, 12 students achieved perfect scores and 5 students achieved a score of 0.5. The students who did not score perfectly also scored lower in the rest of the survey ($0.5 \pm 0.21$ to $0.75 \pm 0.21$ without visualization and $0.63 \pm 0.19$ to $0.69 \pm 0.25$ with visualization) and took a smaller amount of time on the correctly solved questions ($82 \pm 74$ to $114 \pm 71$ seconds without visualization and $86 \pm 57$ to $89 \pm 55$ seconds with visualization) than the students who solved the connectional understanding questions perfectly. 

Average \ac{ecl} ($2.88 \pm 1.67$ with visualization and $2.76 \pm 1.54$ without) was lower than the average \ac{icl} ($3.75 \pm 0.97$ with visualization and $3.64 \pm 1.04$ without) and \ac{gcl} ($3.94 \pm 1.26$ with visualization and $4.13 \pm 1.41$ without).

\subsection{Two-qubit survey}\label{sec:A_results_two}

\begin{figure*}
    \centering
    \includegraphics[width=\linewidth]{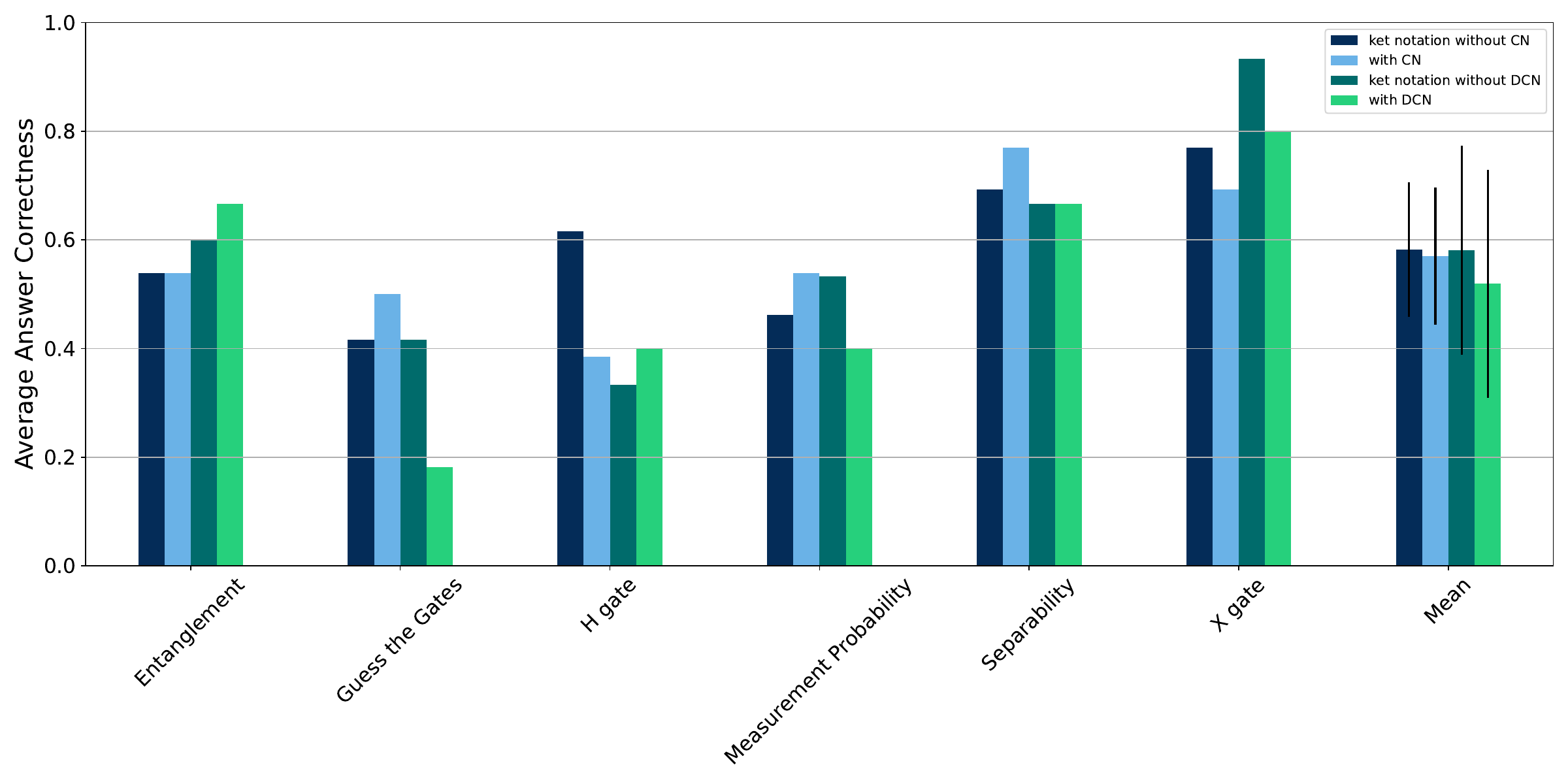}
    \caption{Average score for each question in the two-qubit survey, separated by participants who where presented with \ac{cn} (blue, N=12) and participants who where presented with \ac{dcn} (green, N=15). Darker colors are for questions where participants were not presented with a visualization and lighter colors for questions where participants were presented with a visualization.}
    \label{fig:score_2}
\end{figure*}

\begin{figure*}
    \centering
    \includegraphics[width=\linewidth]{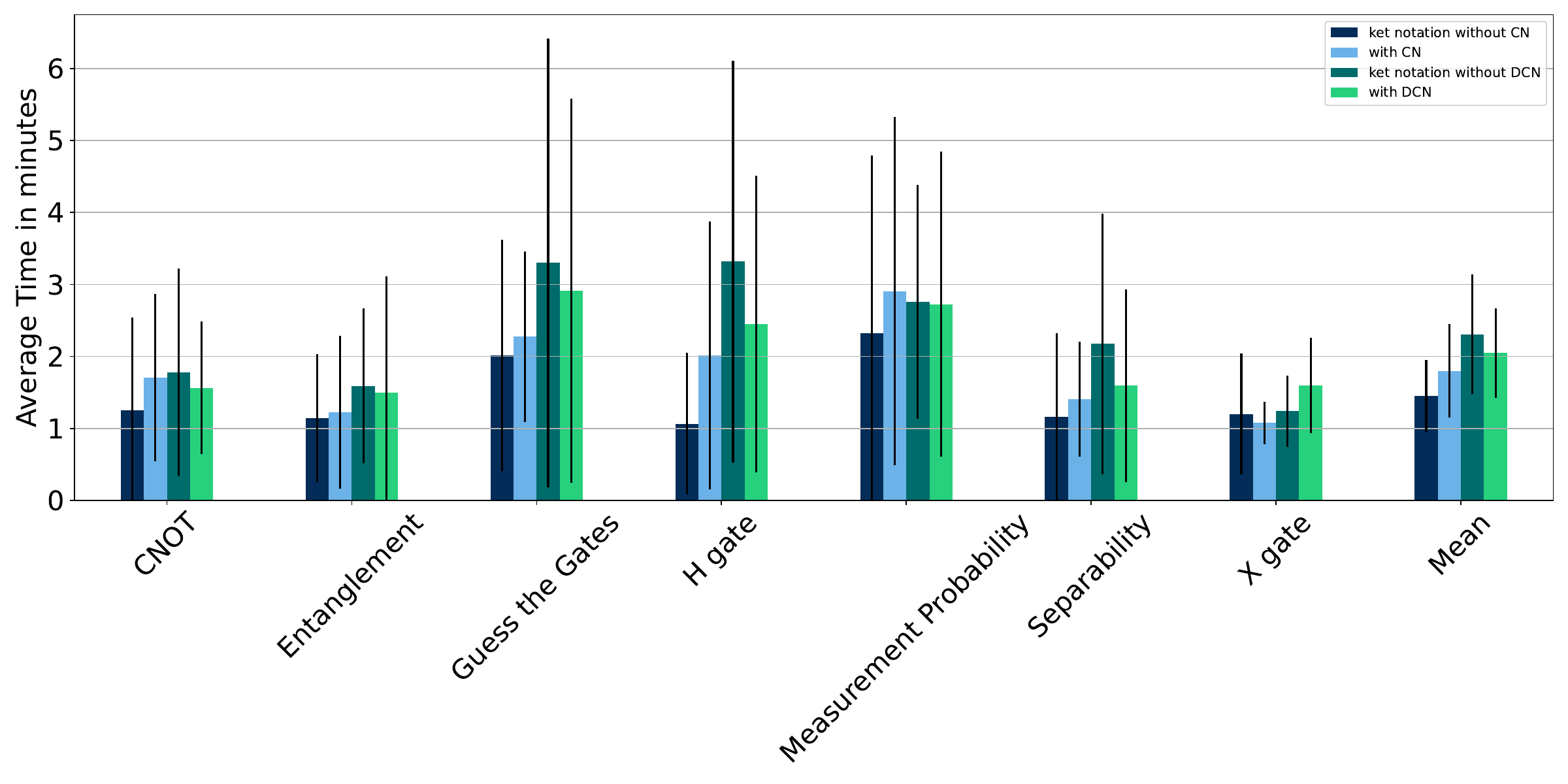}
    \caption{Average time taken for correct answers to each question in minutes in the two-qubit survey, separated by participants who where presented with \ac{cn} (blue, N=11) and participants who where presented with \ac{dcn} (green, N=11). Darker colors are for questions where participants were not presented with a visualization and lighter colors for questions where participants were presented with a visualization.}
    \label{fig:time_2}
\end{figure*}

Table \ref{tab:2qubit_correctness} lists the average scores and Table \ref{tab:2qubit_time} lists the average time taken for correctly solved one-qubit questions, with and without visualization, for surveys that were taken supported by \ac{cn} and \ac{dcn}. Figure \ref{fig:score_2} shows the combined average score per question and Figure \ref{fig:time_2} shows the combined average time taken (for all answers, not just correct ones). Cases where a participant took more than 20 minutes for one question were excluded.

\begin{table*}
\caption{Average scores for two-qubit questions, with and without visualization (N=27).}\label{tab:2qubit_correctness}
\begin{tabular*}{\linewidth}{@{\extracolsep\fill}lcccccc}
\toprule%
 & \multicolumn{6}{@{}c@{}}{\textbf{Circle Notation}} \\
 \midrule%
\textbf{survey part:} &  \multicolumn{2}{@{}c@{}}{\textbf{A}} & \multicolumn{2}{@{}c@{}}{\textbf{B}} & \multicolumn{2}{@{}c@{}}{\textbf{combined}} \\
\cmidrule(r{0.75em}){1-1}\cmidrule(lr{0.75em}){2-3}\cmidrule(lr{0.75em}){4-5}\cmidrule(lr{0.75em}){6-7}
\textbf{question} & \textbf{w/o vis.}\footnotemark[2] & \textbf{with vis.}\footnotemark[2] & \textbf{w/o vis.}\footnotemark[2] & \textbf{with vis.}\footnotemark[2] & \textbf{w/o vis.}\footnotemark[1] & \textbf{with vis.}\footnotemark[1] \\
\midrule%
\textbf{X gate} & 0.86 & 0.5 & 0.67 & 0.86 & 0.77 & 0.69 \\
\textbf{H gate} & 0.86 & 0.17 & 0.33 & 0.57 & 0.62 & 0.38 \\
\textbf{Measurement Prob.} & 0.57 & 0.5 & 0.33 & 0.57 & 0.46 & 0.54 \\
\textbf{Separability} & 0.57 & 1.0 & 0.83 & 0.57 & 0.69 & 0.77 \\
\textbf{Entanglement} & 0.57 & 0.83 & 0.5 & 0.29 & 0.54 & 0.54 \\
\textbf{Guess-the-Gates} & 0.5 & 0.5 & 0.33 & 0.5 & 0.42 & 0.5 \\
\midrule%
\textbf{Mean} & 0.65$\pm$0.15 & 0.58$\pm$0.27 & 0.5$\pm$0.19 & 0.56$\pm$0.17 & 0.58$\pm$0.12 & 0.57$\pm$0.13 \\
\toprule%
 & \multicolumn{6}{@{}c@{}}{\textbf{Dimensional Circle Notation}} \\
 \midrule%
\textbf{survey part:} &  \multicolumn{2}{@{}c@{}}{\textbf{A}} & \multicolumn{2}{@{}c@{}}{\textbf{B}} & \multicolumn{2}{@{}c@{}}{\textbf{combined}} \\
\cmidrule(r{0.75em}){1-1}\cmidrule(lr{0.75em}){2-3}\cmidrule(lr{0.75em}){4-5}\cmidrule(lr{0.75em}){6-7}
\textbf{question} & \textbf{w/o vis.\footnotemark[4]} & \textbf{with vis.\footnotemark[3]} & \textbf{w/o vis.\footnotemark[3]} & \textbf{with vis.\footnotemark[4]} & \textbf{w/o vis.\footnotemark[5]} & \textbf{with vis.\footnotemark[5]} \\
\midrule%
\textbf{X gate} & 1.0 & 0.88 & 0.88 & 0.71 & 0.93 & 0.8 \\
\textbf{H gate} & 0.14 & 0.62 & 0.5 & 0.14 & 0.33 & 0.4 \\
\textbf{Measurement Prob.} & 0.43 & 0.5 & 0.62 & 0.29 & 0.53 & 0.4 \\
\textbf{Separability} & 0.57 & 0.62 & 0.75 & 0.71 & 0.67 & 0.67 \\
\textbf{Entanglement} & 0.43 & 0.88 & 0.75 & 0.43 & 0.6 & 0.67 \\
\textbf{Guess-the-Gates} & 0.71 & 0.2 & 0.0 & 0.17 & 0.42 & 0.18 \\
\midrule%
\textbf{Mean} & 0.55$\pm$0.27 & 0.62$\pm$0.23 & 0.58$\pm$0.29 & 0.41$\pm$0.24 & 0.58$\pm$0.19 & 0.52$\pm$0.21 \\
\botrule%
\end{tabular*}
\footnotetext[0]{Note: Standard deviation is used for confidence intervals.}
\footnotetext[1]{N = 12}
\footnotetext[2]{N = 6}
\footnotetext[3]{N = 7}
\footnotetext[4]{N = 8}
\footnotetext[5]{N = 15}
\end{table*}

\begin{table*}
\caption{Average time taken for correct answers to two-qubit questions in minutes (N = 22).}\label{tab:2qubit_time}
\begin{tabular*}{\linewidth}{@{\extracolsep\fill}lcccccccccccc}
\toprule%
 & \multicolumn{6}{@{}c@{}}{\textbf{Circle Notation}\footnotemark[1]} \\
 \midrule%
\textbf{survey part:} &  \multicolumn{2}{@{}c@{}}{\textbf{A}} & \multicolumn{2}{@{}c@{}}{\textbf{B}} & \multicolumn{2}{@{}c@{}}{\textbf{combined}} \\
\cmidrule(r{0.75em}){1-1}\cmidrule(lr{0.75em}){2-3}\cmidrule(lr{0.75em}){4-5}\cmidrule(lr{0.75em}){6-7}
\textbf{question} & \textbf{w/o vis.\footnotemark[2] }& \textbf{with vis.\footnotemark[3]} & \textbf{w/o vis.\footnotemark[3]} & \textbf{with vis.\footnotemark[2]} & \textbf{w/o vis.\footnotemark[1]} & \textbf{with vis.\footnotemark[1]} \\
\midrule%
\textbf{X gate} & 0.71$\pm$0.16(5) & 1.34$\pm$0.06(2) & 2.05$\pm$1.06(3) & 0.8$\pm$0.35(5) & 0.95$\pm$0.39(6) & 1.08$\pm$0.23(6) \\
H gate & 0.93$\pm$0.96(5) & 2.18(1) & 2.28$\pm$0.83(2) & 2.2$\pm$0.64(3) & 1.15$\pm$0.96(3) & 2.2$\pm$0.64(3) \\
\textbf{Measurement Prob.} & 2.38$\pm$0.33(3) & 4.58$\pm$2.83(3) & 1.38$\pm$0.34(2) & 4.12$\pm$4.42(3) & 2.0$\pm$0.67(4) & 4.39$\pm$3.42(4) \\
\textbf{Separability} & 1.78$\pm$1.34(3) & 1.87$\pm$0.61(5) & 0.95$\pm$0.94(4) & 0.84$\pm$0.25(3) & 1.3$\pm$1.2(7) & 1.38$\pm$0.7(7) \\
\textbf{Entanglement} & 1.47$\pm$0.39(4) & 1.73$\pm$1.21(5) & 1.09$\pm$0.33(3) & 1.24$\pm$0.12(2) & 1.36$\pm$0.43(5) & 1.86$\pm$1.13(5) \\
\textbf{Guess-the-Gates} & 2.34$\pm$2.01(3) & 2.08$\pm$0.61(3) & 2.34$\pm$2.29(2) & 1.17$\pm$1.04(3) & 2.41$\pm$2.37(4) & 1.48$\pm$0.91(4) \\
\midrule%
\textbf{Mean} & 1.6$\pm$0.64 & 2.28$\pm$1.06 & 1.57$\pm$0.63 &  1.78$\pm$1.17 & 1.55$\pm$0.51 & 2.07$\pm$1.1 \\
\toprule%
 & \multicolumn{6}{@{}c@{}}{\textbf{Dimensional Circle Notation}\footnotemark[1]} \\
 \midrule%
\textbf{survey part:} &  \multicolumn{2}{@{}c@{}}{\textbf{A}} & \multicolumn{2}{@{}c@{}}{\textbf{B}} & \multicolumn{2}{@{}c@{}}{\textbf{combined}} \\
\cmidrule(r{0.75em}){1-1}\cmidrule(lr{0.75em}){2-3}\cmidrule(lr{0.75em}){4-5}\cmidrule(lr{0.75em}){6-7}
\textbf{question} & \textbf{w/o vis.\footnotemark[2]} &\textbf{with vis.\footnotemark[3]} & \textbf{w/o vis.\footnotemark[3]} & \textbf{with vis.\footnotemark[2]} & \textbf{w/o vis.\footnotemark[1]} & \textbf{with vis.\footnotemark[1]} \\
\midrule%
\textbf{X gate} & 1.13$\pm$0.38(6) & 1.77$\pm$0.67(4) & 1.35$\pm$0.65(4) & 1.7$\pm$0.66(5) & 1.29$\pm$0.49(9) & 1.73$\pm$0.67(9) \\
\textbf{H gate} & 2.73(1) & 6.12$\pm$7.98(3) & 6.17$\pm$1.65(2) & 5.88(1) & 7.82(1) & 17.4(1) \\
\textbf{Measurement Prob.} & 3.22$\pm$1.7(3) & 2.91$\pm$1.62(3) & 1.71$\pm$0.26(3) & 5.13$\pm$3.27(2) & 1.86$\pm$0.34(4) & 2.65$\pm$1.48(4) \\
\textbf{Separability} & 2.43$\pm$1.85(3) & 3.08$\pm$1.99(2) & 2.48$\pm$2.22(3) & 1.68$\pm$0.9(5) & 2.83$\pm$2.03(5) & 2.14$\pm$1.51(5) \\
\textbf{Entanglement} & 2.27$\pm$0.49(3) & 1.39$\pm$1.4(4) & 1.41$\pm$0.95(3) & 2.77$\pm$2.48(2) & 1.81$\pm$0.95(5) & 2.09$\pm$2.04(5) \\
\textbf{Guess-the-Gates} & 2.53$\pm$0.68(4) & 1.9(1) & (0) & 7.07(1) & (0) & (0) \\
\midrule%
\textbf{Mean} & 2.39$\pm$0.64 & 2.68$\pm$1.65 & 2.44$\pm$1.87 & 3.71$\pm$2.27 &  2.95$\pm$2.44 & 4.64$\pm$6.26 \\
\botrule%
\end{tabular*}
\footnotetext[0]{Note: The number of participants N that solved the question correctly and are therefore included in the average is shown in brackets. Standard deviations are shown where N $>1$. The confidence interval of the mean is the mean of the standard deviations.}
\footnotetext[1]{N = 11}
\footnotetext[2]{N = 6}
\footnotetext[3]{N = 5}
\end{table*} 

\subsubsection{Circle Notation} 

During part A, participants assigned to the \ac{cn} group solved the questions without visualization with an average score of $0.72 \pm 0.14$ with an average time of $1.54 \pm 0.62$ minutes for correctly solved questions, while the other group answered the same questions with visualization with an average score of $0.65 \pm 0.26$ with an average time of $2.18 \pm 1.02$ minutes. During part B, the mean scores with and without visualization were $0.55 \pm 0.17$ and $0.62 \pm 0.12$, respectively, at average times for correctly solved questions of $1.46 \pm 0.67$ and $1.68 \pm 1.12$ minutes. When combining parts A and B of the survey, the average scores were $0.64 \pm 0.13$ without \ac{cn} and $0.64 \pm 0.12$ with \ac{cn}, at $1.48 \pm 0.51$ and $1.98 \pm 1.05$ minutes.

While group math-vis, that was assigned questions without \ac{cn} first, scored higher than group vis-math on the X gate and H gate questions in survey part A with, on average, $0.9 \pm 0.1$ to $0.3 \pm 0.1$, the inverse is apparent for the Separability and Entanglement questions, where group vis-math scored higher than group math-vis 1.0 to $0.7 \pm 0.1$. Group vis-math, with \ac{cn}, took longer to correctly solve the X gate, H gate, Measurement Prob., Separability, Entanglement, and CNOT gate questions with $132 \pm 61$ to $86 \pm 32$ seconds and took less time on the Guess-the-Gates questions with 140 to 125 seconds. 

In survey part B, group math-vis (that was now given questions with \ac{cn}) scored higher than group vis-math on the X gate, H gate, Measurement Prob., and Guess-the-Gates questions with respective average scores $0.65 \pm 0.09$ to $0.45 \pm 0.09$ and group vis-math scored higher than group math-vis on the separability, entanglement, and CNOT gate questions with $0.73 \pm 0.09$ to $0.53 \pm 0.09$. In this survey part, group math-vis, with \ac{cn}, took longer on the Measurement Prob., entanglement, and CNOT gate questions with, on average, $120 \pm 23$ to $65 \pm 9$ seconds while taking less time on the X gate, H gate, separability, and Guess-the-Gates questions with $102 \pm 38$ to $65 \pm 22$ seconds on these questions on average.

Comparing within each group, group math-vis scored worse with \ac{cn} than without \ac{cn} on the X gate, H gate, entanglement, and CNOT gate questions with $0.85 \pm 0.09$ to $0.6 \pm 0.14$. When solving questions correctly, group math-vis took longer on the X gate, H gate, Measurement Prob., and CNOT gate questions when presented with \ac{cn}, with $102 \pm 38$ to $65 \pm 22$ seconds, while taking a shorter amount of time on the separability, entanglement, and Guess-the-Gates questions with $120 \pm 23$ to $65 \pm 9$ seconds.

When presented with \ac{cn}, group vis-math scored higher than when not presented with a visualization on the Measurement Prob., separability, entanglement, and Guess-the-Gates questions with, on average, with $0.8 \pm 0.2$ to $0.55 \pm 0.17$. The group scored higher when solving questions without \ac{cn} on the X gate and H gate questions with $0.5 \pm 0.1$ to $0.3 \pm 0.1$. They took longer with \ac{cn} when solving the Measurement Prob., separability, entanglement, and CNOT gate questions correctly, with an average of $155 \pm 70$ to $65 \pm 11$ seconds, and took less time on the X gate, H gate, and Guess-the-Gates questions when presented with \ac{cn}, with $134 \pm 8$ to $112 \pm 22$ seconds.

\subsubsection{Dimensional Circle Notation} During part A, participants assigned to the \ac{dcn} group solved questions without visualization with an average score of $0.57 \pm 0.23$ with an average time of $2.39 \pm 0.64$ minutes for correctly solved questions, while the same questions with visualization were answered with an average score of $0.63 \pm 0.25$ with an average time of $2.68 \pm 1.65$ minutes. During part B, the average scores were $0.54 \pm 0.26$ and $0.45 \pm 0.26$, respectively, at average times for correctly solved questions of $2.44 \pm 1.87$ and $3.71 \pm 2.27$ minutes. When combining the survey parts A and B, the average scores were $0.56 \pm 0.2$ without visualization and $0.53 \pm 0.2$ with visualization, at $2.95 \pm 2.44$ and $4.64 \pm 6.26$ minutes.

In both survey parts, on the X gate, Separability, and Guess-the-Gates questions, group math-vis scored higher than group vis-math. In part A, this was $0.72 \pm 0.21$ to $0.47 \pm 0.25$ and in part B, with $0.61 \pm 0.31$ to $0.47 \pm 0.34$. On the other hand, group vis-math scored higher than group math-vis on the H gate, Measurement Prob., Entanglement, and CNOT gate questions $0.75 \pm 0.17$ to $0.46 \pm 0.18$ in part A and $0.6 \pm 0.14$ to $0.33 \pm 0.12$ in part B. 

In survey part A, group math-vis took longer on the H gate, Measurement Prob., Entanglement, CNOT gate, and Guess-the-Gates questions with $158 \pm 20$ to $99 \pm 47$ seconds and less time on the X gate and Separability questions with $145 \pm 39$ to $107 \pm 39$ seconds. In survey part B, group math-vis took longer on the X gate, Measurement Prob., Entanglement, and CNOT gate questions with $170 \pm 83$ to $90 \pm 9$ seconds and group vis-math took longer on the H gate and Separability questions with $259 \pm 111$ to $227 \pm 126$ seconds.

Group math-vis had a higher score on the separability questions with \ac{dcn} with $0.83$ to $0.5$ while performing worse with \ac{dcn} on the X gate, Measurement Prob., Entanglement, CNOT, and Guess-the-Gates questions with $0.67 \pm 0.18$ to $0.43 \pm 0.23$. The group took longer with \ac{dcn} when correctly solving the X gate, H gate, Measurement Prob., Entanglement, and Guess-the-Gates questions with $271 \pm 119$ to $143 \pm 42$ seconds and took less time on the Separability and CNOT questions with $146 \pm 0$ to $103 \pm 3$ seconds. 

Supported by \ac{dcn}, group vis-math scored higher on the H gate, Entanglement, CNOT, and Guess-the-Gates questions with $0.65 \pm 0.3$ to $0.45 \pm 0.3$ while scoring worse on the Separability question with $0.6$ to $0.4$. On the X gate, Measurement Prob., Separability, and CNOT questions, group vis-math took longer with \ac{dcn}, with $140 \pm 40$ to $106 \pm 26$ seconds, while taking longer without \ac{dcn} on the H gate and Entanglement questions with $227 \pm 143$ to $56 \pm 28$ seconds.

\subsection{Three-qubit survey}\label{sec:A_results_three}

\begin{figure*}
    \centering
    \includegraphics[width=\linewidth]{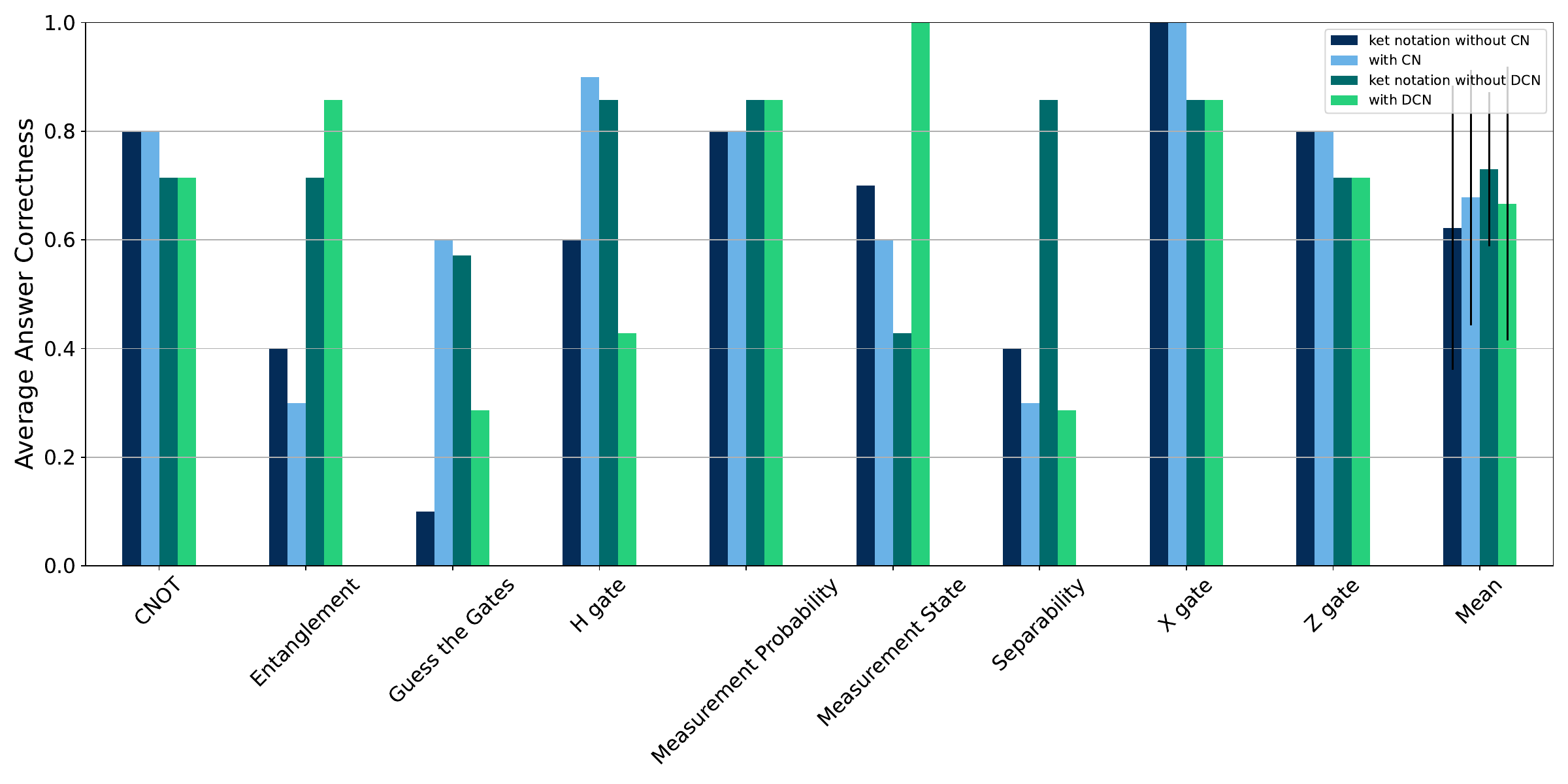}
    \caption{Average score for each question in the three-qubit survey, separated by participants who where presented with \ac{cn} (blue, N=10) and participants who where presented with \ac{dcn} (green, N=7). Darker colors are for questions where participants were not presented with a visualization and lighter colors for questions where participants were presented with a visualization.}
    \label{fig:score_3}
\end{figure*}

\begin{figure*}
    \centering
    \includegraphics[width=\linewidth]{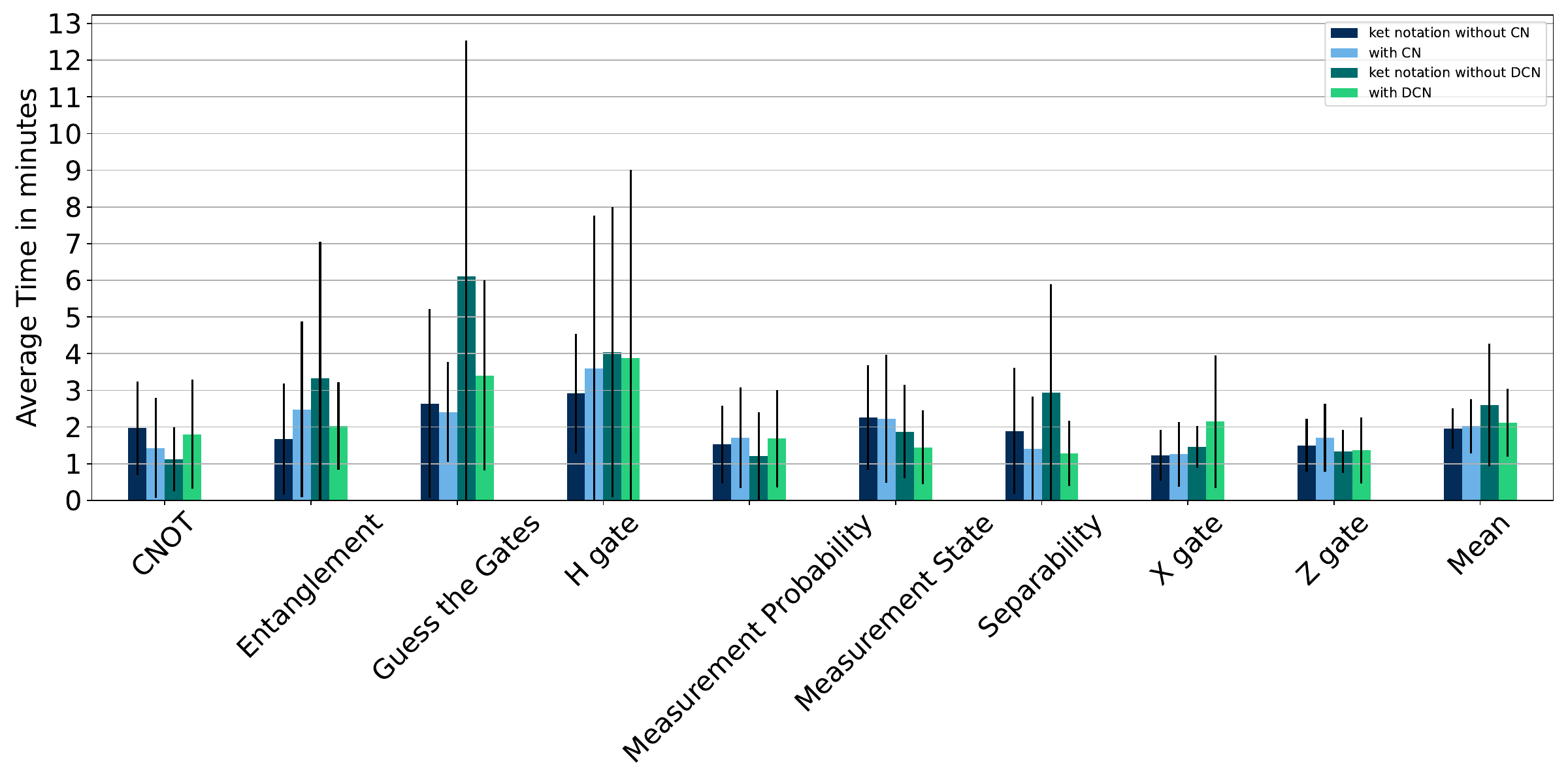}
    \caption{Average time taken for each question in minutes in the three-qubit survey, separated by participants who where presented by \ac{cn} (blue, N=10) and participants who where presented with \ac{dcn} (green, N=5). Darker colors are for questions where participants were not presented with a visualization and lighter colors for questions where participants were presented with a visualization.}
    \label{fig:time_3}
\end{figure*}

Table \ref{tab:3qubit_correctness} lists the average scores and Table \ref{tab:3qubit_time} lists the average time taken for correctly solved one-qubit questions, with and without visualization, for surveys that were taken supported by \ac{cn} and \ac{dcn}. Figure \ref{fig:score_3} shows the combined average score per question and Figure \ref{fig:time_3} shows the combined average time taken (for all answers, not just the correct ones). Cases in which a participant took more than 20 minutes for one question were excluded.

\begin{table*}
\caption{Average scores for three-qubit questions, with and without visualization (N=17).}\label{tab:3qubit_correctness}
\begin{tabular*}{\linewidth}{@{\extracolsep\fill}lcccccc}
\toprule%
 & \multicolumn{6}{@{}c@{}}{\textbf{Circle Notation}\footnotemark[1]} \\
 \midrule%
\textbf{survey part:} &  \multicolumn{2}{@{}c@{}}{A} & \multicolumn{2}{@{}c@{}}{\textbf{B}} & \multicolumn{2}{@{}c@{}}{\textbf{combined}} \\
\cmidrule(r{0.75em}){1-1}\cmidrule(lr{0.75em}){2-3}\cmidrule(lr{0.75em}){4-5}\cmidrule(lr{0.75em}){6-7}
\textbf{question} & \textbf{w/o vis.\footnotemark[2]} & \textbf{with vis.\footnotemark[2]} & \textbf{w/o vis.\footnotemark[2]} & \textbf{with vis.\footnotemark[2]} & \textbf{w/o vis.\footnotemark[1]} & \textbf{with vis.\footnotemark[1]} \\
\midrule%
\textbf{X gate} & 1.0 & 1.0 & 1.0 & 1.0 & 1.0 & 1.0 \\
\textbf{Z gate} & 0.6 & 1.0 & 1.0 & 0.6 & 0.8 & 0.8 \\
\textbf{H gate} & 0.6 & 0.8 & 0.6 & 1.0 & 0.6 & 0.9 \\
\textbf{Measurement Prob.} & 0.6 & 1.0 & 1.0 & 0.6 & 0.8 & 0.8 \\
\textbf{Measurement State} & 0.6 & 0.6 & 0.8 & 0.6 & 0.7 & 0.6 \\
\textbf{Separability} & 0.4 & 0.4 & 0.4 & 0.2 & 0.4 & 0.3 \\
\textbf{Entanglement} & 0.6 & 0.2 & 0.2 & 0.4 & 0.4 & 0.3 \\
\textbf{CNOT gate} & 0.8 & 1.0 & 0.8 & 0.6 & 0.8 & 0.8 \\
\textbf{Guess-the-Gates} & 0.0 & 0.8 & 0.2 & 0.4 & 0.1 & 0.6 \\
\midrule%
\textbf{Mean} & 0.58$\pm$0.26 & 0.76$\pm$0.28 & 0.67$\pm$0.31 & 0.6$\pm$0.25 & 0.62$\pm$0.28 & 0.68$\pm$0.25 \\
\toprule%
 & \multicolumn{6}{@{}c@{}}{\textbf{Dimensional Circle Notation}\footnotemark[2]} \\
 \midrule%
\textbf{survey part:} &  \multicolumn{2}{@{}c@{}}{\textbf{A}} & \multicolumn{2}{@{}c@{}}{\textbf{B}} & \multicolumn{2}{@{}c@{}}{\textbf{combined}} \\
\cmidrule(r{0.75em}){1-1}\cmidrule(lr{0.75em}){2-3}\cmidrule(lr{0.75em}){4-5}\cmidrule(lr{0.75em}){6-7}
\textbf{question} & \textbf{w/o vis.\footnotemark[3]} & \textbf{with vis.\footnotemark[4]} & \textbf{w/o vis.\footnotemark[4]} & \textbf{with vis.\footnotemark[3]} & \textbf{w/o vis.\footnotemark[2]} & \textbf{with vis.\footnotemark[2]} \\
\midrule%
\textbf{X gate} & 1.0 & 0.75 & 0.75 & 1.0 & 0.86 & 0.86 \\
\textbf{Z gate} & 1.0 & 0.75 & 0.5 & 0.67 & 0.71 & 0.71 \\
\textbf{H gate }& 1.0 & 0.25 & 0.75 & 0.67 & 0.86 & 0.43 \\
\textbf{Measurement Prob.} & 1.0 & 0.75 & 0.75 & 1.0 & 0.86 & 0.86 \\
\textbf{Measurement State} & 0.33 & 1.0 & 0.5 & 1.0 & 0.43 & 1.0 \\
\textbf{Separability} & 1.0 & 0.5 & 0.75 & 0.0 & 0.86 & 0.29 \\
\textbf{Entanglement} & 0.67 & 0.75 & 0.75 & 1.0 & 0.71 & 0.86 \\
\textbf{CNOT gate} & 1.0 & 0.75 & 0.5 & 0.67 & 0.71 & 0.71 \\
\textbf{Guess-the-Gates} & 0.33 & 0.25 & 0.75 & 0.33 & 0.57 & 0.29 \\
\midrule%
\textbf{Mean} & 0.81$\pm$0.28 & 0.64$\pm$0.24 & 0.67$\pm$0.12 & 0.7$\pm$0.33 & 0.73$\pm$0.14 & 0.67$\pm$0.25 \\
\botrule%
\end{tabular*}
\footnotetext[0]{Note: Standard deviations are used for confidence intervals.}
\footnotetext[1]{N = 10}
\footnotetext[2]{N = 7}
\footnotetext[3]{N = 3}
\footnotetext[4]{N = 4}
\end{table*}

\begin{table*}
\caption{Average time taken for correct answers to three-qubit questions in minutes (N=15).}\label{tab:3qubit_time}
\begin{tabular*}{\linewidth}{@{\extracolsep\fill}lcccccccccccc}
\toprule%
 & \multicolumn{6}{@{}c@{}}{\textbf{Circle Notation}\footnotemark[1]} \\
 \midrule%
 \textbf{survey part:} &  \multicolumn{2}{@{}c@{}}{\textbf{A}} & \multicolumn{2}{@{}c@{}}{\textbf{B}} & \multicolumn{2}{@{}c@{}}{\textbf{combined}} \\
\cmidrule(r{0.75em}){1-1}\cmidrule(lr{0.75em}){2-3}\cmidrule(lr{0.75em}){4-5}\cmidrule(lr{0.75em}){6-7}
\textbf{question} & \textbf{w/o vis.\footnotemark[2]} & \textbf{with vis.\footnotemark[2]} & \textbf{w/o vis.\footnotemark[2]} & \textbf{with vis.\footnotemark[2]} & \textbf{w/o vis.\footnotemark[1]} & \textbf{with vis.\footnotemark[1]} \\
\midrule%
\textbf{X gate} & 0.94$\pm$0.36(5) & 1.46$\pm$1.16(5) & 1.52$\pm$0.82(5) & 1.05$\pm$0.34(5) & 1.23$\pm$0.7(10) & 1.26$\pm$0.88(10) \\
\textbf{Z gate} & 1.03$\pm$0.04(3) & 2.02$\pm$1.18(5) & 1.53$\pm$0.87(5) & 1.17$\pm$0.29(3) & 1.35$\pm$0.73(8) & 1.7$\pm$1.04(8) \\
\textbf{H gate} & 3.72$\pm$1.74(3) & 6.82$\pm$5.99(4) & 3.64$\pm$1.59(3) & 1.85$\pm$0.53(5) & 3.68$\pm$1.67(6) & 4.79$\pm$5.26(6) \\
\textbf{Measurement Prob.} & 1.42$\pm$0.66(3) & 2.7$\pm$1.17(5) & 1.27$\pm$0.87(5) & 0.44$\pm$0.2(3) & 1.32$\pm$0.8(8) & 1.85$\pm$1.43(8) \\
\textbf{Measurement State} & 1.92$\pm$0.1(3) & 3.53$\pm$1.2(3) & 2.75$\pm$1.71(4) & 1.28$\pm$0.44(3) & 2.14$\pm$1.31(6) & 2.41$\pm$1.44(6) \\
\textbf{Separability} & 1.19$\pm$0.36(2) & 3.1$\pm$1.92(2) & 2.18$\pm$0.28(2) & 1.27(1) & 2.01$\pm$0.46(2) & 1.23$\pm$0.04(2) \\
\textbf{Entanglement} & 2.53$\pm$2.24(3) & 2.1(1) & 2.27(1) & 1.16$\pm$0.14(2) & 3.07$\pm$1.89(3) & 1.47$\pm$0.46(3) \\
\textbf{CNOT gate} & 1.41$\pm$0.46(4) & 1.68$\pm$1.72(5) & 1.6$\pm$0.94(4) & 1.06$\pm$0.04(3) & 1.53$\pm$0.8(7) & 1.48$\pm$1.49(7) \\
\textbf{Guess-the-Gates} & (0) & 3.59$\pm$1.35(4) & 5.13(1) & 1.98$\pm$0.42(2) & 5.13(1) & 5.13(1) \\
\midrule%
\textbf{Mean} & 1.77$\pm$0.94 & 3.0$\pm$1.63 & 2.43$\pm$1.26 & 1.25$\pm$0.45 & 2.38$\pm$1.33 & 2.37$\pm$1.51 \\
\toprule%
 & \multicolumn{6}{@{}c@{}}{\textbf{Dimensional Circle Notation}\footnotemark[2]} \\
\midrule%
\textbf{survey part} &  \multicolumn{2}{@{}c@{}}{\textbf{A}} & \multicolumn{2}{@{}c@{}}{\textbf{B}} & \multicolumn{2}{@{}c@{}}{\textbf{combined}} \\
\cmidrule(r{0.75em}){1-1}\cmidrule(lr{0.75em}){2-3}\cmidrule(lr{0.75em}){4-5}\cmidrule(lr{0.75em}){6-7}
\textbf{question} & \textbf{w/o vis.\footnotemark[3]} & \textbf{with vis.\footnotemark[4]} & \textbf{w/o vis.\footnotemark[4]} & \textbf{with vis.\footnotemark[3]} & \textbf{w/o vis.\footnotemark[2]} & \textbf{with vis.\footnotemark[2]} \\
\midrule%
\textbf{X gate} & 1.83$\pm$0.42(2) & 3.11$\pm$2.38(2) & 1.49$\pm$0.43(2) & 1.94$\pm$0.69(2) & 1.66$\pm$0.45(4) & 2.52$\pm$1.84(4) \\
\textbf{Z gate} & 1.23$\pm$0.55(2) & 0.97$\pm$0.1(2) & 1.95(1) & 1.88$\pm$1.25(2) & 1.47$\pm$0.56(3) & 1.61$\pm$1.09(3) \\
\textbf{H gate} & 8.02$\pm$3.47(2) & (0) & 1.79$\pm$0.51(2) & 1.32(1) & 4.55(1) & 1.32(1) \\
\textbf{Measurement Prob.} & 2.43$\pm$1.0(2) & 1.12$\pm$0.04(2) & 0.52$\pm$0.09(2) & 2.35$\pm$1.9(2) & 1.48$\pm$1.19(4) & 1.74$\pm$1.48(4) \\
\textbf{Measurement State} & 1.05(1) & 1.08$\pm$0.2(3) & 1.77(1) & 2.0$\pm$1.4(2) & 1.41$\pm$0.36(2) & 0.83$\pm$0.23(2) \\
\textbf{Separability} & 3.66$\pm$2.66(2) & 1.57(1) & 3.51$\pm$3.28(2) & (0) & 6.78(1) & 1.57(1) \\
\textbf{Entanglement} & 6.12$\pm$4.09(2) & 1.88$\pm$0.52(2) & 2.1$\pm$1.93(2) & 2.76$\pm$1.43(2) & 4.11$\pm$3.78(4) & 2.32$\pm$1.16(4) \\
\textbf{CNOT gate} & 0.89$\pm$0.18(2) & 1.55$\pm$0.62(2) & 2.68(1) & 4.52(1) & 1.88$\pm$0.81(2) & 2.72$\pm$1.79(2) \\
\textbf{Guess-the-Gates} & 17.48(1) & 7.28(1) & 2.18$\pm$1.78(2) & 1.52(1) & 3.97(1) & 7.28(1) \\
\midrule%
\textbf{Mean} & 4.75$\pm$5.37 & 2.32$\pm$2.12 & 2.0$\pm$0.82 & 2.29$\pm$1.01 & 3.03$\pm$1.91 & 2.44$\pm$1.92 \\
\botrule%
\end{tabular*}
\footnotetext[0]{Note: The number of participants N that solved the question correctly and are therefore included in the average is shown in brackets. Standard deviations are shown where N $>1$. The confidence interval of the mean is the mean of the standard deviations.}
\footnotetext[1]{N = 10}
\footnotetext[2]{N = 5}
\footnotetext[3]{N = 2}
\footnotetext[4]{N = 3}
\end{table*} 

\subsubsection{Circle Notation} During part A, participants assigned to the \ac{cn} group solved questions without visualization with an average score of $0.58 \pm 0.26$ and an average time of $1.77 \pm 0.94$ minutes for correctly solved questions, while the same questions with visualization were answered with an average score of $0.76 \pm 0.28$ and an average time of $3.0 \pm 1.63$ minutes. During part B, the average scores were $0.67 \pm 0.31$ and $0.6 \pm 0.25$, respectively, at average times for correctly solved questions of $2.43 \pm 1.26$ and $1.25 \pm 0.45$ minutes. When combining parts A and B of the survey, the average scores were $0.62 \pm 0.28$ without visualization and $0.68 \pm 0.25$ with visualization, at $2.38 \pm 1.33$ and $2.37 \pm 1.51$ minutes.

In survey part A, group math-vis, that was presented questions without \ac{cn} first, scored higher than group vis-math on the Entanglement question with 0.6 to 0.2. Group vis-math scored higher on the Z gate, the H gate, the Measurement Prob., the CNOT, and the Guess-the-Gates questions with $0.92 \pm 0.1$ to $0.52 \pm 0.27$. In this survey part, group math-vis took longer when solving the Entanglement question correctly, with 152 to 126 seconds, while group vis-math took longer for correct solutions to the X gate, Z gate, H gate, Measurement Prob., Measurement State, Separability, and CNOT gate questions, with $183 \pm 101$ to $100 \pm 53$ seconds.

In survey part B, group math-vis scored higher on the H gate, entanglement, and Guess-the-Gates questions with $0.6 \pm 0.28$ to $0.33 \pm 0.19$, while group vis-math scored higher on the Z gate, Measurement Prob., Measurement State, Separability, and CNOT gate questions with $0.8 \pm 0.22$ to $0.52 \pm 0.16$. Group vis-math took longer on all questions in survey part B with $146 \pm 71$ to $75 \pm 26$ seconds. 

Group math-vis, that was presented questions without \ac{cn} first, scored higher on the H gate and Guess-the-Gates questions with $0.7 \pm 0.3$ to $0.3 \pm 0.3$ with \ac{cn}. The group scored worse with \ac{cn} on the separability, entanglement, and CNOT gate questions by a score difference of $0.6 \pm 0.16$ to$ 0.4 \pm 0.16$. With \ac{cn}. Participants in the group took longer when solving the X gate, Z gate and Separability questions correctly, $70 \pm 5$ to $63 \pm 6$ seconds, while taking less time on the H gate, Measurement Prob., Measurement State, Entanglement, CNOT gate, and Guess-the-Gates questions with$ 132 \pm 52$ to $69 \pm 27$ seconds.

Group vis-math, that was presented questions with \ac{cn} first, scored higher on the H gate, CNOT gate, and Guess-the-Gates questions with \ac{cn} with $0.87 \pm 0.09$ to $0.53 \pm 0.25$. The group scored lower with \ac{cn} on the Measurement State question with 0.6 as compared to 0.8. When presented with \ac{cn}, participants in the group took longer on the Z gate, H gate, Measurement Prob., Measurement State, Separability, and CNOT gate questions, with $198 \pm 101$ to $130 \pm 49$ seconds. They took less time on the X gate, entanglement, and Guess-the-Gates questions, with $178 \pm 93$ to $143 \pm 53$ seconds. 

\subsubsection{Dimensional Circle Notation}
During part A, participants assigned to the \ac{dcn} group solved questions without visualization with an average score of $0.89 \pm 0.21$ with an average time of $4.75 \pm 1.77$ minutes for correctly solved questions, while the same questions with visualization were answered with an average score of 0.6$\pm$0.24 with an average time of $2.32 \pm 0.63$ minutes. During part B, the average scores were $0.56 \pm 0.16$ and $0.72 \pm 0.34$, respectively, at average times for correctly solved questions of $2.0 \pm 1.34$ and $2.29 \pm 1.33$ minutes. When combining parts A and B of the survey, the average scores were $0.69 \pm 0.15$ without visualization and $0.62 \pm 0.29$ with visualization, at $3.03 \pm 0.78$ and $2.44 \pm 0.9$ minutes.

Group math-vis, that was assigned questions without \ac{dcn} first, scored higher than group vis-math on the X gate, Z gate, H gate, Measurement Prob., separability, entanglement, CNOT, and Guess-the-Gates questions in survey part A, with $0.94 \pm 0.17$ to $0.5 \pm 0.24$. The inverse is apparent for the Measurement State question, where group vis-math scored higher than group math-vis with 1 to 0.5. Group vis-math, with \ac{dcn}, took longer to correctly solve the Z gate, Measurement Prob., Separability, Entanglement, and Guess-the-Gates questions, with $371 \pm 353$ to $154 \pm 143$ seconds, whereas group math-vis took longer for correctly solving the X gate, Measurement State, and CNOT gate questions, $115 \pm 52$ to $76 \pm 25$ seconds.

In survey part B, group math-vis (that now was presented questions with \ac{dcn}) scored higher than group vis-math on the X gate, Z gate, Measurement Prob., Measurement State, Entanglement, and CNOT gate questions with $0.92 \pm 0.19$ to $0.5 \pm 0.17$ and group vis-math scored higher than group math-vis on the H gate, separability, and Guess-the-Gates questions with $0.67 \pm 0.0$ to $0.33 \pm 0.24$. In this survey part, group math-vis, with \ac{dcn}, took longer on the X gate, Measurement Prob., Measurement State, Entanglement, and CNOT gate questions, with $163 \pm 57$ to $103 \pm 43$ seconds, while taking less time on the Z gate, H gate, and Guess-the-Gates questions, with $118 \pm 10$ to $94 \pm 14$ seconds.

Comparing within each group, group math-vis scored higher with \ac{dcn} on the Measurement State question with 1 to 0.5, and scored lower with \ac{dcn} than without \ac{dcn} on the H gate, Separability, and CNOT gate questions by averages of $1.0 \pm 0.0$ to $0.33 \pm 0.24$. When solving questions correctly, group math-vis took longer on the X gate, Z gate, Measurement State, and CNOT gate questions when presented with \ac{dcn}, with $155 \pm 67$ to $75 \pm 21$ seconds, while taking less time on the H gate, Measurement Prob., Entanglement, and Guess-the-Gates questions, $511 \pm 333$ to $119 \pm 35$ seconds.

When presented with \ac{dcn}, group vis-math scored higher than itself on the Z gate, Measurement State, and CNOT gate questions by, on average, $0.78 \pm 0.16$ to $0.33 \pm 0.0$. The group scored higher when solving questions without \ac{dcn} on the H gate, Separability, and Guess-the-Gates questions with $0.67 \pm 0.0$ to $0.22 \pm 0.16$. They took longer with \ac{dcn} when solving the X gate, Measurement Prob., and Guess-the-Gates questions correctly, $230 \pm 154$ to $84 \pm 41$ seconds, and took less time on the Z gate, Measurement State, separability, Entanglement, and CNOT gate questions when presented with \ac{dcn}, on average $144 \pm 38$ to $85 \pm 20$ seconds.

\clearpage

\end{document}